\documentclass[aps,prd,twocolumn,preprintnumbers,showpacs,nofootinbib,amssymb]{revtex4}
\usepackage{graphicx}
\usepackage{amsmath}
\usepackage{amssymb}
\usepackage{amsfonts}
\usepackage{bm}
\usepackage{color}

\def\be{\begin{equation}}
\def\ee{\end{equation}}
\def\bea{\begin{eqnarray}}
\def\eea{\end{eqnarray}}

\begin{document}

\title{Caloric curves of classical self-gravitating systems in general
relativity}
\author{Giuseppe Alberti}
\affiliation{Laboratoire de Physique
Th\'eorique, Universit\'e de Toulouse,
CNRS, UPS, France}
\affiliation{
Living Systems Research, Roseggerstra\ss e 27/2, A-9020 Klagenfurt am
W\"{o}rthersee,
Austria}
\author{Pierre-Henri Chavanis}
%\email{chavanis@irsamc.ups-tlse.fr}
\affiliation{Laboratoire de Physique
Th\'eorique, Universit\'e de Toulouse,
CNRS, UPS, France}

\begin{abstract} 

We determine the caloric curves of classical self-gravitating
systems at statistical equilibrium in general relativity.  In the
classical limit, the caloric curves of a self-gravitating gas
depend on a
unique parameter $\nu=GNm/Rc^2$, called the compactness parameter, where $N$ is
the particle number and
$R$ the system's size. Typically, the caloric curves have the
form
of a double spiral.
The ``cold spiral'', corresponding to weakly relativistic configurations, is a
generalization of the caloric curve of nonrelativistic classical
self-gravitating
systems. The ``hot spiral'', corresponding to  strongly relativistic
configurations, is similar (but not identical) to the caloric
curve of the ultrarelativistic self-gravitating black-body radiation. We
introduce two types of
normalization of  energy and temperature in order
to obtain asymptotic  caloric curves
describing respectively the cold and the hot spirals in the limit
$\nu\rightarrow
0$. As the number of particles increases, the cold and the
hot spirals approach each other, merge at $\nu'_S=0.128$, form a loop above
$\nu_S=0.1415$, reduce to a point at $\nu_{\rm max}=0.1764$, and finally
disappear. Therefore, the double spiral shrinks when the compactness parameter
$\nu$ increases, implying that
general relativistic effects render the system more unstable. We discuss the
nature of the gravitational collapse at low and high energies
with respect to a dynamical (fast) or a thermodynamical (slow) instability.

\end{abstract}


\maketitle

\section{Introduction}

The statistical mechanics of self-gravitating systems has a rich history. In
this introduction (see also \cite{rgb}),
we review the main developments of this subject, restricting
ourselves to classical self-gravitating systems (the case
of self-gravitating
fermions is reviewed in \cite{acf,rgf} and the case of
self-gravitating bosons is reviewed in \cite{prd1}). This review is useful
because no history of this important topic has been given previously. We
successively consider
nonrelativistic
stellar systems and general relativistic star clusters.

The statistical mechanics of nonrelativistic stellar systems started with
the work of Chandrasekhar \cite{chandrabook} (see Appendix
\ref{sec_hgc} for the pre-history of the subject). He
developed a
kinetic theory   of  stellar
systems in order to determine their time of relaxation due to
gravitational two-body close
encounters. He obtained an expression of the form $t_{\rm coll}\sim
(N/\ln N)t_{\rm D}$, where $N$ is the number of stars in the system
and $t_{\rm D}$ is the dynamical time.\footnote{An estimate of the time of
relaxation of a star cluster was
previously obtained by Jeans 
\cite{jeanskin,jeanskin2,jeansotherbook,jeansbook}, Eddington \cite{book1914},
Charlier
\cite{charlier}, Schwarzschild 
\cite{schwarzschild}, Rosseland \cite{rosseland},
Smart \cite{smart}, Ambartsumian \cite{ambartsumian}, Mineur
\cite{mineur}, and Spitzer \cite{spitzer}.}
For galaxies, containing a large number of
stars ($N\sim 10^{12}$), the relaxation time is much larger than the
age of the Universe so
these systems are essentially collisionless, described by the
Vlasov (or collisionless Boltzmann) equation.\footnote{This equation was
introduced by Jeans \cite{jeansth} in the context of stellar
systems and by Vlasov \cite{vlasov} for Coulombian plasmas. See 
H\'enon \cite{henonvlasov} for some comments about the name that should be given
to this equation.} By contrast, for globular clusters
containing a small number of stars ($N\sim 10^6$), the time of relaxation is of
the order of their age so
they should be close to a state of statistical
equilibrium.\footnote{The important difference between the
relaxation
time of galaxies and globular clusters was first pointed
out by Jeans  \cite{jeanskin}.} Statistical
equilibrium may also be established in the central regions of galaxies where the
density is high. 
Therefore, Chandrasekhar \cite{chandrabook} assumed that globular clusters
(and the core of galaxies) are
in a statistical equilibrium state described by the Maxwell-Boltzmann
distribution in which the gravitational potential is determined by the Poisson
equation. This leads
to the Boltzmann-Poisson equation which is equivalent to Emden's equation
describing the hydrostatic equilibrium of an isothermal gas sphere
\cite{emden,chandrabook2}.
However, Chandrasekhar \cite{chandrabook} noted that the isothermal
approximation cannot be used for a complete description of a globular cluster
because it predicts an infinite mass.\footnote{This problem was previously
pointed out by Eddington 
\cite{eddington2,eddington3}, Jeans  \cite{jeanskin2}, 
Heckmann and Siedentopf \cite{hs}, Dicke
\cite{dicke}, and Spitzer \cite{spitzer}.}
  Indeed,
the density of a self-gravitating isothermal sphere decreases
as $r^{-2}$ at large distances \cite{zollner,ritteriso}. Therefore, the
isothermal approximation is
inadequate for describing the outer regions of a globular cluster. Statistical
equilibrium does not hold anymore at low densities because the relaxation time
increases so the system does not reach a statistical equilibrium state. On the
other hand, there is a continual loss of stars by escape and
this phenomenon is important in the outer regions (stars with an energy
larger than the escape energy leave the cluster). The continuous loss of
stars
from the system leads to a contraction and to a gradual disintegration of the
system as pointed out by Jeans \cite{jeanskin2}, Ambartsumian
\cite{ambartsumian}, Dicke \cite{dicke}, and Spitzer
\cite{spitzer}. In this sense, there is
no
statistical
equilibrium state for self-gravitating systems.   However, evaporation occurs on
a
long timescale, much larger
than the relaxation time. Therefore, the escape of stars is a slow process and
the system can reach a quasiequilibrium state, close to the Maxwell-Boltzmann
distribution, on an intermediate timescale. The effect of evaporation is to
change
the Maxwell-Boltzmann distribution close to the escape
energy.\footnote{The modification of the distribution
function due to the
escape of stars was considered later by Woolley \cite{woolley}
(following Eddington \cite{eddington3}),
Michie \cite{michie} and King
\cite{kingVP}.} Using an analogy between
stellar
dynamics and Brownian motion, Chandrasekhar
\cite{chandra1,chandra2,chandra3} derived a kinetic equation for stellar systems
of the
Fokker-Planck form\footnote{The kinetic theory of stellar systems is
reviewed in \cite{aakin}.}  and used it to study in greater detail the
relaxation of the
system towards statistical equilibrium and the escape of stars from globular
clusters (see also White \cite{white}, Spitzer and H\"arm
\cite{sh}, von Hoerner \cite{hoernercommeKing}, King
\cite{king1,king2,king3,king4,king5,kingVP,king}, H\'enon
\cite{henon60,henon61,henon65,henon69}, Michie
\cite{michieA,michie,mb,michieZ},  Miller and Parker \cite{mp}, and Spitzer
and
Saslaw \cite{ss}). King \cite{king2} suggested that a globular cluster, as a
result of
evaporation and contraction, may end as a binary
star.

Ogorodnikov \cite{ogo1,ogo2} looked for the most probable distribution of
stars in a galaxy by using methods of statistical
mechanics.
This amounts to maximizing the Boltzmann entropy at fixed energy
and particle number.\footnote{Ogorodnikov \cite{ogo1,ogo2} argued that the
relaxation time
of galaxies must be of the order of the dynamical time. As a result, in his
point of view, the
Maxwell-Boltzmann distribution (most probable state) cannot be established by
two-body encounters (like in Ref. \cite{chandrabook}) but by a relaxation
process of a
different
nature that he left unspecified. This timescale problem was solved later by
Lynden-Bell \cite{lb} in his theory of  violent collisionless relaxation.} He
only
considered the first order variational problem (extremization of entropy) and
derived the
mean
field
Maxwell-Boltzmann distribution in which the gravitational potential is
determined self-consistently by the Poisson equation. Like previous authors, he
was
careful to note that this distribution function is
not valid at large distances (outside the main body) and for high velocities
(greater than the escape velocity) where the phase-space density is low. As a
result, the Maxwell-Boltzmann distribution should be truncated at large energies
in order to take into account the escape of high energy stars. This is how
he proposed to solve the infinite mass problem.

The complete statistical mechanics problem initiated by Ogorodnikov
\cite{ogo1,ogo2} was solved by Antonov \cite{antonov}. He neglected the escape
of stars, proceeding as if the
stars were confined within a sphere of radius $R$ with reflecting
boundary.   The physical reason advocated is that the escape of
stars is
a slow process so that, on intermediate timescales, everything happens as if
the system were confined within a box. Thus, Antonov \cite{antonov} considered
the
problem of maximizing the Boltzmann entropy at fixed energy and particle
number within a box.   He showed that the Boltzmann entropy has no global
maximum (there is no fully stable equilibrium
state)\footnote{The entropy diverges if we concentrate a part of
the system and redistribute the released
potential energy in another part of the system under the form of
kinetic energy.} but that it may have a local maximum
(corresponding to a metastable equilibrium 
state) with the Maxwell-Boltzmann distribution. He showed that
a local entropy maximum is necessarily spherically
symmetric. Furthermore, by calculating the second order variations of
entropy, he showed that the density contrast of a star system with
Maxwellian distribution cannot exceed $709$ otherwise that
Maxwellian distribution is not a local entropy maximum. In that
case, the system evolves away from the Maxwell distribution and becomes
non-stationary.

Lynden-Bell \& Wood \cite{lbw} 
confirmed and extended the results of Antonov
\cite{antonov} by calculating the series of equilibria of
self-gravitating isothermal spheres using the results  known
in the context of stellar structure and described in the books of
Emden \cite{emden} and Chandrasekhar \cite{chandrabook2}.
Lynden-Bell \& Wood \cite{lbw} showed that there is no equilibrium state
if the energy is lower than a minimum value $E_c=-0.335\, GM^2/R$
corresponding to the density contrast of $709$ found by Antonov. 
In that case,
they argued that the system takes a core-halo structure and evolves away from
equilibrium: the center of the system contracts
and achieves very high temperatures and densities while the halo remains
cool. They explained this
runaway (Antonov's instability) in terms of the negative specific heat
($C=dE/dT<0$) of
self-gravitating systems: by losing heat, the core grows hotter, contracts, and
loses heat again to the profit of the halo in an unstoppable process. They
called this phenomenon the
``gravothermal
catastrophe''
(another proposed name was ``thermal runaway'' \cite{lbcnrs}).\footnote{Their
results prove that the escape of stars from a
cluster is not necessary for its evolution (as was believed before
their work) but rather that extended systems naturally grow a core-halo
structure reminiscent of the internal constitution of a red giant star. Even
when the system is confined within a box,
 there can be an evolution
away from the Boltzmann distribution
provided that the system is sufficiently centrally condensed.
This can lead to the formation of a small dense nucleus which
is to some extent independent of the outer parts of the system. This
corroborates previous works by von Hoerner \cite{hoerner1,hoerner2,hoernerTH},
H\'enon
\cite{henon61,henon65} and Aarseth \cite{aarseth}. In particular, H\'enon
\cite{henon61,henon65} (see also von Hoerner \cite{hoernerTH}) developed
an homologous model of globular clusters leading to an infinite
central density in finite time and suggested that the formation of
binary stars would occur at very high densities and that this would
produce a new energy source.}
They studied the thermodynamical stability of isothermal spheres by using the
Poincar\'e theory of linear series of equilibria
\cite{poincare}.\footnote{Poincar\'e \cite{poincare}
invented a powerful method for separating stable from unstable
equilibria. It is based on series of equilibrium configurations. The method was
used by Poincar\'e and many others (see for instance Jeans
\cite{jeansbook}, Lyttleton \cite{lyttleton} and Ledoux
\cite{ledouxbook}) to find stable equilibria of
rotating liquid
masses and rotating systems of rigid bodies.} This theory
tells us that the
instability arises at a turning point. In the microcanonical
ensemble
(entropy $S$ maximum at fixed energy $E$, mass $M$ and volume $V$), the series
of equilibria
becomes
unstable when the density contrast is larger than $709$, corresponding to
the first energy peak $E_c$. This is the
point where the specific heat vanishes ($C=0^{\mp}$), passing from 
negative to positive values. In the
canonical
ensemble (Helmholtz free energy $F=E-TS$ minimum at fixed temperature $T$,
mass $M$ and volume $V$), they showed that there is no
equilibrium state if the temperature is lower than a minimum value
$T_c=GMm/2.52k_B R$ corresponding to a density
contrast of $32.1$.\footnote{This result was actually discovered by Emden
\cite{emden} (see Chap. XI).} On the other hand, using the Poincar\'e criterion,
they showed that the series of equilibria
becomes
unstable when the density contrast is larger than $32.1$, corresponding to
the first temperature peak $T_c$. This is the point
where the specific heat becomes infinite ($C=\pm\infty$), passing from positive
to negative values.
They also considered other ensembles making the
connection with the earlier works of Ebert \cite{ebert,ebert2}, Bonnor
\cite{bonnor} and McCrea \cite{crea} (Gibbs free
energy $G=E-TS+PV$ minimum at fixed
temperature $T$, mass $M$ and  pressure $P$) 
and with the Sch\"onberg-Chandrasekhar \cite{sch} limit (corresponding
approximately to the canonical
ensemble). Therefore, the
onset of instability is different depending
whether the energy or the temperature are held fixed. This
corresponds to what is now called ensembles inequivalence for
systems with long-range interactions in statistical
mechanics. Lynden-Bell and Wood \cite{lbw} mentioned this inequivalence  (see
their footnote p. 509) and connected it with
the existence of negative specific heats that are allowed in the microcanonical
ensemble but not in the canonical ensemble (see Appendix
\ref{sec_nsh}).
 Finally, they related the absence
of global entropy maximum with
the formation of a subset of particles very closely bound together by
gravity, like a binary star. However, they mentioned that the formation of
binary stars is a rare event so we can consider ``frozen
equilibria''
(corresponding to metastable states) in which the number of binaries is
unchanged in the time available.

Thirring \cite{thirring} studied the statistical mechanics of
self-gravitating systems without being aware
of previous works on the subject (he mentioned the work of Lynden-Bell and Wood
\cite{lbw} as a {\it Note added in proof}). He emphasized the fact that
self-gravitating systems may have negative specific heats in the microcanonical
ensemble while the specific heat is necessarily positive in the canonical
ensemble (see Appendix \ref{sec_nsh}). He concluded therefore that the
ensembles are not equivalent.
He showed that the microcanonical entropy, defined as the logarithm of the
density of states, diverges if one particle is sent to infinity or if
two of them are approached at infinitely close distances. It is therefore
necessary to introduce a box and a small distance repulsion. Using a saddle
point approximation, he derived the
mean field Boltzmann distribution in which the gravitational potential is
produced by the system as a whole.

Horwitz and Katz \cite{hkpart1,hk3} (independently from
Thirring
\cite{thirring}) developed a field theory based on path
integrals to study the equilibrium statistical mechanics of stellar systems. 
Their approach uses a rigorous microcanonical formulation which
produces an exact functional integral expression for the
density of states and the entropy.\footnote{True
statistical equilibrium for particles interacting
gravitationally with an $r^{-1}$ law is impossible because the
statistical
integral diverges both when a particle moves out to infinity, as well as when
two particles approach one another indefinitely. The ultimate configuration,
reached for $t\rightarrow +\infty$, consists in a hard binary plus $N-2$ high
velocity stars. However, the evaporation of particles proceeds slowly and the
formation of binaries is a rare event. We can therefore consider near
equilibrium states of the system with given energy and particle number. 
Horwitz and Katz \cite{hkpart1,hk3} confined the
stars to a finite volume (box) in order to eliminate evaporation and 
introduced an
appropriate short distance cut-off in the interparticle interaction so as to
eliminate tightly bound pairs. This is necessary to make their
integrals convergent. However, at the level of the mean field
approximation, they showed that the short distance
cutoff can finally be taken to zero.} This integral is evaluated by
steepest-descent methods, the saddle-point value
giving the mean field entropy (Boltzmann entropy) and the analysis of quadratic
fluctuations yielding
conditions of stability. This provides an approximation which is mean field plus
fluctuations. Horwitz and Katz \cite{hkpart1,hk3}  determined sequences of
equilibrium states which  are
presumed to  simulate slowly evolving,
near-equilibrium, configurations of real star clusters. 
They considered different ensembles (microcanonical,
canonical and grand canonical)
and obtained stability criteria in the form
of eigenvalue equations involving the Schr\"odinger operator introduced by
Lynden-Bell and Sanitt \cite{lbs}. They also showed that nonradial
perturbations are stable in all ensembles. Using the Poincar\'e
\cite{poincare} criterion, they showed that instability occurs at the turning
point of an appropriate thermodynamic potential. Therefore,
the onset of
instabilities for spherical
perturbations can be associated with the sign change of standard thermodynamic
functions such as the heat capacity.\footnote{Generally, constraints act to
stabilize the system. The microcanonical
which constrains both
the energy and the mass is the most stable (with a critical density
constrast of $709$), while the grand canonical
ensemble
which allows both the energy and the mass to fluctuate (at fixed temperature
and
chemical potential) is the least stable (with a critical density
constrast of $1.58$). The canonical ensemble which
allows the
energy to fluctuate (at fixed temperature) but constrains the mass lies between
the two (with a critical density
constrast of $32.1$).}  Finally, they noted that
a system which is thermodynamically stable is also dynamically stable.

Katz \cite{katzpoincare1,katz2} generalized the turning point criterion of
Poincar\'e when there are more than one turning point and applied it
to thermodynamical problems thereby recovering the results of Lynden-Bell and
Wood
\cite{lbw} and Horwitz and Katz \cite{hk3}. The Poincar\'e criterion allows one
to determine the
thermodynamical stability of the system from topological properties of
continuous series of equilibria (provided stability conditions are
known for one configuration) without having to solve an eigenvalue equation. In
this connection, Katz \cite{katzpoincare1}  plotted
for the first time (by
hand) the caloric curve 
$\beta(E)$ giving the inverse temperature as a function of the energy
($\beta$
is the variable conjugate to $E$ with respect to $S$). This curve
has a striking
spiralling  (snail-like) shape. The series of equilibria is
parametrized in terms of the
density contrast
which grows as one spirals inwards. The series of
equilibria becomes unstable at the first energy minimum in the microcanonical
ensemble and at the first
temperature minimum in the canonical ensemble as previously showed by
Lynden-Bell and Wood
\cite{lbw}. However,
Katz \cite{katzpoincare1} showed that more and more modes of stability are lost 
(more and more
eigenvalues become
negative)
as one rotates
clockwise along the spiralling series of equilibria.

Lecar and Katz
\cite{lecarkatz} introduced a new ensemble, called the grand microcanonical
ensemble which constrains the energy but allows the mass to fluctuate (at
fixed chemical potential). Using the turning point
method they found that the
series of
equilibria is successively stable, unstable, stable again, and finally
unstable.

Padmanabhan \cite{paddy} wrote the first review on the statistical
mechanics of self-gravitating systems. In his review, and in Ref.
\cite{paddyapj}, he
presented a simplified derivation of the Antonov instability and gravothermal
catastrophe by explicitly solving the zero
eigenvalue equation associated with the second variations of the
entropy.\footnote{See also the previous works of Ebert \cite{ebert2},
Yabushita \cite{yabushitaNG,yabushitaNGE}, Taff and van Horn \cite{taffY,taff},
Nakada \cite{nakada}, Hachisu and Sugimoto \cite{hsu}, Hachisu {\it et al.}
\cite{hetal} and Inagaki \cite{inagaki}.}
This
solution allowed him to study the nature of the mode that triggers the
instability in the microcanonical ensemble. He proposed a very elegant graphical
method to obtain the
point of marginal stability and the form of the perturbation at the
critical point. He found that that this mode has a
``core-halo'' structure (the density perturbation $\delta \rho$ has two nodes).

Chavanis \cite{aaiso,grand} applied the approach of  Padmanabhan
\cite{paddy,paddyapj}
to the canonical, grand-canonical and
grand-microcanonical ensembles,  thereby recovering and extending
the results previously obtained by Lynden-Bell and Wood \cite{lbw}, Horwitz and
Katz \cite{hk3}, Katz \cite{katzpoincare1,katz2}, and Lecar and Katz
\cite{lecarkatz}. He also  
considered the isobaric ensembles previously studied by
Ebert \cite{ebert,ebert2}, Bonnor \cite{bonnor} and McCrea \cite{crea}. In
the canonical ensemble, he showed that thermodynamical stability
coincides with dynamical stability (with respect to the Euler-Poisson equations)
and that the mode of marginal
instability  has a ``core'' structure (the density perturbation $\delta \rho$
has just one node) contrary to the ``core-halo'' structure found by Padmanabhan
\cite{paddy,paddyapj} in the microcanonical ensemble. As a consequence, in the
microcanonical ensemble the
gravothermal catastrophe \cite{lbw} below $E_c$ ultimately leads to a binary
star
surrounded by a hot halo (this structure has an infinite
entropy $S\rightarrow +\infty$ at fixed energy) while in the canonical ensemble
the isothermal
collapse \cite{aaiso}
below $T_c$
ultimately leads to a Dirac peak containing all the particles (this structure
has an infinite free energy $F\rightarrow -\infty$). This is
another manifestation of ensembles inequivalence.  For
isolated self-gravitating systems described by an
$N$-body Hamiltonian system,  like galaxies and globular clusters, only the
microcanonical ensemble is relevant. The other ensembles have no physical
meaning.\footnote{For systems with long-range interactions one cannot
deduce the canonical ensemble (and the other ensembles) from the microcanonical
ensemble by considering a subsystem of a large ensemble because
the energy is nonadditive (the sum of energies of all the small sub-systems is
not equal to the total
energy of the system). Nevertheless, a
mathematical interest of considering less constrained ensembles is that they
are simpler to study and that they provide {\it sufficient} conditions of
thermodynamical stability.} However, the canonical
ensemble is rigorously
justifed for the model of
self-gravitating Brownian particles introduced and studied by Chavanis and
Sire \cite{crs,sc,post,tcoll,virial1,virial2,zero}.

Katz \& Okamoto \cite{katzokamoto} (see also the review of Katz
\cite{found}) studied temperature fluctuations
in self-gravitating isothermal spheres and showed that the onset of
gravitational instability (gravothermal catastrophe) is advanced
because of finite $N$ effects. The critical density contrast taking into
account the finite number of particles is ${\cal
R}_c=709\times {\rm exp}(-3.30N^{-1/3})$. This may explain why observations
reveal that a greater number
of globular clusters than is normally believed may already be in an advanced
stage of core collapse.

Chavanis \cite{metastable} (see also the review \cite{ijmpb}) argued
that the lifetime of a self-gravitating system trapped in a  metastable state
(local entropy maximum) scales as $e^{N\Delta s}$, where
$\Delta s$ is the
barrier of entropy per particle.\footnote{In the canonical ensemble the
lifetime of a
metastable state is given by the Kramers formula $e^{N\Delta f/k_B T}$, where
$\Delta f$ is the
barrier of free energy  per particle.}
For $N\gg 1$, the lifetime of a
metastable state
is exponentially large making it of
extreme physical relevance.\footnote{For $t\rightarrow +\infty$, a 
self-gravitating system trapped in a metastable state is expected to
ultimately collapse
and form binaries since there is no global entropy maximum. However, the system
may find itself
``blocked'' in a local entropy maximum for a very long time, of the order of
$e^N t_D$, much larger than the age of the Universe. Only a
large random
fluctuation can drive the system out of this local maximum
of entropy. This is a rare event. This makes local entropy
maxima extremely  important on the  time scales relevant in astrophysics.}
As a result, globular clusters that lie on the series of equilibria 
before the point of instability can be considered as long-lived
metastable
states. Chavanis \cite{metastable}  calculated the barrier
of entropy close to the critical point $E_{c}$ and recovered the threshold
of gravitational collapse due to
finite $N$ effects obtained by Katz \& Okamoto \cite{katzokamoto}.

de Vega and Sanchez \cite{dvs1,dvs2} (see also
\cite{dvsc,semelin0,semelin})  studied the statistical mechanics of the
self-gravitating gas using field theory, thereby confirming and complementing
the former works
of Horwitz and Katz \cite{hkpart1,hk3}. The statistical mechanics approach
(writing the
density of states or the partition function as path integrals and
evaluating them by making a saddle point approximation) is more rigorous than
the thermodynamical method (maximizing or minimizing a relevant
thermodynamic potential to obtain the most probable state) but it is
considerably more formal and complicated. It can be shown that the statistical
mechanics
approach gives exactly the same results as the thermodynamic approach in a
proper thermodynamic limit where the number of particles $N\rightarrow +\infty$
keeping $\Lambda=-ER/GM^2$ and $\eta=\beta GMm/R$ fixed \cite{ijmpb}. We refer
to \cite{ms,kiessling,joyce,alastuey} for rigorous mathematical results on this
subject.

The works that we have reviewed so far consider box-confined isothermal
stellar
systems. 
However, real stellar systems like globular clusters, are not in boxes and
particles with high enough energy can escape from the system. As a result, the
distribution function is not exactly given by the
Boltzmann distribution (especially for high energies) and truncated models have
been introduced, notably the Woolley model \cite{woolley} and the Michie-King
model \cite{michie,king} (see footnote 5). The
thermodynamical
stability of
these models has been studied by analogy with the thermodynamics of box-confined
systems \cite{lbw,khd,katzking,cn,clm1}. The caloric curves  of the Woolley
and Michie-King models have the form of spirals. The stability limits can be
determined from the Poincar\'e criterion by identifying the turning point of
energy in the series of equilibria. As first suggested by Lynden-Bell and
Wood \cite{lbw} and Horwitz and Katz \cite{hkpart1}, the
sequences of equilibrium states are presumed to simulate slowly
evolving near-equilibrium configurations of real stellar
systems.

The physical picture that emerges from these studies is the following. Because
of stellar encounters and evaporation, the globular clusters slowly evolve along
a series of equilibria corresponding to the King model. The evolution is such
that the central density
increases and the energy decreases.\footnote{This can be
understood as
follows. Under the effect of close encounters,
stars leave the system with an energy positive or close to zero. Therefore, the
energy of the cluster decreases or remains approximately constant. Since the
number of stars in the cluster decreases, the cluster contracts (according to
the virial theorem) and becomes more and more concentrated. Therefore,
the central density increases with time. Another argument,
related to the $H$-theorem, explaining why the central density
naturally increases with time is developed in \cite{rgb}.} In the region of
positive specific heat, the
temperature decreases while it increases in the region of negative specific
heat. The evolution continues until the point of instability,
corresponding to the turning point of energy (energy minimum). This is when the
specific heat vanishes. At that point, the system undergoes the Antonov
instability \cite{antonov}, also known as the gravothermal catastrophe
\cite{lbw}, and collapses. This instability can be followed by
using dynamical models based either on moment equations derived from the
Fokker-Planck equation \cite{larson1,larson2}, Monte Carlo models
\cite{hmc}, $N$-body simulations \cite{aarsethN}, heuristic fluid equations
\cite{hetal,lbe}, or kinetic equations
such as the orbit-averaged-Fokker-Planck equation \cite{cohn}. During the
collapse the system takes a
core-halo structure (reminiscent of a red giant) in which
the cluster develops a dense and ``hot'' core and a diffuse envelope. The
dynamical evolution of the system is due to
the gradient of temperature (velocity dispersion) between the core and the halo
and the fact that the core has a negative specific heat (see Appendix
\ref{sec_nsh}). The core loses heat to
the profit of the halo, becomes hotter, and
contracts. If the temperature increases more rapidly in the
core than in the halo there is no possible equilibrium
and we get a thermal runaway: this is the gravothermal catastrophe.
As a result, the core collapses
and reaches higher and higher densities and higher and higher temperatures while
the halo is not sensibly affected by the collapse of the
core and maintains its initial structure (it remains cool). The collapse of the
core is
self-similar and leads to a finite time
singularity: the central density and the temperature become infinite in
a finite time while the core radius shrinks to nothing \cite{lbe,cohn}. This
is called core
collapse. The mass
contained in the core tends to zero at the
collapse time. 
In the case of globular clusters, the
evolution continues in a self-similar postcollapse regime \cite{inagakilb} with
the
formation of a binary star  containing a significant fraction
of the cluster energy (as previously found by Aarseth \cite{aarsethN} in his
$N$-body simulations).  The energy released by
the binary
can
stop the collapse and induce a reexpansion of the system. Then, a series of
gravothermal oscillations is expected to follow
\cite{oscillations,hr}.\footnote{Analogously, the isothermal collapse of
self-gravitating Brownian particles in the canonical ensemble when $T<T_c$ can
be followed by solving the Smoluchowski-Poisson equations \cite{crs}. The
collapse of
the system is self-similar and leads to a finite time singularity: the central
density becomes infinite in a finite time while the core radius and the core
mass vanish \cite{sc}. The evolution continues in a self-similar postcollapse
regime with the formation of a Dirac peak progressively accreting all the mass
\cite{post}.}

It has to be stressed that the gravothermal catastrophe is a
thermodynamical instability, not a dynamical instability. Indeed, it has been
shown that {\it all} isotropic stellar
systems with a distribution function of the form  $f(\epsilon)$ with
$f'(\epsilon)<0$, including the truncated Maxwell-Boltzmann (isothermal)
distribution, are dynamically stable with respect to the collisionless
Vlasov-Poisson equations 
\cite{doremus71,doremus73,gillon76,sflp,ks,kandrup91}. In particular,
all the isothermal configurations on
the series of equilibria are
dynamically stable, even those deep into the spiral that are thermodynamically
unstable.
Therefore, dynamical and thermodynamical stability do not coincide in
Newtonian
gravity (thermodynamical stability implies dynamical stability but the converse
is wrong \cite{ih,cc}). This implies
that the gravothermal catastrophe is a very long (secular) process, occurring on
a collisional relaxation timescale of the order of the age of
the Universe, not on a fast dynamical timescale.

The statistical mechanics of relativistic star clusters\footnote{They may
be clusters of stars like white dwarfs, neutron stars or stellar mass black
holes in galactic nuclei.} 
started with the seminal work of Zel'dovich and Podurets \cite{zp}. They
took
into account collisions and evaporation and
studied star clusters described by the truncated Maxwell-Boltzmann distribution 
in full general relativity.\footnote{Truncated isothermal distributions of
relativistic star clusters  have been studied independently by Fackerell
\cite{fackerell66}.} This
distribution has a finite mass so there is no need to
introduce an artificial box to confine the system. They considered a series of
quasi-equilibrium states and plotted the temperature $T_{\infty}$ measured by an
infinitely-remote observer as a
function of the central density $\rho_0$.  They found that the
function $T_{\infty}(\rho_0)$ rises up to a maximum temperature
$k_B(T_\infty)_{\rm max}/mc^2=0.273$ then undergoes
damped oscillations.\footnote{They mentioned that the damped oscillations of
$T(\rho_0)$ are similar to the damped oscillations of $M(\rho_0)$ for neutron
stars discovered by Dmitriev and Kholin \cite{dk}.
This is because, in the
ultrarelativistic limit where the density is large, the equation of state of a
classical
isothermal gas takes the form $P=\epsilon/3$, where $\epsilon$ is the energy
density, like the ultrarelativistic equation of state of a Fermi gas at $T=0$.}
As a result,
equilibrium states can exist only below a maximum
temperature $(T_\infty)_{\rm max}$. They
heuristically argued that the series of equilibria becomes unstable after the
first turning point of temperature and that a new mode of instability
appears at each turning
point of temperature. These
considerations led them to the following scenario. Because of collisions and
evaporation, a star cluster slowly (secularly) evolves
 through a sequence of quasiequilibrium states. During this
evolution both the temperature and the central density increase. At a certain
point the system has high temperatures, large velocities $v\sim c$, and is
therefore general relativistic even if during the initial stage it was
Newtonian ($v\ll c$).
When the temperature reaches $(T_\infty)_{\rm max}$ the system
becomes unstable and
undergoes a catastrophic gravitational
collapse. This is what
they called an
``avalanche-type catastrophic contraction of the system.'' The
mechanism of the collapse proposed by Zel'dovich and
Podurets \cite{zp} is the following. The orbits of highly relativistic
particles become unstable and the corresponding particles start falling in
spirals towards the center. The
collapse of the orbits of some particles leads to an increase of the field
acting on the other particles, whose orbits collapse in turn etc.  This
catastrophic collapse occurs rapidly, on a dynamical timescale.   A large
fraction of the system (the main mass)
 rapidly contracts to its gravitational radius and forms what is
now called a black hole.\footnote{The name ``black hole'' was
popularized by Wheeler
\cite{wheelerBH,wheelerBH2} but it appeared earlier \cite{ewing,rosenfeld},
being probably introduced by Dicke in analogy with the
Black Hole prison of Calcutta (see Ref. \cite{bhhl}).}
However, only
the core of the system collapses. There remains a
cloud
surrounding the main mass.  The particles in the cloud, following the laws of
slow evolution, gradually fall into the collapsed mass.

Ipser \cite{ipser69b}   studied the
dynamical stability with respect to the Vlasov-Einstein equations of
isothermal relativistic star clusters with heavily truncated Maxwell-Boltzmann
velocity distribution\footnote{Following Zel'dovich
and Podurets
\cite{zp}, Ipser \cite{ipser69b} assumed a certain relation between
the energy cutoff and the
temperature. This relatively {\it ad hoc} choice was later criticized. This led
to several
generalizations of the problem by Katz {\it et al.} \cite{khk},
Suffern and Fackerell \cite{sf}, Fackerell and Suffern \cite{fs},
Merafina and Ruffini \cite{mr,mreuro,mr90}, and Bisnovatyi-Kogan {\it et al.}
\cite{bmrv,bmrv2} that we do not review here.} by using the equation of
pulsations derived by Ipser
and Thorne \cite{ipserthorne,ipser69}. He showed that the clusters
are unstable against
gravitational collapse if the redshift $z_0$ of a photon emitted from its center
and
received at infinity is larger than $0.516$. The clusters are likely to be
stable
if $z_0\le
0.516$.  Ipser \cite{ipser69b} also plotted 
the fractional binding energy
$E/Nmc^2=(M-Nm)/Nm$
as a function of the central redshift $z_0$ and showed that it reaches a
minimum before undergoing damped
oscillations. Interestingly, the point
of onset of gravitational collapse ($z_c=0.516$) appears to coincide with the
first turning point of fractional binding energy
(minimum). At
that point $z_c=0.516$, $[(M-Nm)/Nm]_{c}=0.0357$, $(Rc^2/2GNm)_{c}=4.42$ and
$(k_BT_{\infty}/mc^2)_{\rm
c}=0.23$. This is
different from the turning point of temperature reported by Zel'dovich and
Podurets \cite{zp} corresponding to $z_0=1.08$, $(M-Nm)/Nm=0.0133$,
$Rc^2/2GNm=3.92$, and
$k_B(T_{\infty})_{\rm max}/mc^2=0.27$. In particular, the gravitational
instability occurs sooner than predicted by  Zel'dovich and Podurets \cite{zp}.

Based on these results, Fackerell {\it et al.}
\cite{fit} developed a scenario for the evolution of spherical relativistic star
clusters improving the original picture of Zel'dovich and Podurets
\cite{zp}. A protocluster is expected to relax towards a relativistic isothermal
distribution,\footnote{They noted that if Newtonian
stellar systems evolve only through
evaporation they will never reach, on a relevant timescale, relativistic
densities (as assumed by  Zel'dovich and Podurets \cite{zp}). However, they
added
that recent
studies by Antonov \cite{antonov} and Lynden-Bell and Wood \cite{lbw}
indicate there is a rapid evolution by a ``thermal runaway'' in which the
cluster develops a dense and hot core and a diffuse envelope on a timescale
that
could be less than $10^{10}$ years for the nuclei of some
galaxies.} after which it might evolve  quasistatically  along a series of
equilibria by means of stellar
collisions and by the evaporation of stars. Both collisions and evaporation
should drive the cluster towards states of tighter and tighter
binding. Indeed, when a star is
ejected from a cluster, it carries away nonzero kinetic energy as measured by an
observer at infinity, and thereby decreases the fractional binding energy of
the cluster; when two stars collide and stick they increase the cluster's rest
mass and hence decrease its binding energy. When the
cluster reaches the point of minimum fractional binding energy it can no longer
evolve quasistatically and a catastrophic relativistic gravitational collapse
ensues: the stars spiral inward through the gravitational radius of the
cluster towards its center leaving behind a ``black hole'' in space with perhaps
some stars orbiting
it.\footnote{For a
fluid sphere (star) the collapse is a radial infall of all the fluid. For a star
cluster it
is an inward spiralling of all the stars.} They speculated that violent events
in the nuclei
of galaxies and
in quasars might be associated with the onset of such a collapse or with
encounters between an already collapsed cluster (black hole) and surrounding
stars.

The thermodynamics of relativistic truncated isothermal star clusters
(relativistic Woolley model) and relativistic isothermal clusters in a box was
also studied by Katz and Horwitz \cite{kh2} (see also \cite{kh,khk}) who
extended their path integral approach and steepest descent techniques
\cite{hkpart1,hk3,khd}
to general relativity. They showed that extrema (saddle points) of the action
give the Einstein equations (including the condition of mechanical
equilibrium) and the Tolman-Klein relations (uniformity of global temperature
and global chemical potential). For box-confined systems, the mean field action
is equal to the Boltzmann entropy and the mean field equilibrium distribution
is the Maxwell-Boltzmann distribution. They derived a criterion of
thermodynamical stability in terms of
a relativistic ``Schr\"odinger operator'' extending the one
introduced by Lynden-Bell and Sanitt \cite{lbs} in Newtonian
gravity.\footnote{Their paper
\cite{kh2} (and \cite{hkpart1} in Newtonian gravity) contains
mistakes due to the inequivalence of statistical ensembles that were corrected
in \cite{hk} (and \cite{hk3,khd} in Newtonian gravity).} Using
the Poincar\'e turning
point criterion,
Horwitz and Katz \cite{hk} showed that the change of
thermodynamical stability
in the microcanonical ensemble occurs at the turning point (minimum) of
binding energy  with respect to the central redshift
along the series of equilibria at
fixed $N$. This is when
the heat capacity becomes equal to zero.  Therefore, 
truncated isothermal spheres are thermodynamically stable before the turning
point of energy ($z_0<0.516$) and
thermodynamically  unstable after the turning point of energy ($z_0>0.516$).
This coincides with the dynamical
stability results of  Ipser \cite{ipser69b}.

Ipser \cite{ipser80} (see also \cite{ipser74,ih} in Newtonian gravity)
considered
the maximization of an arbitrary ``entropic'' functional $S$ at fixed
mass-energy $Mc^2$ and particle number $N$.  This variational principle
describes
a large class of relativistic clusters that are not necessarily isothermal. He
showed that the extermization problem determines an isotropic stationary
solution of the Vlasov-Einstein equations (for maxwellian clusters the
Tolman-Klein relations are
automatically
satisfied in this formulation). He also showed that a
distribution function that is a maximum of entropy $S$ at fixed mass-energy
$Mc^2$
and particle number $N$ is
necessarily dynamically stable with respect to the Vlasov-Einstein equations.
Therefore thermodynamical
stability  (in a general sense) implies
dynamical stability. Using the Poincar\'e criterion, he
obtained a binding energy stability theorem: ``an isotropic relativistic star
cluster is
dynamically
stable at least  up to the turning point of fractional binding energy.'' This
theorem is valid both for Newtonian clusters (as previously showed in \cite{ih})
and general relativistic clusters. Considering
isothermal clusters as a particular case, it implies that the
configurations with $z_0<0.516$ are both thermodynamically and
dynamically
stable. On
the other hand, configurations with $z_0>0.516$ turn out to be
dynamically unstable  (as shown numerically in \cite{ipser69b})
in addition of being thermodynamically unstable. Ipser  \cite{ipser80}
therefore
concluded 
that, in general relativity, thermodynamical stability coincides with dynamical
stability. He conjectured that this result remains valid for an arbitrary
``entropic'' functional so that, in general relativity, all isotropic star
clusters become dynamically unstable
after the turning point of binding energy. This is in sharp contrast
with the Newtonian case where it has been
shown \cite{doremus71,doremus73,gillon76,sflp,ks,kandrup91} that all
isotropic models are dynamically stable with respect to the Vlasov-Poisson
equations, even those that lie after the turning point of energy.

These theoretical results have been confirmed by Shapiro and Teukolsky
\cite{st1,st2,st3,st4,strevue} who
numerically solved the relativistic Vlasov-Einstein equations
governing the dynamical evolution of a collisionless spherical gas of
particles in general relativity. They followed the series of equilibria
of truncated isothermal distributions (assumed to result from the
gravothermal catastrophe of initially Newtonian clusters) and showed from direct
numerical
simulations that above a critical redshift
$z_c\sim 0.516$, corresponding to the turning point of fractional binding
energy,
the relativistic star cluster becomes dynamically unstable and undergoes a 
catastrophic collapse to a supermassive black hole on a dynamical time
scale.

Sorkin {\it et al.} \cite{sorkin} and more
recently Chavanis \cite{aarelat2} studied the thermodynamics of a
self-gravitating black-body radiation confined within a cavity in general
relativity. Black-body radiation is equivalent to an
ultrarelativistic gas of massless bosons (photons) with a linear equation of
state
$P=\epsilon/3$, where $\epsilon$ denotes the energy
density.\footnote{This
equation of
state also corresponds to the ultrarelativistic
limit of an ideal gas of any kind of massive
particles, classical, fermionic or bosonic.} The equilibrium state of the
self-gravitating black-body radiation is obtained by
maximizing the entropy $S$ (proportional to the particle number $N$) at fixed
mass-energy $Mc^2$. This leads to the Tolman-Oppenheimer-Volkoff
\cite{tolman,ov} equations
expressing the condition of
hydrostatic equilibrium and to the Tolman \cite{tolman}
relation. The caloric curve
$T_{\infty}({\cal E})$, where $T_{\infty}$ denotes the Tolman temperature and
${\cal E}=Mc^2$ the mass-energy, forms a
spiral
so that no equilibrium state exists
above a maximum energy ${\cal E}_{\rm max}=0.24632\, Rc^4/G$ for an isolated
system or
above a maximum
temperature  $k_B(T_\infty)_{\rm max}=0.445\,(\hbar^3c^7/GR^2)^{1/4}$ for a 
system in contact with a heat bath (see Fig. 15 of
\cite{aarelat2}).
Using different methods, Sorkin {\it et al.} \cite{sorkin}
and Chavanis \cite{aarelat2} showed
that thermodynamical stability coincides with dynamical stability with respect
to the Euler-Einstein equations and that the series of
equilibria becomes unstable
after the first turning point of
energy in agreement with the  Poincar\'e criterion. This is when the
specific heat $C=d{\cal E}/dT_{\infty}$ vanishes, passing from negative to
positive values. This
corresponds to an energy
density contrast ${\cal R}_{\rm MCE}=22.4$ \cite{aarelat2}.
 The system becomes unstable when it is ``too hot'' because
energy is mass so
that it gravitates.  This is what Tolman \cite{tolman} called the ``weight of
heat''. The mode of marginal instability at ${\cal
E}_{\rm max}$
has a ``core''  structure (the energy density perturbation $\delta\epsilon$ has
just one node) \cite{aarelat1,aarelat2}. Therefore, gravitational
collapse is expected to lead to the formation of a black
hole.

The statistical mechanics of general relativistic classical self-gravitating
systems confined within a box was reconsidered recently by Roupas \cite{roupas}
and,
independently, by us (our study was made during the PhD thesis of
G.
Alberti from 2014 to 2017). In this paper, we
report our results which confirm and complete the results obtained by
Roupas \cite{roupas}. A comparison between the two studies is made in the
conclusion. The paper is organized as follows. In Sec. \ref{sec_b} we recall the
main equations governing the structure of a general relativistic
classical gas at statistical equilibrium. In Sec. \ref{sec_pl} we recall  the
caloric curve of a nonrelativistic classical self-gravitating gas (cold spiral)
and
the caloric curve of the self-gravitating black-body radiation (hot spiral). In
Sec. \ref{sec_generic} we discuss a typical example where the caloric curve of a
general relativistic classical gas presents a double spiral connecting
the cold and  hot spirals found previously.  In Sec. \ref{sec_gc} we treat
the general case
and show how the caloric curve changes as we increase the number of particles.
In Sec. \ref{sec_nzero} we consider the limit $N\rightarrow 0$. We introduce two
types of normalization of energy and temperature, appropriate to the
nonrelativistic and ultrarelativistic limits, in order to obtain
asymptotic caloric curves
describing the cold and hot spirals respectively. We discuss the analogies and
the differences between the caloric curve of an ultrarelativistic classical gas
and the caloric curve of the self-gravitating black-body radiation. In Sec.
\ref{sec_ev}
we study the evolution of the critical points of the caloric curves as a
function of $N$ and obtain explicit asymptotic results. The
extension of our results to self-gravitating fermions in general relativity is
made in our companion paper \cite{acf} (see also \cite{bvr,rc}). A summary of
our main results is presented in \cite{ca}.

\section{Basic equations of a general relativistic classical gas}
\label{sec_b}

In this section, we recall the basic equations describing the structure of a
general relativistic classical gas at statistical equilibrium (see
\cite{roupas,rgf,rgb} for their
derivation). In order to make the connection with our companion paper
\cite{acf},
we assume that this classical gas corresponds to the nondegenerate limit of a
gas of fermions. At statistical equilibrium it is described by the
Maxwell-Juttner distribution 
\begin{equation}
\label{mal1}
f({\bf r},{\bf p})=\frac{g}{h^3}e^{\alpha}e^{-E(p)/k_B T(r)}, 
\end{equation}
where $E(p)=\sqrt{p^2c^2+m^2c^4}$ is the energy of a particle. The
temperature and the chemical potential are space-dependent. They are given by
the Tolman-Klein relations $T(r)=T_{\infty} e^{-\nu(r)/2}$ and
$\mu(r)=\mu_{\infty} e^{-\nu(r)/2}$, where $T_{\infty}$ and $\mu_{\infty}$ 
are the temperature and chemical potential measured by an observer at infinity
and $\nu(r)$ is the metric coefficient. $T_{\infty}$ will be called the Tolman
(global) temperature \cite{tolman} and $\mu_{\infty}$ will be called the Klein
(global) chemical
potential \cite{klein}. Since the local temperature $T(r)$ and the local
chemical potential $\mu(r)$ are red-shifted in the same manner, their
ratio
\begin{equation}
\label{b5}
\alpha=\frac{\mu(r)}{k_B T(r)}=\frac{\mu_{\infty}}{k_B T_{\infty}}
\end{equation}
is uniform throughout the system.

The number density $n(r)=\int f\, d{\bf p}$, the
energy density $\epsilon(r)=\int f E(p)\, d{\bf p}$ and
the
pressure $P(r)=(1/3)\int f p E'(p)\, d{\bf p}$ are related to the local
temperature $T(r)$ and to the
local chemical
potential $\mu(r)$ by
\begin{equation}
\label{b1}
n(r)=\frac{4\pi gm^3c^3}{h^3}e^{\alpha}\frac{1}{b(r)}K_2(b(r)),
\end{equation}
\begin{equation}
\label{b2}
\epsilon(r)=\frac{4\pi gm^4c^5}{h^3}e^{\alpha}\frac{1}{b(r)}K_2(b(r))\left
\lbrack
\frac{K_1(b(r))}{K_2(b(r))}+\frac{3}{b(r)}\right \rbrack,
\end{equation}
\begin{equation}
\label{b3}
P(r)=n(r)\frac{mc^2}{b(r)},
\end{equation}
where $K_n(z)$ are the modified Bessel functions. In the foregoing equations, we
have introduced the normalized local inverse temperature 
\begin{equation}
b(r)=\frac{mc^2}{k_B T(r)}
\end{equation}
and the
parameter $\alpha$ defined by Eq. (\ref{b5}). Eqs. (\ref{b1})-(\ref{b3}) define
the equation
of state of a relativistic classical gas in parametric form. By construction,
$b(r)\ge 0$ and $-\infty<\alpha<+\infty$.

The Tolman-Oppenheimer-Volkoff (TOV) equations \cite{tolman,ov}, which
correspond to the
equation of hydrostatic equilibrium in general relativity, can be written as
\begin{equation}
\label{b7}
\frac{dM}{dr}=\frac{\epsilon(r)}{c^2}4\pi r^2,
\end{equation}
\begin{equation}
\label{b6}
\frac{1}{b(r)}\frac{{d}b}{{d}r}=\frac{1}{c^2}\frac{\frac{GM(r)}{r^2}+\frac{4\pi
G}{c^2}P(r)r}{1-\frac{2GM(r)}{r c^2}},
\end{equation}
where $M(r)$ is the mass-energy
contained within the sphere of radius $r$. They
have to
be solved with the boundary conditions
\begin{equation}
\label{b8}
M(0)=0,\qquad b(0)=b_0\ge 0.
\end{equation}
We assume that the system is confined within a spherical box of radius $R$ (as
recalled
in the Introduction, a box is necessary to prevent the evaporation of the gas
and have a
well-defined statistical equilibrium state). The
mass-energy of the gas and the particle number are then given by
\begin{equation}
\label{b9}
M=M(R)=\frac{1}{c^2} \int_0^R \epsilon(r) 4\pi r^2\, dr,
\end{equation}
\begin{equation}
\label{b10}
N=\int_0^R n(r)\left \lbrack 1-\frac{2GM(r)}{rc^2}\right
\rbrack^{-1/2}4\pi
r^2\, dr.
\end{equation}
The Tolman temperature can be obtained from the relation
\begin{equation}
\label{b11}
T_{\infty}=T(R) \sqrt{1-\frac{2GM}{Rc^2}},
\end{equation}
where $T(R)$ is the temperature of the system on the box. Finally, the
entropy is given
by
\begin{equation}
\label{b12}
S=\int_0^R \frac{\epsilon(r)}{T(r)}\left \lbrack
1-\frac{2GM(r)}{rc^2}\right
\rbrack^{-1/2}4\pi
r^2\, dr+k_B N-\alpha k_B N,
\end{equation}
and the free energy by
\begin{equation}
\label{b13}
F=E-T_{\infty} S,
\end{equation}
where  
\begin{equation}
\label{emn}
E=(M-Nm)c^2
\end{equation}
is the binding energy.\footnote{The binding energy is usually
defined with the opposite sign, i.e., $E_b=(Nm-M)c^2$. We shall, however,
call $E$ the binding energy or,
simply, the energy. In the nonrelativistic limit $c\rightarrow +\infty$, it
reduces to the Newtonian energy $E=E_{\rm kin}+W$ (kinetic $+$ potential).}

In the
following, in order to be consistent with our companion
paper \cite{acf}, we shall express the results in terms of the relativistic
gravitational potential
$\Phi(r)$ defined
by\footnote{The gravitational potential is denoted
$\varphi(r)$ in \cite{rgb}.}
\begin{equation}
\label{b14}
\frac{k_B
T(r)}{mc^2}=\frac{1}{b(r)}=\frac{1}{|\alpha|}\sqrt{1+\frac{\Phi(r)}{c^2}},
\end{equation}
instead of the temperature $T(r)$. By construction, $\Phi(r)\ge -c^2$. The
method
used
to obtain the caloric curve
$T_{\infty}(E)$ is explained in Appendix \ref{sec_cons}. In order to make
contact with the
nonrelativistic results, we shall use the dimensionless energy $\Lambda$ and the
dimensionless inverse Tolman temperature $\eta$ defined by
\begin{equation}
\label{ble}
\Lambda=-\frac{E R}{GN^2m^2}\qquad  \eta=\frac{\beta_{\infty} GNm^2}{R},
\end{equation}
In the microcanonical ensemble, a stable equilibrium state corresponds to a
maximum of entropy $S$ at fixed energy $E$ and particle number $N$. In the
canonical ensemble, a stable equilibrium state corresponds to a
minimum of free energy  $F$ at fixed particle number $N$. The stability of the
system can be settled by plotting the caloric curve (or series
of equilibria) $\eta(\Lambda)$ and using the Poincar\'e \cite{poincare}
turning point criterion.

As shown in \cite{roupas,rgb}, the caloric curve of the general relativistic
classical
gas depends on a single control parameter
\begin{equation}
\nu=\frac{GNm}{Rc^2},
\end{equation}
called the compactness parameter. It can be interpreted as the ratio
$\nu=R_S^*/R$
between an effective Schwarzschild radius $R_S^*=GNm/c^2$, defined in terms of
the rest mass $Nm$, and the box radius $R$. Alternatively, $\nu=Nm/M_S^*$
where $M_S^*=Rc^2/G$ is an effective Schwarzschild mass. Using the 
normalized variables introduced in Appendix B of \cite{acf} we can take
$\hbar=c=G=m=g/2=R=1$ without restriction of generality. In that case, we get
\begin{equation}
\Lambda=-\frac{E}{N^2}\qquad  \eta=\beta_{\infty}N\qquad \nu=N.
\end{equation}

\section{Particular limits}
\label{sec_pl}

In this section, we recall  the
caloric curve of a nonrelativistic  classical self-gravitating gas and
the caloric curve of the self-gravitating black-body radiation. They will help
us
interpreting the nonrelativistic and ultrarelativistic limits of the general
relativistic classical gas.

\subsection{Nonrelativistic  classical self-gravitating gas}
\label{sec_nr}

\begin{figure}
\begin{center}
\includegraphics[clip,scale=0.3]{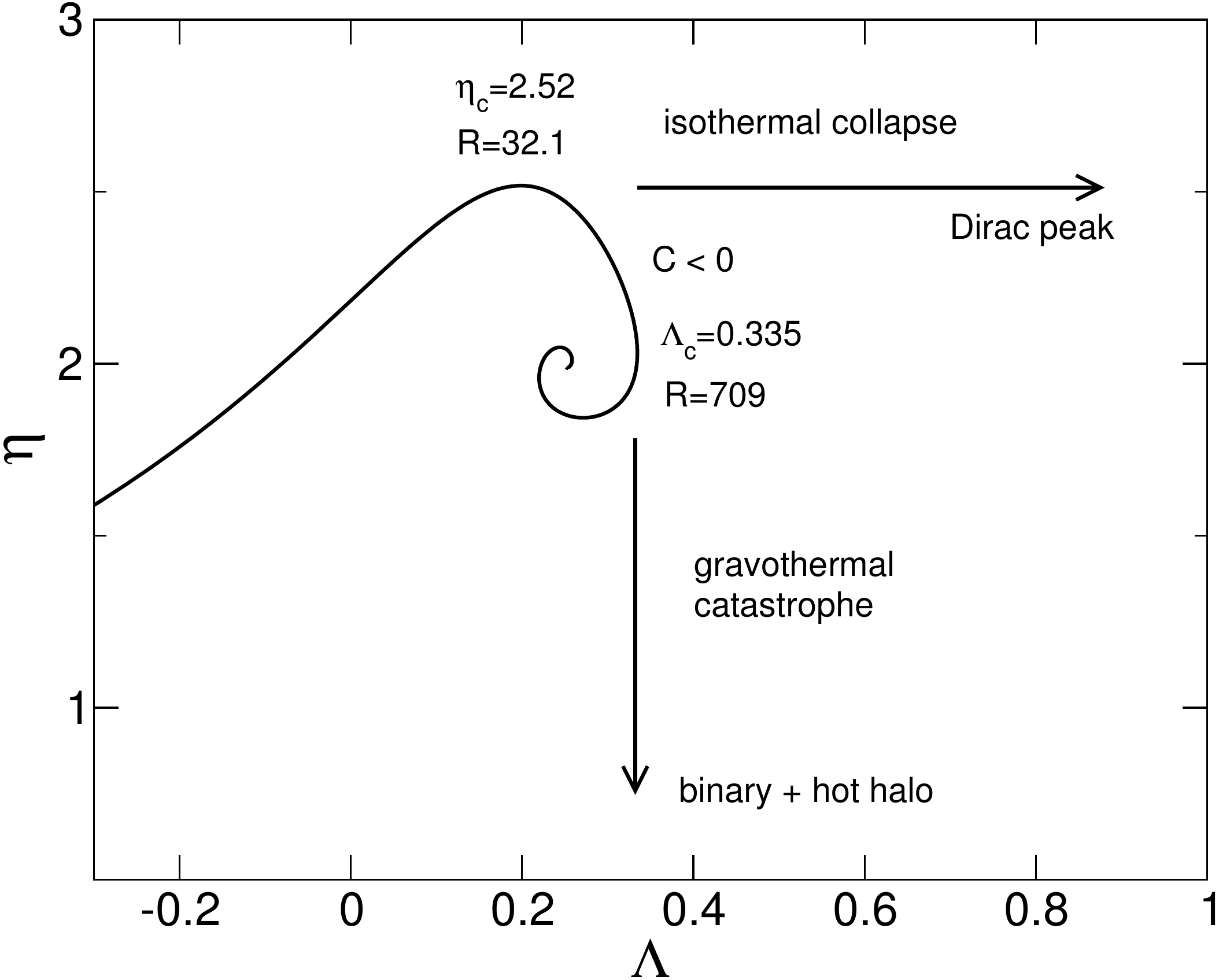}
\caption{Caloric curve of the nonrelativistic classical self-gravitating
gas.}
\label{etalambda}
\end{center}
\end{figure}

The thermodynamics of a nonrelativistic  classical  self-gravitating gas
confined within a box has
been studied in detail in
\cite{antonov,lbw,thirring,hkpart1,hk3,katzpoincare1,
katz2,lecarkatz,paddyapj,paddy,aaiso,grand,katzokamoto,
found,metastable,dvsc,dvs1,dvs2}. It is
described by a linear equation of state $P(r)=\rho(r) k_B T/m$ where $T$ is
uniform throughout the system. The natural dimensionless
energy and inverse temperature are
\begin{equation}
\label{nr1}
\Lambda=-\frac{ER}{GM^2}\qquad  \eta=\frac{\beta GMm}{R},
\end{equation}
where we recall that $M=Nm$ for nonrelativistic systems. 
The caloric curve (or series of
equilibria) $\eta(\Lambda)$ has the form of a spiral (see Fig.
\ref{etalambda}) parametrized by the density contrast ${\cal
R}=\rho(0)/\rho(R)$. It was first plotted by Katz 
\cite{katzpoincare1}. This
spiral is implicit in the papers of Antonov \cite{antonov} and Lynden-Bell and
Wood \cite{lbw}. It also appears (plotted in terms of
other variables) in earlier works on
isothermal
stars \cite{emden,chandrabook2,ebert,ebert2,bonnor,crea,sch}.

In the microcanonical ensemble there is no equilibrium state below a minimum
energy $E_c$ given by
\cite{lbw}
\begin{equation}
\Lambda_{\rm c}=-\frac{E_{c}R}{GM^2}=0.335.
\end{equation}
Using the Poincar\'e criterion \cite{poincare}, one can show that
the series of equilibria becomes and remains unstable after the
first turning point of energy $\Lambda_{\rm c}$ (a new mode of stability is
lost at each turning point of energy). This is when the specific heat $C=dE/dT$
vanishes. This corresponds to a density
contrast ${\cal R}_{\rm MCE}=709$ \cite{antonov}. At
that point, the system undergoes a
gravothermal catastrophe \cite{lbw}.  In the case of globular
clusters, this
gravitational collapse ultimately leads to the
formation of a binary star surrounded by a hot halo \cite{lbe,inagakilb,cohn}.
This structure has an infinite entropy $S\rightarrow +\infty$ at
fixed energy (see Appendix A of \cite{sc}).

In the canonical ensemble there is no equilibrium state below a minimum
temperature $T_c$ given by
\cite{emden}
\begin{equation}
\eta_{c}\equiv \frac{\beta_c GMm}{R}=2.52.
\end{equation}
Using the Poincar\'e criterion \cite{poincare}, one can show
that the series of
equilibria becomes and remains unstable after the first turning point of
temperature $\eta_c$ (a new mode of stability is
lost at each turning point of temperature). This is when the specific heat
$C=dE/dT$ is infinite. This
corresponds to a density contrast ${\cal R}_{\rm CE}=32.1$ \cite{emden}. At
that point, the
system undergoes an isothermal collapse \cite{aaiso}.  In
the case of
self-gravitating
Brownian particles, this gravitational collapse ultimately leads to a Dirac peak
containing all the mass \cite{post}.  This structure has an infinite
free energy $F\rightarrow -\infty$ (see Appendix B of \cite{sc}).

There is a region of
ensemble inequivalences in the first region of negative specific heat between 
${\cal R}_{\rm CE}=32.1$ and ${\cal R}_{\rm MCE}=709$ (see Fig.
\ref{etalambda}). The system becomes canonically unstable when the specific
heat becomes infinite, passing from positive to negative values, and it becomes 
microcanonically unstable when the specific heat vanishes, passing from negative
to positive values (see Appendix \ref{sec_nsh}).

\subsection{Self-gravitating black-body
radiation}
\label{sec_ur}

The  thermodynamics of the self-gravitating black-body radiation (photon
star)
confined
within a cavity in general relativity has been
studied in \cite{sorkin,aarelat2}.  Black-body radiation is
equivalent to
an ultrarelativistic gas of massless bosons (photons) with a linear equation of
state $P(r)=\epsilon(r)/3$, where $\epsilon$ denotes the energy
density.\footnote{The equilibrium states of a general
relativistic gas described by a linear equation of state of the form
$P=q\epsilon$
were considered by Oppenheimer and Volkoff \cite{ov}, Dmitriev and
Kholin \cite{dk}, Misner and Zapolsky \cite{mz}, Harrison \cite{harrison},
Harrison {\it et al.} \cite{htww65}, Bisnovatyi-Kogan and Zel'dovich \cite{bkz},
Bisnovatyi-Kogan and Thorne \cite{bkt}, Chandrasekhar \cite{chandra72},
Yabushita \cite{yabushitaR1,yabushitaR2,yabushitaR3}, Sorkin {\it et al.}
\cite{sorkin}, Schmidt and Homann \cite{shps}, Chavanis \cite{aarelat1}, Banks
{\it et al.} \cite{banks}, Pesci \cite{pesci} and  Chavanis \cite{aarelat2}. The
coefficient $q=1/3$ describes the ultrarelativistic
core of neutron stars or the black-body radiation; the coefficient $q=1$
describes stiff stars \cite{zeldovichstiff}; the nonrelativistic
isothermal
gas is recovered in the limit $q\rightarrow 0$.} The natural dimensionless
energy and temperature are 
\begin{equation}
\label{ur1}
{\cal M}=\frac{GM}{Rc^2},\qquad {\cal T}=\frac{k_B T_{\infty}
G^{1/4}R^{1/2}}{\hbar^{3/4}c^{7/4}},
\end{equation}
where ${\cal E}=Mc^2$ is the mass-energy and $T_{\infty}$ is the Tolman
temperature.

\begin{figure}
\begin{center}
\includegraphics[clip,scale=0.3]{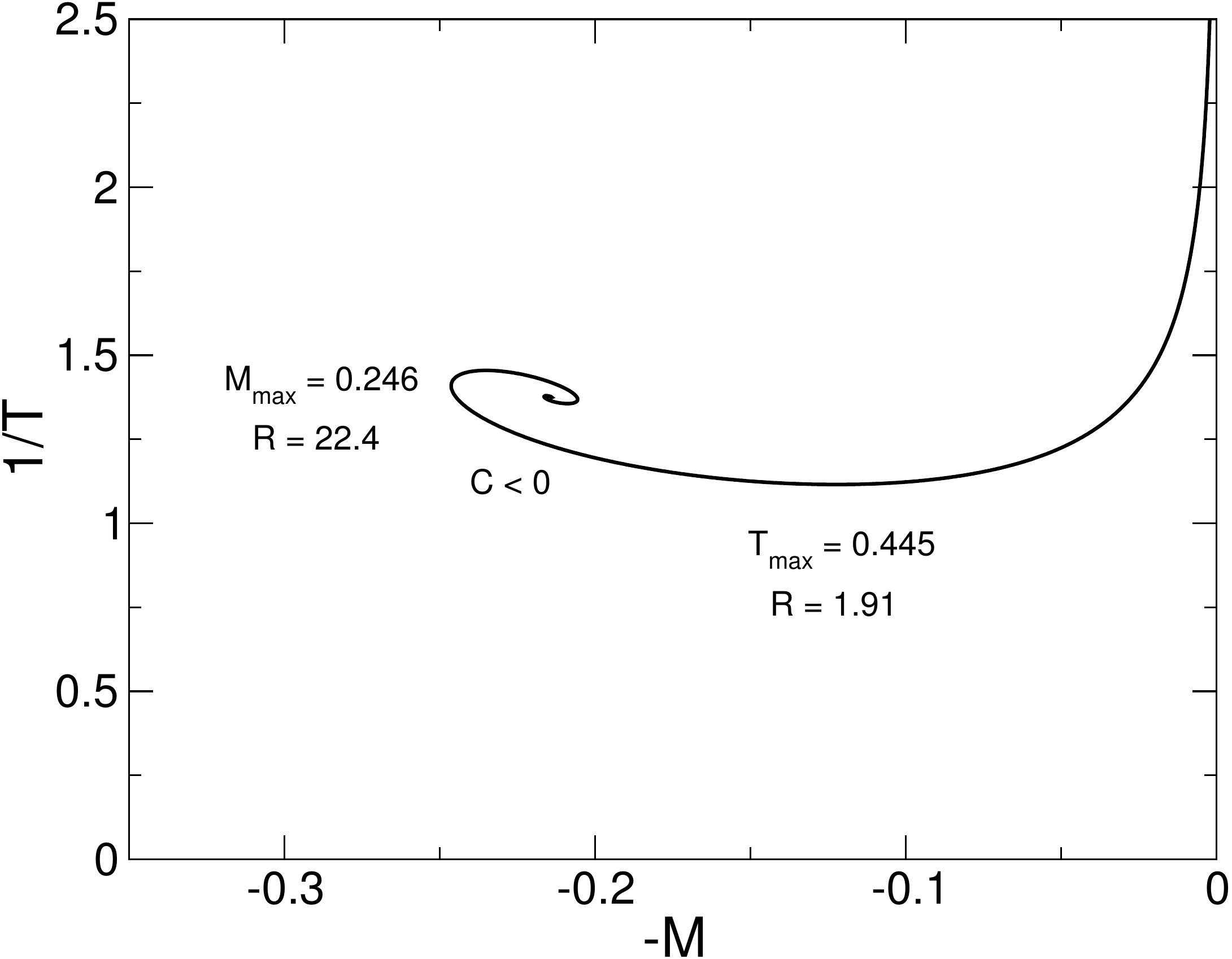}
\caption{Caloric curve of the
self-gravitating black-body 
radiation (ultrarelativistic gas) with the normalization from Eq. (\ref{ur1}).}
\label{radiation}
\end{center}
\end{figure}

The caloric curve (or series of equilibria) $T_{\infty}({\cal E})$  has the form
of a
spiral (see Fig. \ref{radiation}) parametrized by the
energy density contrast ${\cal
R}=\epsilon(0)/\epsilon(R)$.\footnote{Basically, the spiral
arises
because the equation of state of the black-body radiation $P=\epsilon/3$ is
linear, similarly to
the linear equation of state  $P=\rho k_B T/m$ of a nonrelativistic classical
gas which also leads to a spiralling caloric curve (see Fig. \ref{etalambda}).}
It
was first plotted by  Chavanis (see Fig. 15 of
\cite{aarelat2}). It is implicit in the work of Sorkin {\it et al.}
\cite{sorkin} who plotted $\epsilon(R)$ as a
function of $M$ in their Fig. 2.

In the microcanonical ensemble there is no equilibrium state above a maximum
mass-energy ${\cal E}_{\rm
max}=M_{\rm
max}c^2$ given by \cite{sorkin,aarelat2}
\begin{equation}
{\cal M}_{\rm max}=\frac{G M_{\rm max}}{Rc^2}=0.24632.
\end{equation}
Using the Poincar\'e criterion \cite{poincare}, one can show
that the series of equilibria becomes and remains unstable after the first
turning point of energy ${\cal E}_{\rm max}$ (a new mode of stability is
lost at each turning point of energy). This is when the specific heat
$C=d{\cal E}/dT_{\infty}$
vanishes. This corresponds to a density
contrast ${\cal
R}_{\rm MCE}=22.4$ \cite{aarelat2}. In that case, the system is expected to
collapse towards a black hole.

In the canonical ensemble there is no equilibrium
state above a maximum temperature $(T_\infty)_{\rm max}$  given
by \cite{aarelat2}
\begin{equation}
{\cal T}_{\rm max}=\frac{k_B (T_{\infty})_{\rm max}
G^{1/4}R^{1/2}}{\hbar^{3/4}c^{7/4}}=0.445,
\end{equation}
corresponding to a density
contrast ${\cal
R}_{\rm CE}=1.91$ \cite{aarelat2}. In that case, the system is expected to
collapse towards a black hole. As discussed in \cite{aarelat2}, it is not clear
whether the
turning point of temperature at $(T_{\infty})_{\rm max}$ along the series of
equilibria signals an instability in the present context.

\section{A typical example: $N=0.1$}
\label{sec_generic}

We now consider the thermodynamics of a general relativistic
classical gas. Before considering the
general case of an arbitrary number of particles, we first treat a typical
example where $N$ is neither ``too small'' nor ``too large''. We take $N=0.1$
for illustration.

\subsection{Caloric curve}

\begin{figure}
\begin{center}
\includegraphics[clip,scale=0.3]{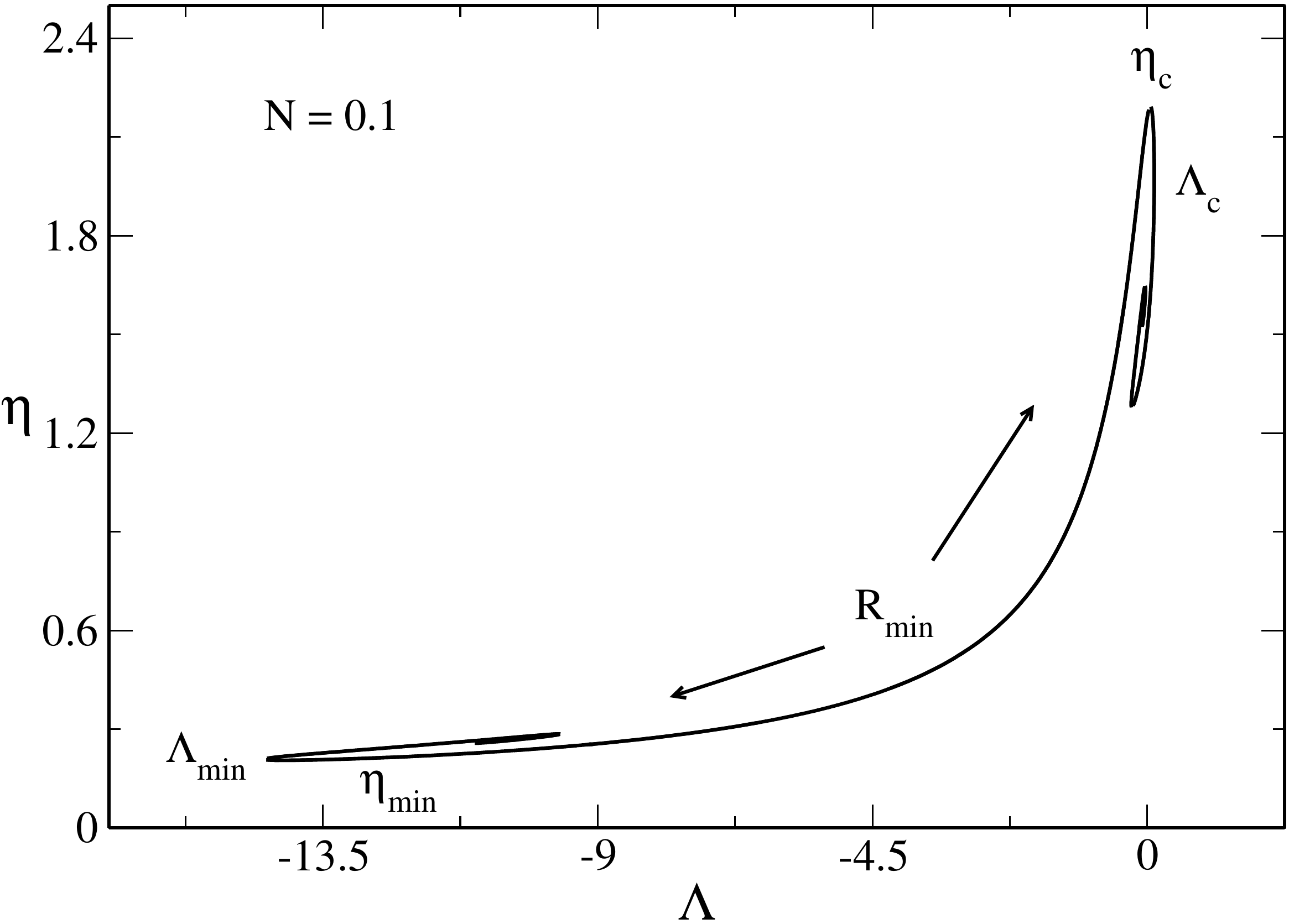}
\caption{Typical caloric curve of a 
classical
self-gravitating gas in general relativity (specifically $N = 0.1$). The
arrows indicate the increase of the density contrast ${\cal
R}=\epsilon(0)/\epsilon(R)$ along the series of
equilibria. The
system undergoes a gravitational collapse at
low energies and low temperatures  as in Newtonian gravity (cold spiral, right).
It
also undergoes a gravitational collapse at high
energies and high temperatures as in the case of the
self-gravitating radiation (hot spiral, left). }
\label{kcal_N01_linked_colorsPH}
\end{center}
\end{figure}

The caloric curve $\eta(\Lambda)$ is plotted in Fig.
\ref{kcal_N01_linked_colorsPH}. It has the
form of a double spiral. The spiral on the right, corresponding to low
temperatures, will be called the ``cold
spiral''. It occurs at low energies, positive and negative, close to $E=0$. It
is a generalization, when the system is relativistic, of the
nonrelativistic caloric curve discussed in Sec. \ref{sec_nr}.
It
corresponds to weakly relativistic configurations ($k_B T/mc^2$ small).  The
spiral on the left,
corresponding to high temperatures,  will
be called the ``hot spiral''. It occurs at large positive energies (positive
energies are allowed because the system is confined within a box). It is
related (but not equivalent)
to the caloric curve of the ultrarelativistic self-gravitating black-body
radiation discussed in Sec. \ref{sec_ur}.
It
corresponds to strongly relativistic configurations ($k_B T/mc^2$ large).
For a general relativistic classical gas these two spirals appear at the
extremities of 
the same caloric
curve $\eta(\Lambda)$.

The caloric curve presents alternating regions of positive and negative specific
heat $C=dE/dT_{\infty}$. This is due to the long-range nature of
gravity. The specific heat
must be defined with $T_{\infty}$ (the variable conjugate to the energy), not
with
another temperature like $T(R)$.  On the cold spiral, the specific heat is
positive before $\eta_c$, becomes
infinite at $\eta_c$, is negative between $\eta_c$ and $\Lambda_c$, vanishes
at $\Lambda_c$, is positive again after $\Lambda_c$ etc. Similarly, on the hot
spiral, the specific heat is
positive before $\eta_{\rm min}$, becomes
infinite at $\eta_{\rm min}$, is negative between $\eta_{\rm min}$ and
$\Lambda_{\rm min}$, vanishes
at $\Lambda_{\rm min}$, is positive again after $\Lambda_{\rm min}$ etc.

\subsubsection{Microcanonical ensemble}

Let us first consider the microcanonical ensemble where the control parameter is
the energy $E$ (or $\Lambda$). As in Newtonian gravity \cite{paddy,found,ijmpb},
there
is no global maximum of entropy $S$ at fixed energy $E$ and particle
number $N$.  However, there exist
long-lived metastable states, whose lifetime scales as $e^N$,  corresponding to
local
entropy maxima. These metastable states are stable in practice.
 From
Poincar\'e's theory of linear series of equilibria, we can
show that these metastable states are located on the main branch of the caloric
curve
between $\Lambda_{\rm min}$ (maximum energy) and $\Lambda_{c}$ (minimum
energy). After these turning
points of energy (as we progress along the spirals), the equilibrium states are
unstable saddle points of entropy. Since the spirals rotate clockwise, a mode
of stability is lost at each turning point of energy.

The caloric curve presents the following features (we
consider
only stable equilibrium states):

(i) Between $\eta_c$ and $\Lambda_{c}$ and between $\eta_{\rm min}$ and
$\Lambda_{\rm min}$, the specific heat $C=dE/dT_{\infty}$ is negative. This is
possible in the microcanonical ensemble  for systems with long-range
interactions because the energy is nonadditive. The specific
heat diverges at $\eta_c$ and $\eta_{\rm min}$ and is equal to zero at 
$\Lambda_{c}$ and $\Lambda_{\rm min}$. We note that the series of equilibria
becomes unstable when the specific heat passes from negative to positive
values for the
first time.\footnote{In the microcanonical ensemble the specific heat of
stable equilibrium states may be positive
or negative. Similarly, unstable configurations may have positive or negative
specific heat. Therefore, for systems with long-range interactions like
self-gravitating systems, the sign of $C$ 
 is not a criterion of stability in the microcanonical ensemble contrary
to the case of systems with short-range interactions which necessarily have
a positive
specific heat.}

(ii) For $\Lambda>\Lambda_c$, i.e. below a minimum energy, there is no
equilibrium state. This is already the case for nonrelativistic systems (see
Sec. \ref{sec_nr}).
When its energy is too low, the system experiences a gravothermal catastrophe
leading to a binary star surrounded by a hot halo. For sufficiently relativistic
systems (corresponding to large
values of $N$), this collapse may lead to the formation of a black
hole. It would be interesting to determine the crossover between these two
behaviors.

(iii) For $\Lambda<\Lambda_{\rm min}$, i.e. above a maximum energy,
there is no equilibrium state. This is specific to strongly relativistic
systems, like the self-gravitating radiation (see Sec. \ref{sec_ur}). When
its energy
is too high, the system undergoes a catastrophic collapse towards a black hole.

\subsubsection{Canonical ensemble}

We now consider the canonical ensemble where the control parameter is the
Tolman temperature $T_{\infty}$ (or $\eta$). As in Newtonian
gravity \cite{paddy,found,ijmpb}, there is no global minimum of free energy $F$
at
fixed
particle
number $N$. However, there exist long-lived metastable states, whose lifetime
scales as $e^N$, 
corresponding to local minima of free energy. These metastable states are stable
in practice. From
Poincar\'e's theory of
linear series of equilibria, we can show that these metastable states are
located on
the main branch of the caloric curve between $\eta_{\rm min}$ (maximum
temperature) and $\eta_c$ (minimum temperature). After these turning points of
temperature (as we progress along the spirals), the
equilibrium states are unstable saddle points of free
energy. Since the spirals rotate clockwise, a mode
of stability is lost at each turning point of temperature.

The caloric curve presents the following features (we
consider only stable equilibrium states):

(i) The specific heat is always positive in the  canonical ensemble as it
should. The specific
heat diverges at $\eta_c$ and $\eta_{\rm min}$. We note that the series of
equilibria
becomes unstable when the specific heat passes from positive to negative values 
for the
first time.\footnote{In the canonical ensemble, the specific heat of stable
equilibrium states is necessarily positive since it measures the variance of
the fluctuations of energy. By contrast, unstable configurations may have
positive or negative specific heats.}

(ii) For $\eta>\eta_c$, i.e. below a minimum temperature,  there is no
equilibrium
state.  This is already the case for nonrelativistic systems (see Sec.
\ref{sec_nr}).
When the
system is too cold, it experiences an isothermal collapse leading to a Dirac
peak. For sufficiently relativistic systems, the Dirac peak should be replaced
by a black
hole. It would be interesting to determine the crossover between these two
behaviors.

(iii) For $\eta<\eta_{\rm min}$, i.e. above a maximum temperature,  there is no
equilibrium state. This is specific to strongly relativistic
systems, like the self-gravitating radiation (see Sec. \ref{sec_ur}). When
the system
is too hot, it undergoes a
catastrophic collapse towards a black hole.

\subsection{Energy density contrast}
\label{sec_edc}

\begin{figure}
\begin{center}
\includegraphics[clip,scale=0.3]{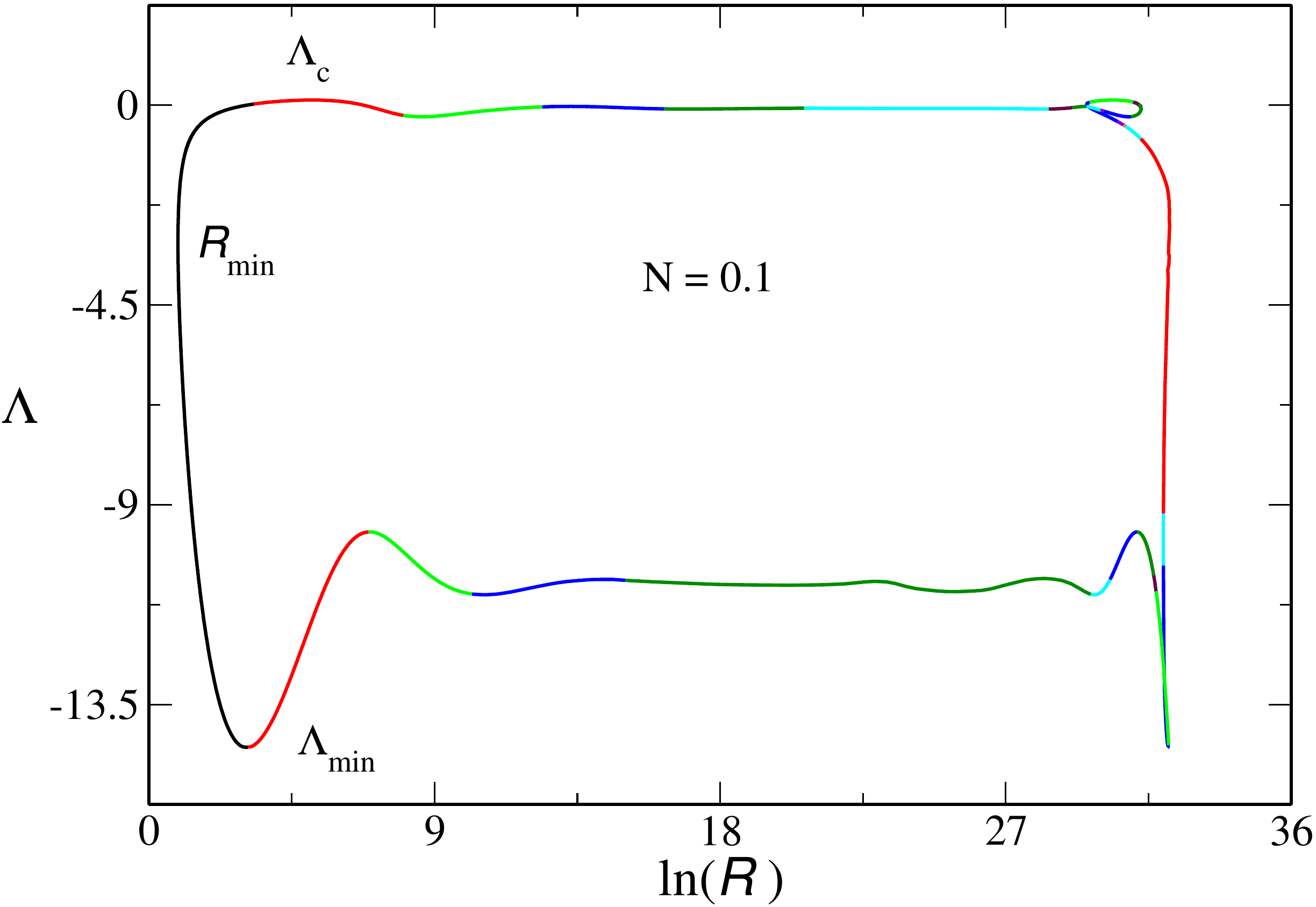}
\caption{Dimensionless
energy $\Lambda$ as a function of the energy density
contrast  ${\cal R}=\epsilon(0)/\epsilon(R)$ for $N=0.1$.}
\label{ZLambda_R_N01a_unified}
\end{center}
\end{figure}

\begin{figure}
\begin{center}
\includegraphics[clip,scale=0.3]{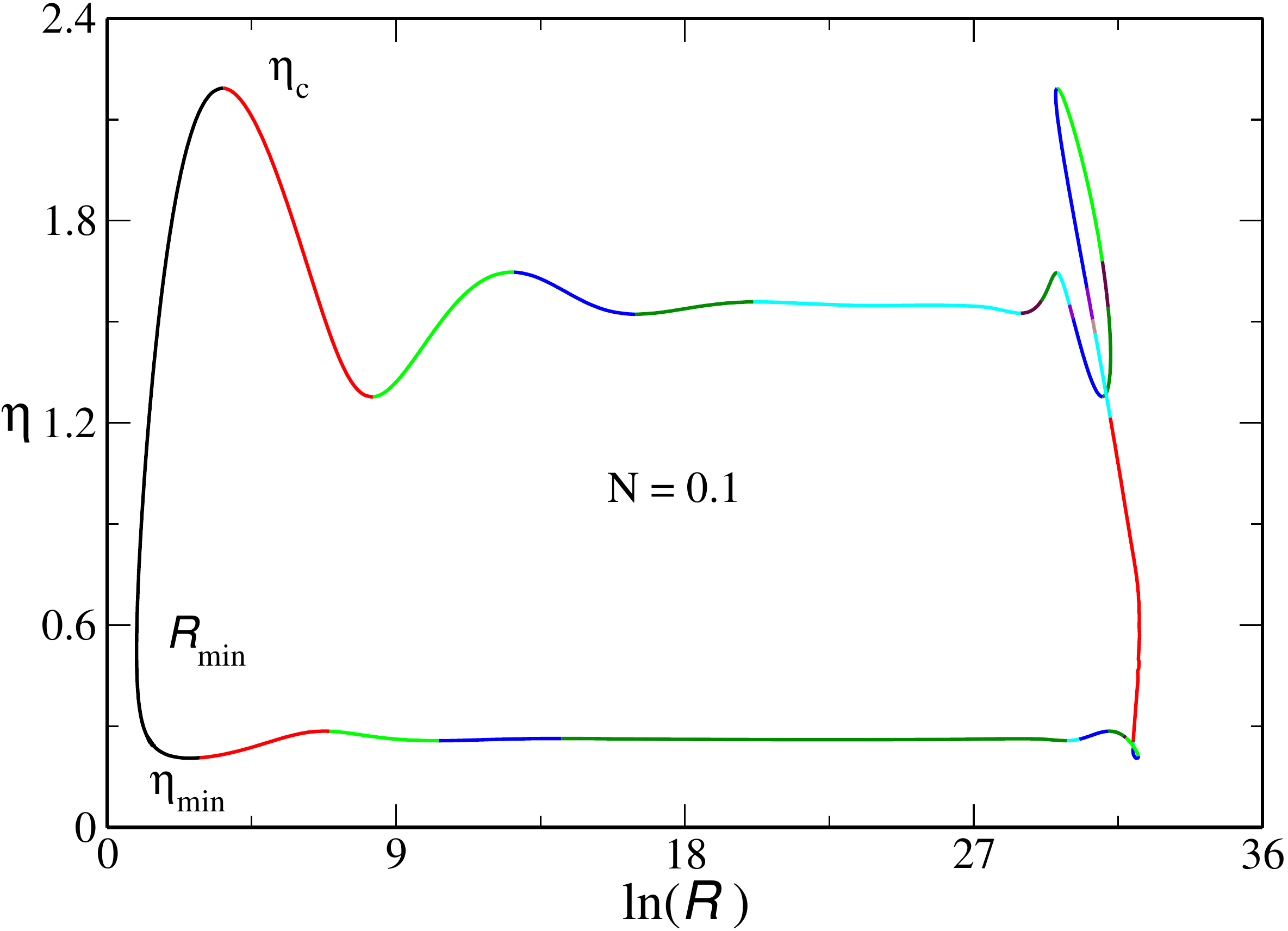}
\caption{Dimensionless inverse
temperature $\eta$ as a function of the energy
density contrast  ${\cal R}=\epsilon(0)/\epsilon(R)$ for
$N=0.1$.}
\label{Zeta_R_N01a_unified}
\end{center}
\end{figure}

In Figs. \ref{ZLambda_R_N01a_unified} and \ref{Zeta_R_N01a_unified} we have
plotted the dimensionless energy $\Lambda$ and the
dimensionless temperature $\eta$ as a function of the energy density contrast
${\cal R}=\epsilon(0)/\epsilon(R)$ along the series of equilibria. As for
the nonrelativistic gas and for the self-gravitating radiation, the 
density contrast parametrizes the series of equilibria.

We first observe that there exist a point ${\cal R}_{\rm
min}$  at which the density contrast is minimum. Starting from that point,
the curves
$\Lambda({\cal R})$ and $\eta({\cal R})$  present two branches which both
display damped oscillations at sufficiently high density contrasts. These two
branches, and their oscillations, correspond to the two spirals on the caloric
curve of Fig. \ref{kcal_N01_linked_colorsPH}. The upper branch (low energies and
low temperatures) corresponds to the cold spiral and the lower branch (high
energies and high temperatures) corresponds to the hot
spiral.\footnote{We see that, for very high density contrasts, the oscillations
are revived and the curves $\Lambda({\cal R})$ and $\eta({\cal R})$ have a
complicated behaviour. Furthermore, the two branches merge at a maximum density
contrast ${\cal R}_{\rm
max}$. This corresponds to the fact that, deep into the spirals, the caloric
curve finally goes backwards and superimposes the original curve. Therefore, an
equilibrium state on the caloric curve with a given value of energy and
temperature $(\Lambda,\eta)$ may correspond to two (or more) configurations
with different density contrasts (see Fig. \ref{branches_new2} for an
illustration of this phenomenon). However, the configurations with the highest
density contrasts are unstable so we will not consider these states in the
sequel. For that reason, the left parts of Figs.  \ref{ZLambda_R_N01a_unified}
and
\ref{Zeta_R_N01a_unified}  will be called ``relevant'' while the right parts
will
be called ``irrelevant'' (see footnote 51). We just
consider the relevant parts.}

\begin{figure}
\begin{center}
\includegraphics[clip,scale=0.3]{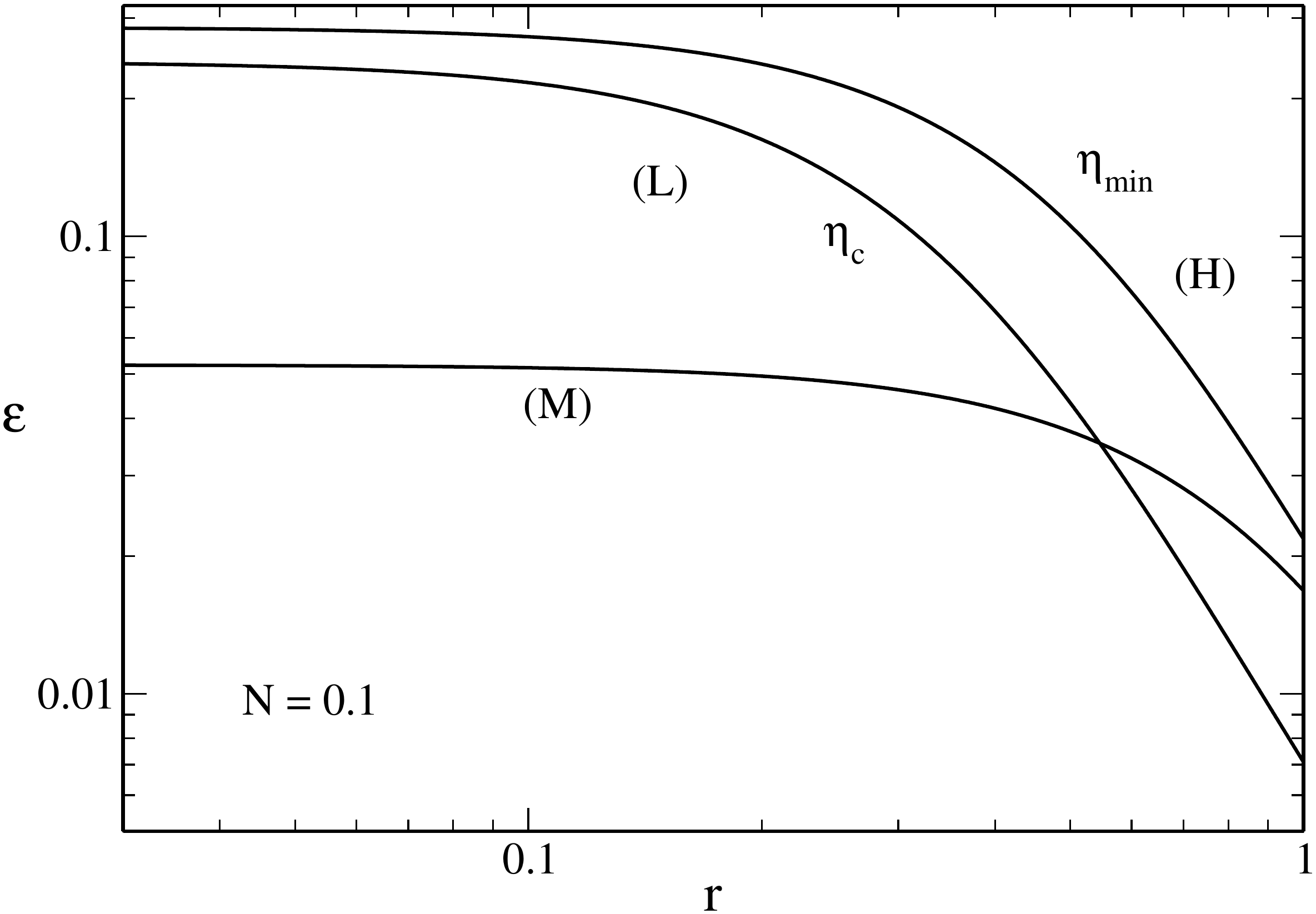}
\caption{Energy density profiles $\epsilon(r)$ corresponding to different
equilibrium states on
the caloric curve of Fig. \ref{kcal_N01_linked_colorsPH}. The curve
(M) corresponds to an equilibrium state with a medium energy and a
medium temperature. 
The curve (L) corresponds to
an equilibrium 
state with a low energy and a low temperature (specifically $\eta_c$). The 
curve (H)
corresponds to an equilibrium  state with a high energy and a high temperature
(specifically $\eta_{\rm min}$).}
\label{profiles_RML_N01}
\end{center}
\end{figure}

Actually, it is more physical to study how the  density contrast changes
with the energy or with the temperature. Starting from $\Lambda({\cal R}_{\rm
min})$ in Fig. \ref{ZLambda_R_N01a_unified} we see that
the density contrast increases
monotonically when $\Lambda$ increases
towards $\Lambda_c$ (i.e. when we reduce the energy) or when $\Lambda$
 decreases towards $\Lambda_{\rm min}$ (i.e. when we increase the energy).
Similarly, starting
from $\eta({\cal R}_{\rm
min})$ in Fig. \ref{Zeta_R_N01a_unified} we see that the density
contrast increases
monotonically when $\eta$ increases
towards $\eta_c$ (i.e. when we reduce the
temperature) or when $\eta$ decreases towards  $\eta_{\rm
min}$ (i.e. when we increase the the temperature).
This behavior is illustrated in Fig. \ref{profiles_RML_N01} showing the energy
density profile $\epsilon(r)$ of different equilibrium states with low (L),
medium (M), and high (H) energies and temperatures. This figure illustrates the
fact that the
density
contrast increases when, starting from the ``middle'' point of the caloric
curve with the minimum
density contrast ${\cal R}_{\rm min}$, we either increase or decrease the
energy and the temperature (i.e. we move towards the hot or the cold spiral).

In the microcanonical
ensemble, the
system becomes unstable after the first turning point of energy on each
branch  ($\Lambda_c$ and $\Lambda_{\rm
min}$). In the canonical
ensemble, the
system becomes unstable after the first turning point of temperature on each
branch  ($\eta_c$ and $\eta_{\rm
min}$). Therefore, instability occurs when the system reaches a certain critical
concentration: ${\cal R}(\Lambda_c)$ and ${\cal R}(\Lambda_{\rm min})$ in the
microcanonical ensemble;  ${\cal R}(\eta_c)$ and ${\cal R}(\eta_{\rm
min})$ in the  canonical ensemble.

\subsection{Central temperature}

\begin{figure}
\begin{center}
\includegraphics[clip,scale=0.3]{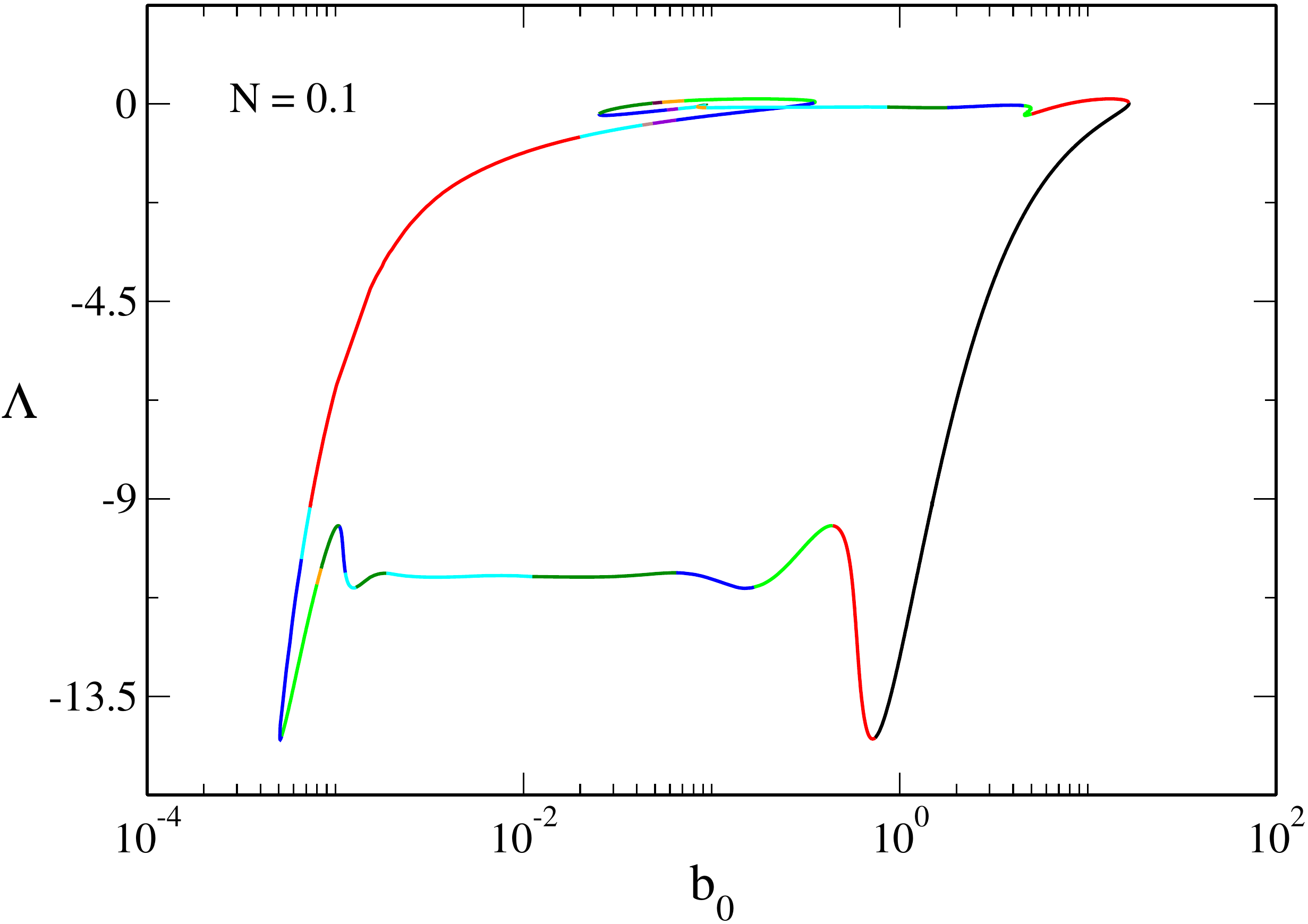}
\caption{Normalized
energy $\Lambda$ as a function of the inverse central temperature  $b_0$ for
$N=0.1$.}
\label{ZLambda_b0_N01c_unified}
\end{center}
\end{figure}

\begin{figure}
\begin{center}
\includegraphics[clip,scale=0.3]{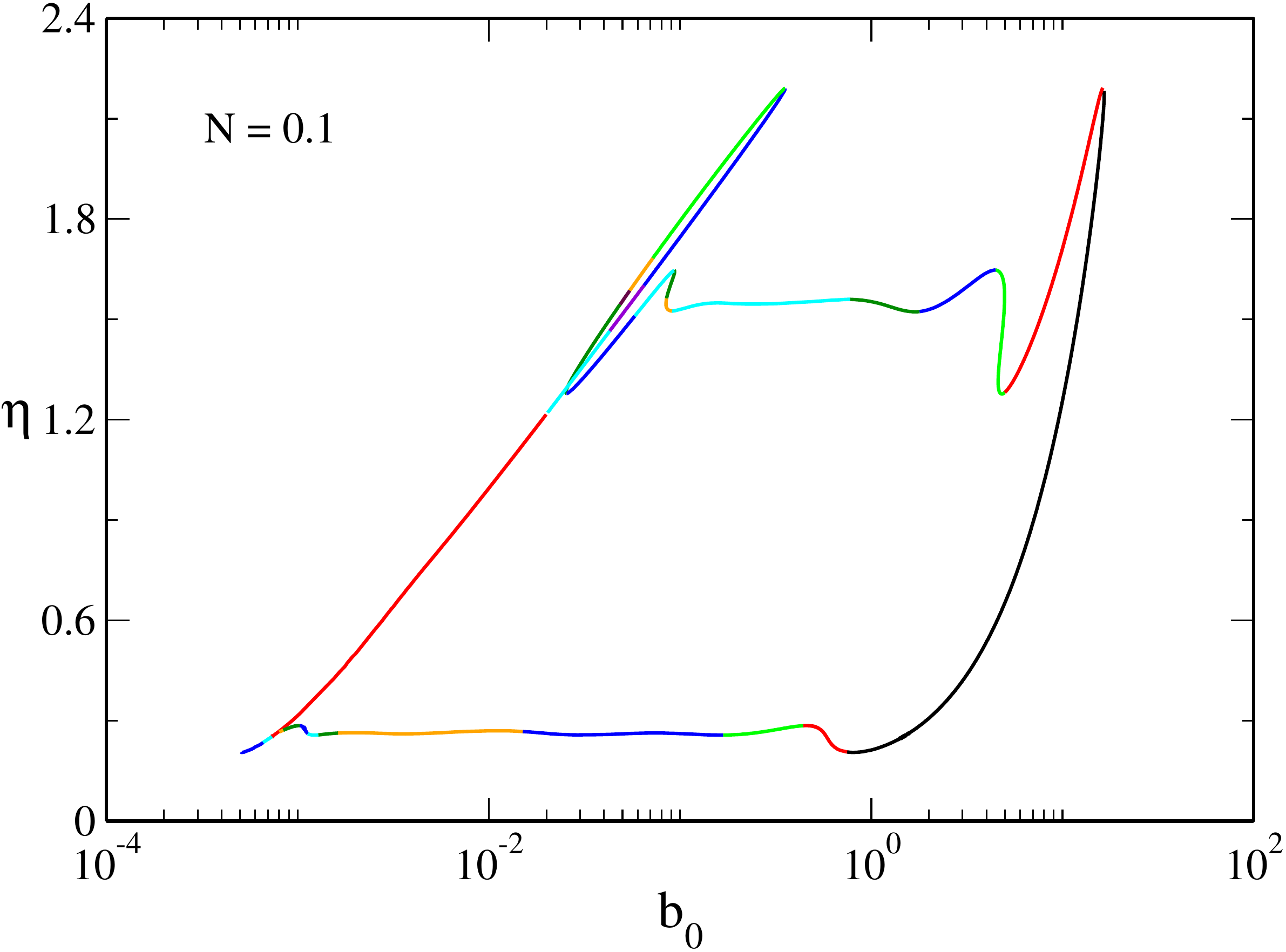}
\caption{Normalized inverse
temperature $\eta$ as a function of the
inverse central temperature  $b_0$ for
$N=0.1$.}
\label{Zeta_b0_N01c_unified}
\end{center}
\end{figure}

In Figs. \ref{ZLambda_b0_N01c_unified} and \ref{Zeta_b0_N01c_unified}  we
have plotted the normalized energy and the normalized inverse Tolman temperature
as a function of the normalized inverse central temperature. We note that the
curve $b_0(\Lambda)$ is very different from the caloric curve $\eta(\Lambda)$
from Fig. \ref{kcal_N01_linked_colorsPH}. In particular, it does not form
spirals. Actually, the curves from Figs.
\ref{ZLambda_b0_N01c_unified} and \ref{Zeta_b0_N01c_unified} are more
closely related to the 
curves from  Figs. \ref{ZLambda_R_N01a_unified} and \ref{Zeta_R_N01a_unified}
showing that $b_0$ plays a role similar to the density contrast (it parametrizes
the series of equilibria). On the other hand,  if we restrict ourselves to the
stable equilibrium
states, we observe on Fig. \ref{Zeta_b0_N01c_unified}  that the central
temperature $T_0$ ``follows'' the Tolman temperature $T_{\infty}$,
i.e., $T_0$
increases
monotonically with
$T_{\infty}$ (except in very small intervals close to the critical
points).\footnote{This can be understood as follows. We recall that
$b_0=mc^2/k_B T_0$
and $\eta=GNm^2/k_B T_{\infty}R$. In the nonrelativistic limit where
$T_0=T_{\infty}$, we have $\eta=\nu b_0$. For $\nu\rightarrow 0$
(see Sec. \ref{sec_nzero}), this linear relationship is valid almost everywhere
except
for very small values of $\eta$ and $b_0$ corresponding to large temperatures
where the system is strongly relativistic. For larger values of $\nu$, the
relation between  $\eta$ and $b_0$ is not linear anymore (see Fig.
\ref{Zeta_b0_N01c_unified}) but it
still remains monotonic in the range of stable states.} This is different
from the energy density contrast that presents a minimum value in the
``middle'' of the caloric curve.
The relation between the
evolution of $T_0$ and 
$T_{\infty}$ is more complicated for the unstable
equilibrium states but these states are not physically relevant.

\subsection{Entropy and free energy}

\begin{figure}
\begin{center}
\includegraphics[clip,scale=0.3]{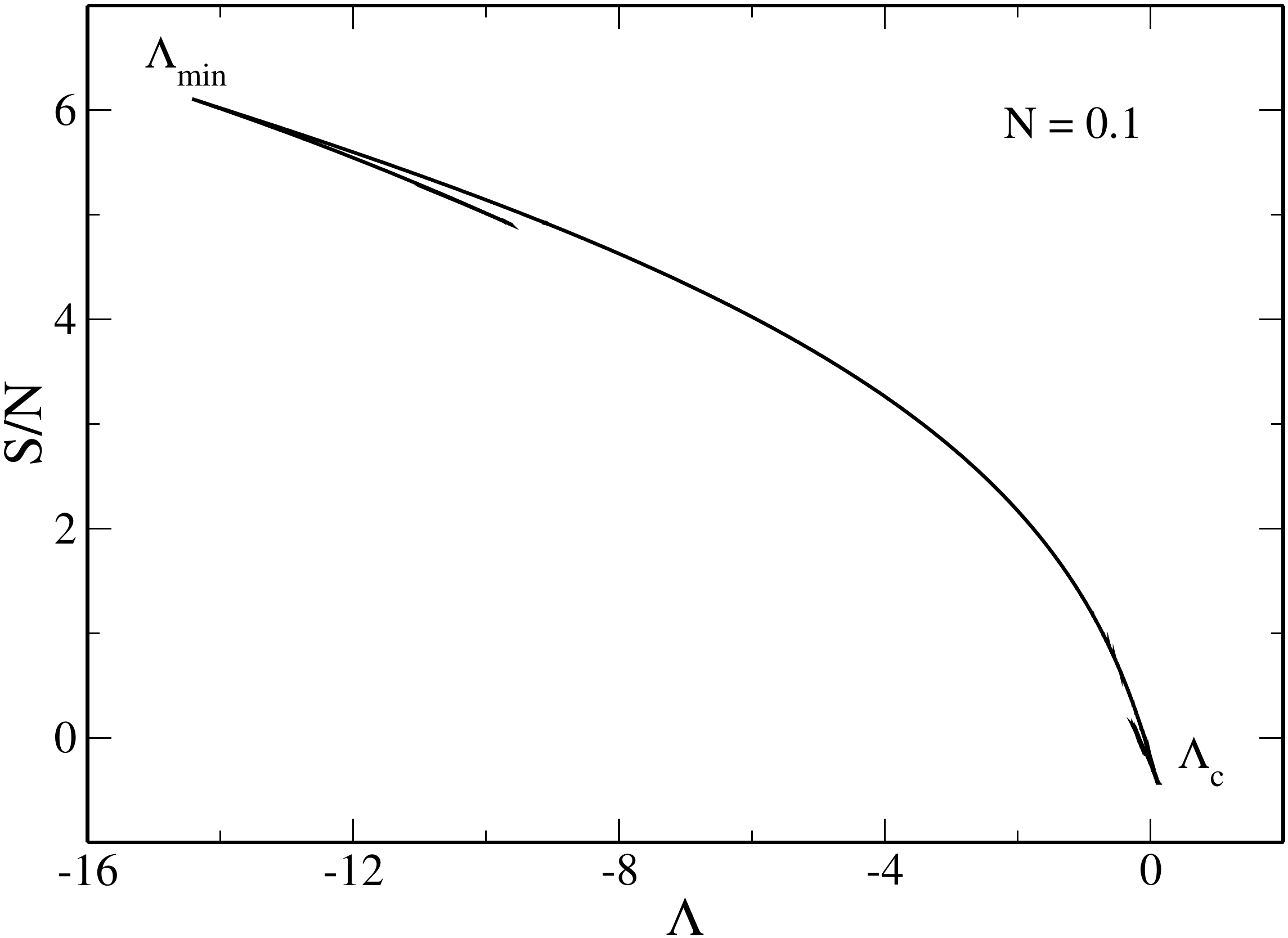}
\caption{Entropy per particle $S/N$ as a function of the
inverse normalized energy $\Lambda$ for $N = 0.1$. We note that the
unstable equilibrium states
have a lower entropy than the stable equilibrium
states. This is of course
expected
since the stable equilibrium states in the microcanonical
ensemble correspond to maxima of entropy at fixed energy
and particle number.}
\label{SLambda_N01b_unified}
\end{center}
\end{figure}

In Fig. \ref{SLambda_N01b_unified} we have plotted the entropy $S/N$ as a
function of
the normalized energy $\Lambda$. As explained in
\cite{pt,rgf}, the curve presents
some spikes at
the extremal
energies $\Lambda_{\rm min}$ and $\Lambda_c$. This is because $\delta
S=\beta_{\infty}
\delta E$ (for a fixed value of $N$)  so
that the
curves
$S(\epsilon_0)$ and $\Lambda(\epsilon_0)$ reach their extremal values at
the same points,  specified by their central density $\epsilon_0$, in the
series of equilibria. We also
note the presence of two tiny convex dips in the curve $S(\Lambda)$ associated
with
the two regions of negative specific heats appearing on the caloric curve of
 Fig. \ref{kcal_N01_linked_colorsPH}.

\begin{figure}
\begin{center}
\includegraphics[clip,scale=0.3]{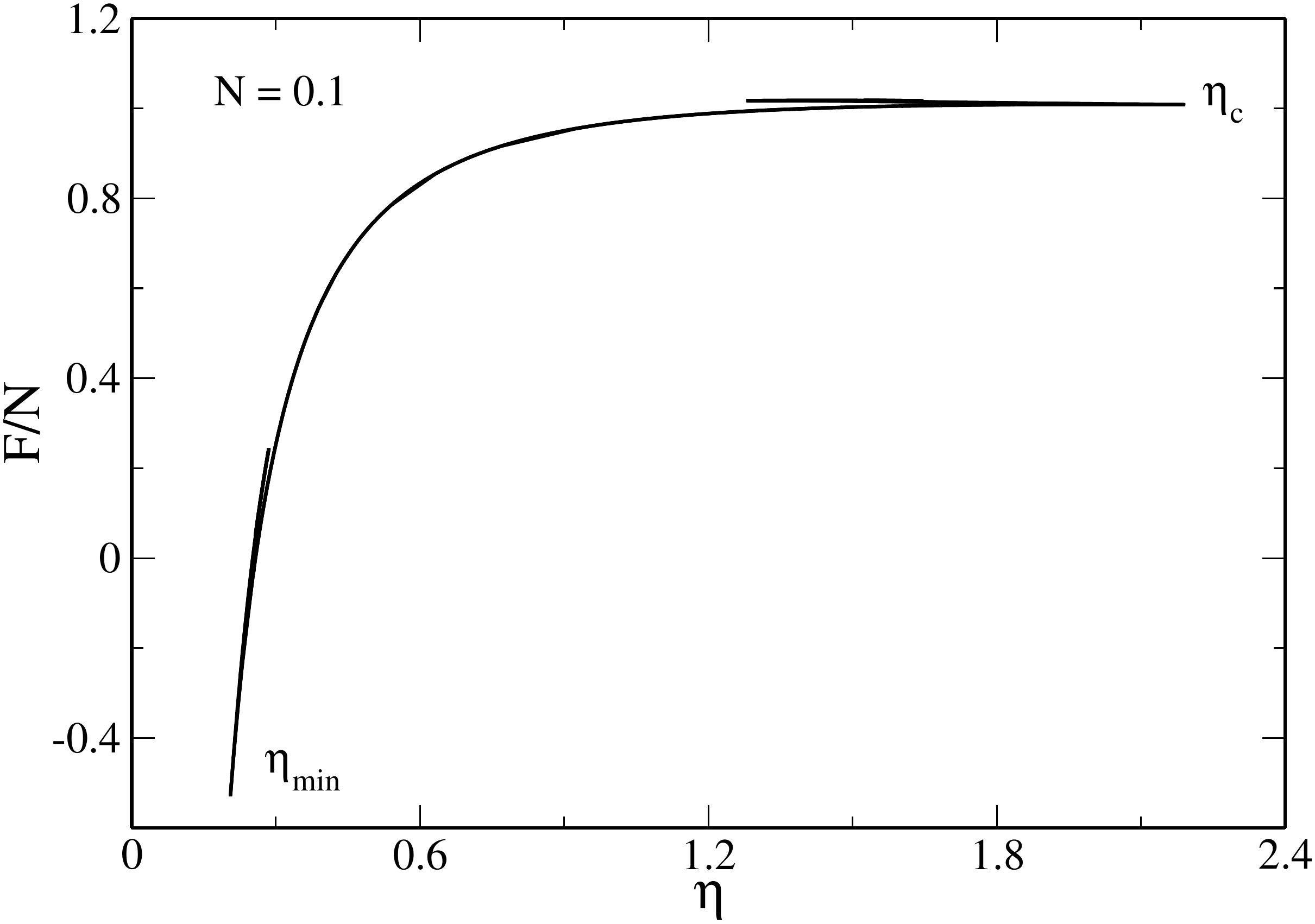}
\caption{Free energy per particle $F/N$ as a function of the
inverse normalized temperature $\eta$ for $N = 0.1$. We note that the
unstable equilibrium states
have a higher free energy than the stable equilibrium
states. This is of course
expected
since the stable equilibrium states in the canonical ensemble correspond
to
minima of free
energy at fixed particle number.}
\label{Feta_N01c_unified}
\end{center}
\end{figure}

In Fig. \ref{Feta_N01c_unified} we have plotted the free energy $F/N$ as a
function of
the normalized inverse temperature $\eta$. As explained
in \cite{pt,rgf}, the curve
presents some spikes at the
extremal inverse temperatures $\eta_{\rm min}$ and $\eta_c$. This is
because $\delta F=-S \delta T_{\infty}$
(for
a fixed value of $N$) so
that the curves $ F(\epsilon_0)$ and $\eta(\epsilon_0)$ reach their
extremal values at the
same points, specified by their central density $\epsilon_0$, in the
series of equilibria.

\subsection{Ensembles inequivalence}

From the preceding results, we note that the thermodynamical instabilities
occur at different points in the microcanonical and canonical ensembles.
Therefore,  the statistical ensembles are inequivalent for 
self-gravitating systems. We can check on the present example the general
result according to which ``canonical stability implies microcanonical
stability'' \cite{cc}.
Generally speaking, this is due to the fact that the microcanonical ensemble is
more constrained than the canonical ensemble (because of the energy
conservation).
In the present example, this
manifests itself  (in relation to the Poincar\'e criterion) by the fact  that
the first
turning point of temperature occurs before the first turning point of energy in
the series of equilibria.

{\it Remark:} Statistical ensembles may be inequivalent
for systems with long-range interactions because their energy
is nonadditive
so that the canonical ensemble cannot be deduced from the microcanonical one
by considering a subpart of a large system. However, ensembles inequivalence is
not compulsory for all systems with long-range
interactions. For example, for the HMF model \cite{houches,cdr,campabook},  the
statistical
ensembles are equivalent.

\subsection{Dynamical versus thermodynamical stability}
\label{sec_dvts}

A statistical equilibrium state is thermodynamically stable in the
microcanonical ensemble if it is a  maximum of entropy at fixed
energy and particle number. This corresponds to the ``most probable''
equilibrium configuration accounting for the constraints of energy and particle
number \cite{rgf,rgb}. As we have seen, the equilibrium states on the main
branch between
$\Lambda_{\rm min}$
and $\Lambda_{c}$ are thermodynamically stable. They are also dynamically stable
with respect to the Vlasov-Einstein equations (governing the evolution of a
collisionless star cluster) in view of the general result
according to which ``thermodynamical stability
implies dynamical stability'' \cite{ih,ipser80,cc}.

Let us now consider the equilibrium states that are not thermodynamically
stable.

In Newtonian gravity, it has been shown
\cite{doremus71,doremus73,gillon76,sflp,ks,kandrup91} that all isotropic stellar
systems with a distribution function of the form $f=f(\epsilon)$ with
$f'(\epsilon)<0$ are dynamically stable with respect to the Vlasov-Poisson
equations. This implies, in particular, that all the equilibrium
states on the caloric curve of Fig. \ref{etalambda} are dynamically stable, even
those that
are thermodynamically unstable (e.g., those deep in the spiral).

By contrast, in general
relativity, there is a conjecture by Ipser \cite{ipser80} according to
which thermodynamically unstable
states are also dynamically unstable, so that thermodynamical and dynamical
stability actually coincide. If this conjecture is
confirmed,\footnote{As recalled in the Introduction the equivalence between
dynamical and thermodynamical stability in general relativity has been checked
numerically for several distribution functions including heavily truncated
isothermal distributions. It would be interesting to know if
it remains valid for box-confined isothermal spheres.} we can conclude that the
equilibrium states located after the
turning points of energy (as we progress along the spirals) in the caloric
curve of Fig. \ref{kcal_N01_linked_colorsPH} are both thermodynamically and
dynamically
unstable.

In order to reconcile these apparently contradictory results we have to
consider the timescale of the instability. It is very likely that the time to
develop the dynamical instability diverges as the level of relativity tends to
zero. Therefore, it is very long for weakly relativistic systems (becoming
comparable to the collisional relaxation time) and decreases for strongly 
relativistic systems. Since, for small or moderate values of $N$, the cold
spiral is made of weakly relativistic configurations, we conclude that the
instability at the minimum energy $E_c$ is essentially a
thermodynamical (slow/secular) instability. By contrast,  since the hot spiral
is 
made of strongly relativistic configurations, we conclude that the instability
at the maximum energy $E_{\rm max}$ is essentially a dynamical
(fast) instability.

The previous results apply to isolated systems in the microcanonical ensemble.
If we now consider the case of systems in contact with a thermal bath in the
canonical ensemble we get different results. A statistical equilibrium state is
thermodynamically stable in the
canonical ensemble if it is a minimum of free energy at fixed particle
number. On the other hand, it can be shown that thermodynamical stability in
the canonical ensemble coincides with dynamical stability with respect to the
Euler-Poisson equations in Newtonian gravity  \cite{aaiso,aaantonov,rgf} and
with respect
to the Euler-Einstein equations in general
relativity \cite{roupas1,roupas1E,gsw,fhj} (governing the evolution of a gaseous
star). Therefore, the equilibrium states on the
main branch between $\eta_{\rm min}$
and $\eta_{c}$ are both thermodynamically and dynamically stable. They cease to
be stable (thermodynamically and dynamically) after the turning points of
temperature.

\section{General case}
\label{sec_gc}

We now describe the caloric curves of a general relativistic self-gravitating
classical gas for an arbitrary number of particles $N$. As the sphere becomes
more compact ($N$ or $\nu$ increases) general relativistic effects are more
intense.

\subsection{The case $N<N_S'$}

The case $N<N_S'=0.128$ corresponds to the typical
example treated in Sec. \ref{sec_generic}.\footnote{The existence and the
value of the critical particle numbers $N_S'$, $N_S$ and $N_{\rm max}$ (see
below) can be understood from simple graphical constructions as explained in
Appendix \ref{sec_cons}.} 
The
caloric curve has the form of a double spiral (see Fig.
\ref{kcal_N01_linked_colorsPH}).

\begin{figure}
\begin{center}
\includegraphics[clip,scale=0.3]
{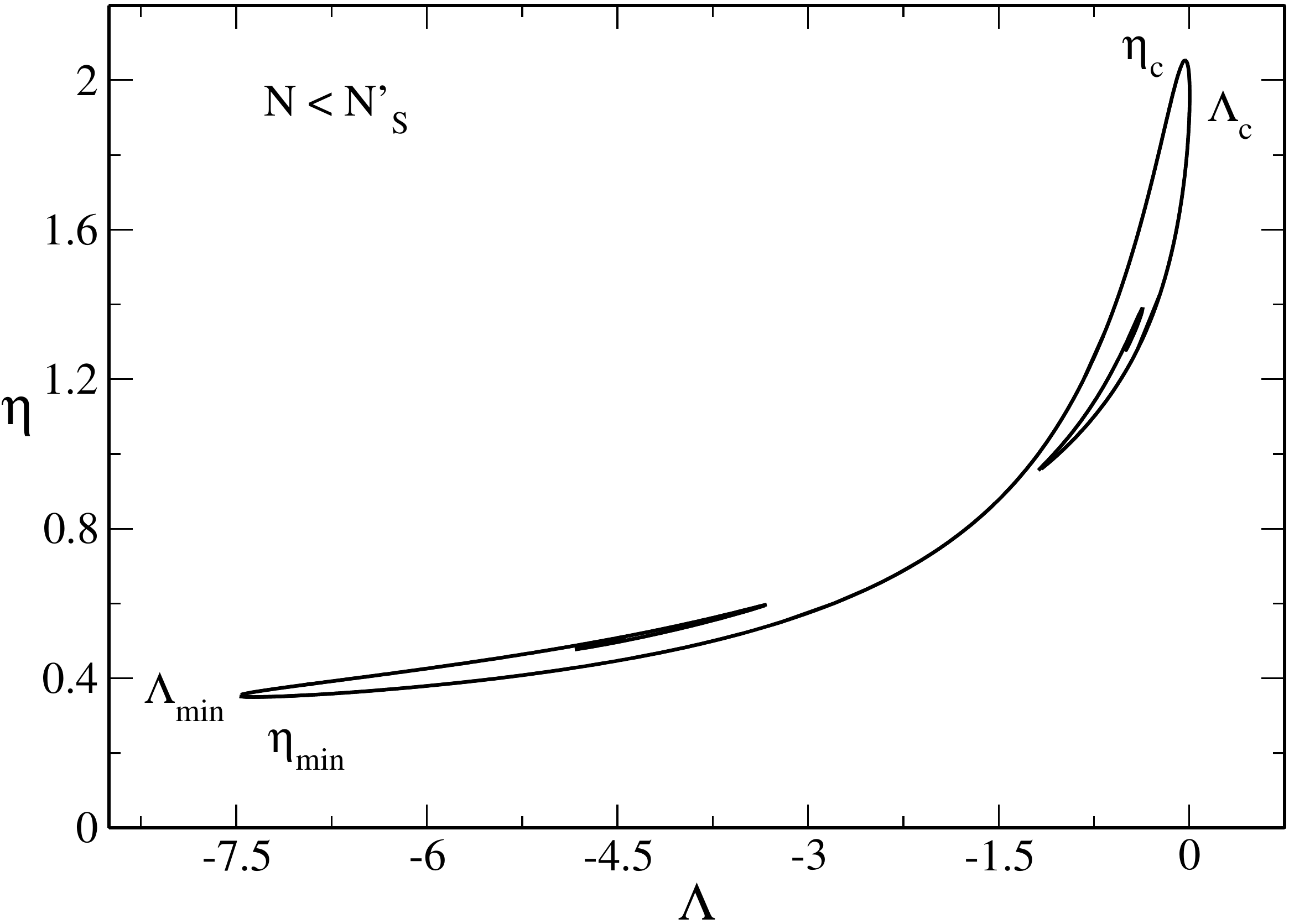}
\caption{Caloric curve for $N<N_S'=0.128$ (specifically  $N = 0.125$).
}
\label{kcal_N0125_linked_colors}
\end{center}
\end{figure}

When $N\rightarrow 0$, the hot
spiral  is pushed towards infinity ($\Lambda_{\rm min}\rightarrow -\infty$ and
$\eta_{\rm min}\rightarrow 0$)  and the rest of the caloric curve, consisting in
the main branch and the cold spiral, tends to a limit curve
corresponding to the nonrelativistic caloric curve of Fig. \ref{etalambda}. The
limit $N\rightarrow 0$ is studied in detail in Sec. \ref{sec_nzero}.

Inversely, as $N$ increases, the two spirals approach each other. For
$N<N_S'$, they remain clearly separated (see Fig.
\ref{kcal_N0125_linked_colors}).

\subsection{The case $N_S'<N<N_S$}

\begin{figure}
\begin{center}
\includegraphics[clip,scale=0.3]{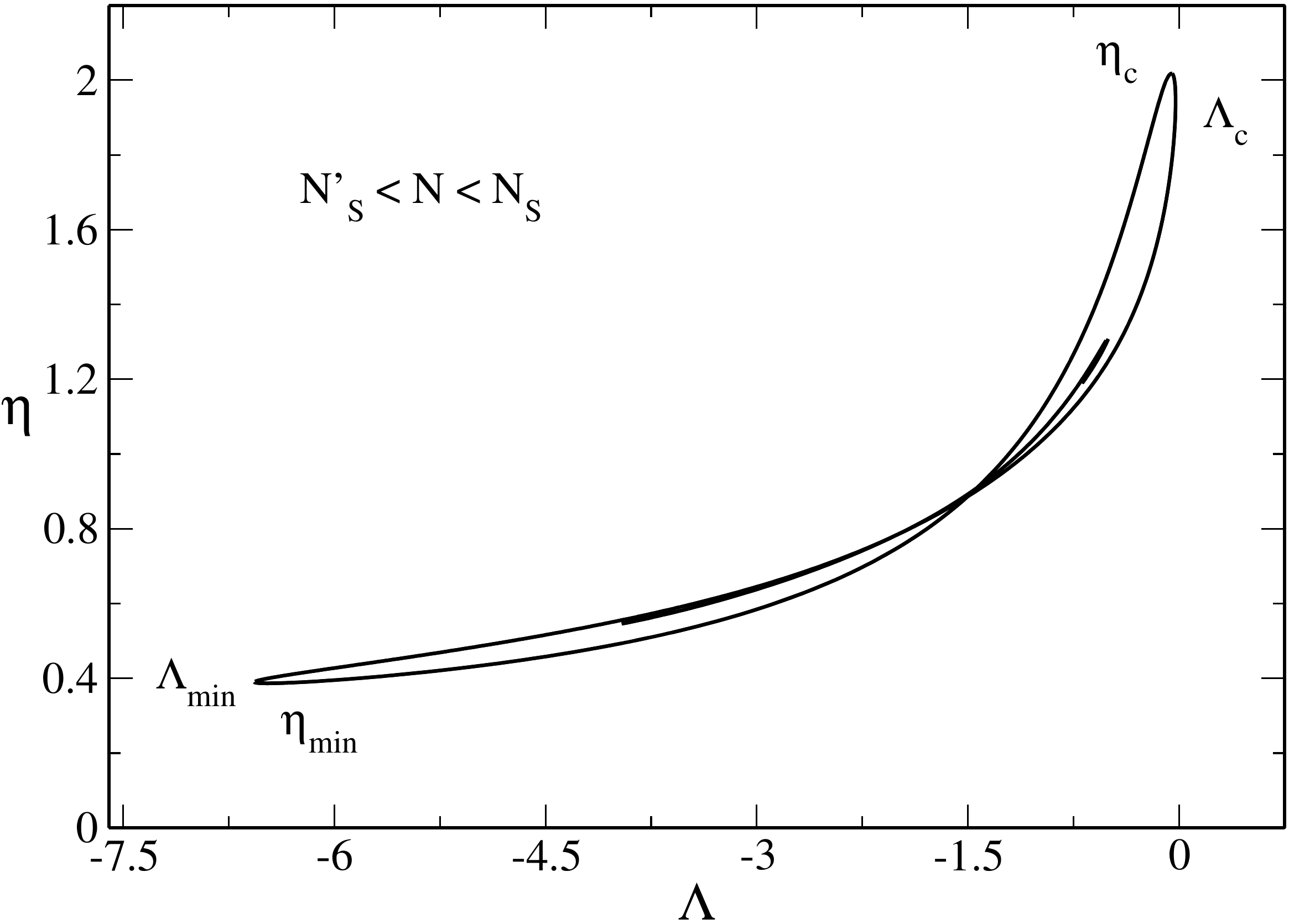}
\caption{Caloric curve for $N_S'=0.128<N<N_S=0.1415$
(specifically $N = 0.13$).}
\label{kcal_N0130_linked}
\end{center}
\end{figure}

For $N'_S=0.128<N<N_S=0.1415$, the caloric curve still
presents
two spirals but the
difference with the previous case is that the spirals are amputed (truncated)
and  touch each
other (see Fig. \ref{kcal_N0130_linked}).

\subsection{The case $N_S<N<N_{\rm max}$}

\begin{figure}
\begin{center}
\includegraphics[clip,scale=0.3]{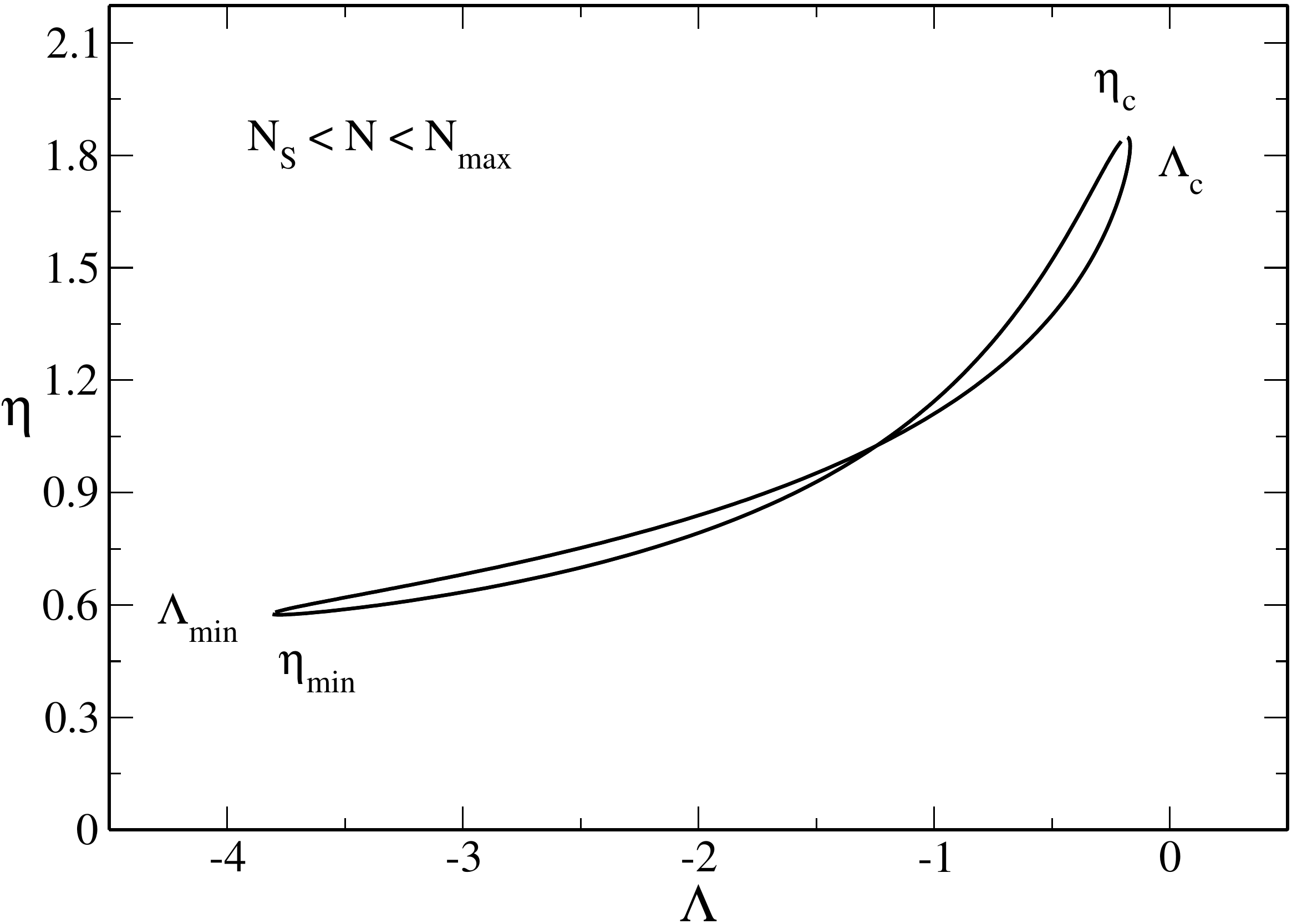}
\caption{Caloric curve for
$N_S=0.1415<N<N_{\rm max}=0.1764$ (specifically $N =
0.15$).}
\label{cal0p15_new}
\end{center}
\end{figure}

For $N_S=0.1415<N<N_{\rm max}=0.1764$, the caloric curve has the form of a
single
loop  resembling the symbol $\infty$ (see Fig. \ref{cal0p15_new}). As $N$
increases further, the loop shrinks more and more until it reduces to a point
$(\Lambda_*,\eta_*)=(-0.9829,1.2203)$ when $N=N_{\rm max}$. For
$N>N_{\rm max}$, no
equilibrium state is possible.

\subsection{The critical point  $N=N_{\rm max}$}

\begin{figure}
\begin{center}
\includegraphics[clip,scale=0.3]{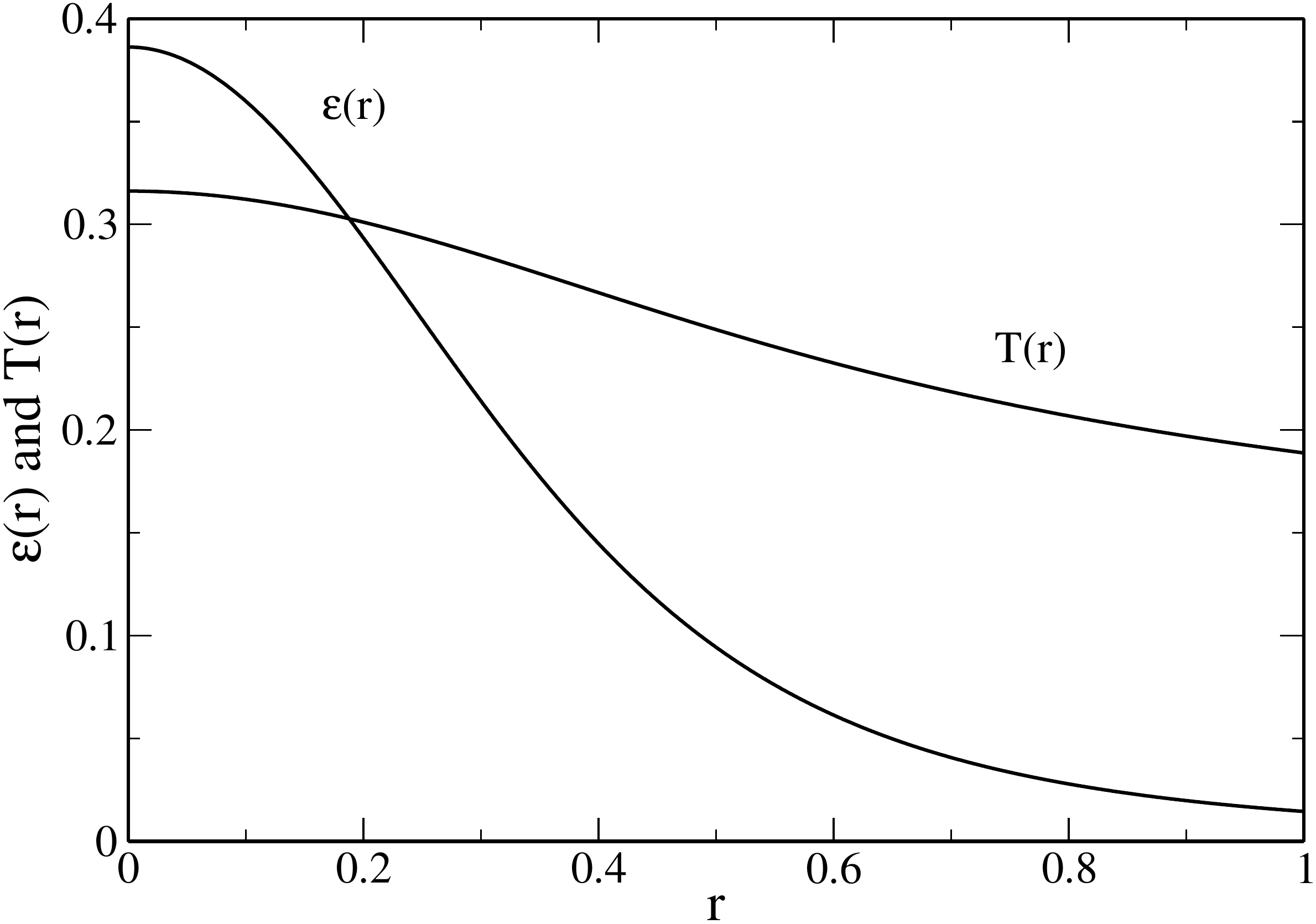}
\caption{Energy density profile $\epsilon(r)$ and temperature  profile
$T(r)$ at the
critical point $N = N_{\rm max}=0.1764$.}
\label{profile_epsilonNmax}
\end{center}
\end{figure}

The case $N=N_{\rm max}=0.1764$,
is very special because an
equilibrium
state exists only at a unique energy $\Lambda_*=-0.9829$ and a unique
temperature $ \eta_*=1.2203$. This equilibrium state is probably unstable or, at
best,
marginally stable.
Despite this very peculiar circumstance, the energy density and
temperature profiles
are regular (see Fig. \ref{profile_epsilonNmax}).

\subsection{The case  $N>N_{\rm max}$}

Equilibrium states may exist only below a maximum particle
number $N_{\rm max}=0.1764$. For $N>N_{\rm max}$ the system is expected to
collapse and form a black hole whatever the value of its energy and
temperature. Coming back to dimensional variables, the inequality $N\le N_{\rm
max}=0.1764$ required for the existence of an equilibrium state can be
written as
\begin{equation}
\label{a11}
\nu=\frac{GNm}{Rc^2}\le \nu_{\rm max}=0.1764.
\end{equation}
For a given box radius $R$, there is no statistical equilibrium state if $N$
is larger than 
\begin{equation}
N_{\rm max}=0.1764\, \frac{Rc^2}{Gm}.
\end{equation}
Alternatively, for a given particle
number $N$ there is no equilibrium state if the radius is smaller than  
\begin{equation}
R_{\rm min}=5.67\, \frac{GNm}{c^2}.
\label{buch}
\end{equation}
The minimum radius $R_{\rm min}$ is similar to the Schwarzschild radius
$R_S=2GM/c^2$ except that it is defined in terms of the
rest mass $Nm$ instead of the mass-energy $M$. Similarly, the maximum particle
number $N_{\rm max}$ is similar to a Schwarzschild
particle number $N_S=Rc^2/2Gm$, where $R$ is the box radius. We note that,
formally, Eq. (\ref{buch}) satisfies the  Buchdahl \cite{buchdahl}  inequality
$R\ge (9/8)R_S$.

\subsection{Physical caloric curve}

As we approach $N_{\rm max}=0.1764$ the series of equilibria  $\eta(\Lambda)$
becomes
very
complex. However, this complexity is only apparent as it essentially  concerns
the
unstable equilibrium states that are not
physically relevant. If we focus on the stable (actually
metastable) equilibrium states,
we just
need to consider the main branch of the series of equilibria. This defines the
physical caloric curve. This curve always has the
same shape. In the microcanonical ensemble it extends from $\Lambda_{\rm min}$
to $\Lambda_c$. In
the canonical ensemble it extends from $\eta_{\rm min}$ to $\eta_c$.
The only effect of increasing $N$ is to make this branch smaller and smaller
until
it completely disappears at $N_{\rm max}$. In this sense, the interpretation of
the caloric curve given in Sec. \ref{sec_generic} is valid for any $N\le N_{\rm
max}$. Since the physical caloric curve shrinks when the compactness parameter
$\nu$ increases, this implies that general relativistic effects render the
system more unstable. This feature will be studied in more detail in the
following sections.

\section{The limit $N\rightarrow 0$}
\label{sec_nzero}

In this section, we study the form of the caloric curve in the limit
$N\rightarrow 0$.  If we come back to dimensional
variables, this corresponds to $\nu=GNm/Rc^2\rightarrow 0$ which can be written
as $N\ll N_{\rm max}$ or $R\gg R_{\rm min}$. We
successively consider the cold spiral and the hot
spiral.

\subsection{The cold spiral}
\label{sec_nzerocold}

When $N\rightarrow 0$, the caloric curve $\eta(\Lambda)$ tends to a limit
curve corresponding to the caloric curve of a nonrelativistic classical
self-gravitating gas (see Fig. \ref{etalambda}).
The hot
spiral is rejected at infinity ($\Lambda_{\rm min}\rightarrow -\infty$ and
$\eta_{\rm min}\rightarrow 0$) as discussed in Sec. \ref{sec_ev}.

\begin{figure}
\begin{center}
\includegraphics[clip,scale=0.3]{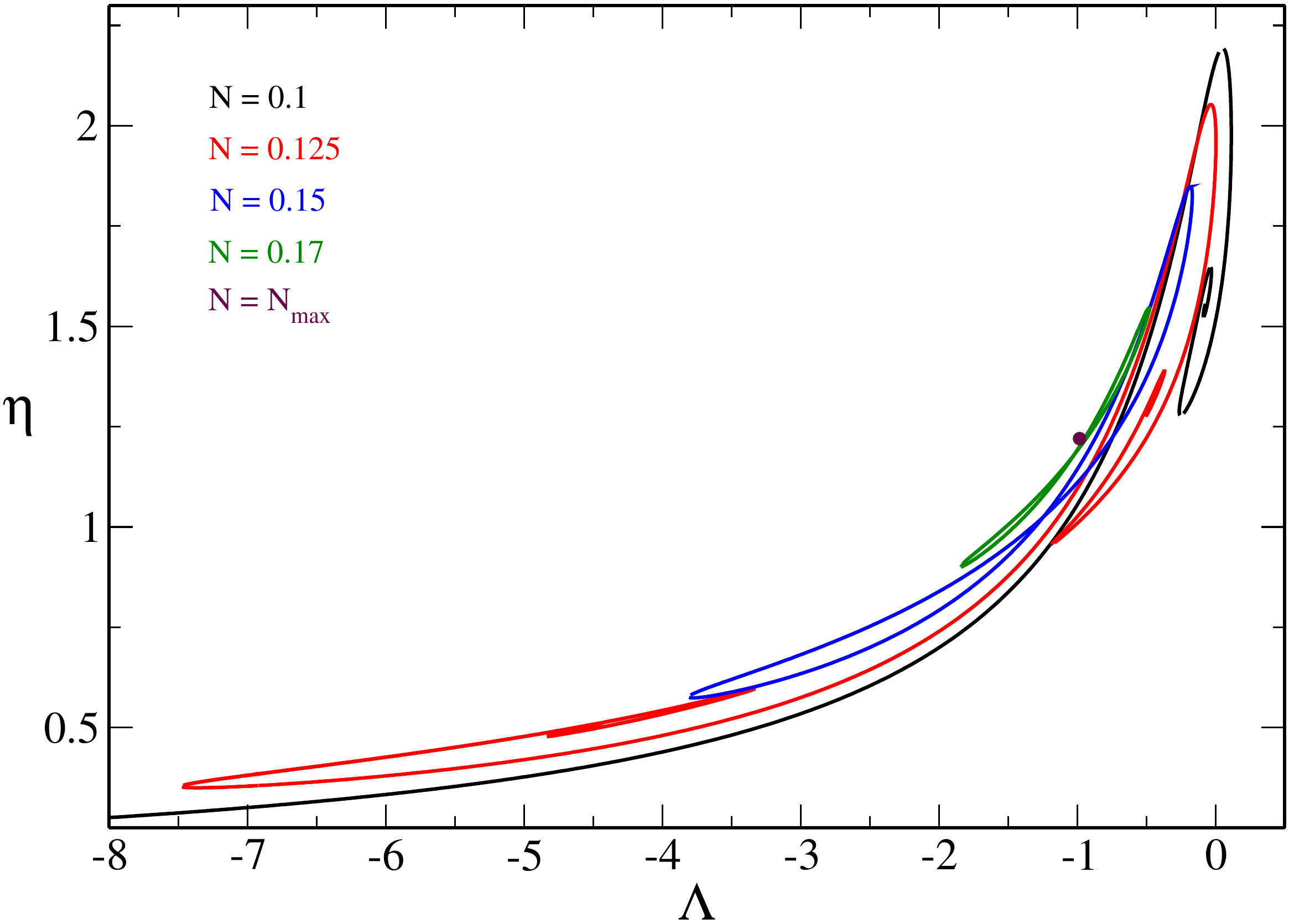}
\caption{Caloric curve
$\eta(\Lambda)$ for different values of $N$ (specifically
$N=0.1, 0.125, 0.15, 0.17, N_{\rm max}$). A zoom of the cold spiral for
small values of $N$ is provided in Fig. \ref{kcal_1PN_corrections_new_bis}.}
\label{total_droite}
\end{center}
\end{figure}

\begin{figure}
\begin{center}
\includegraphics[clip,scale=0.3]
{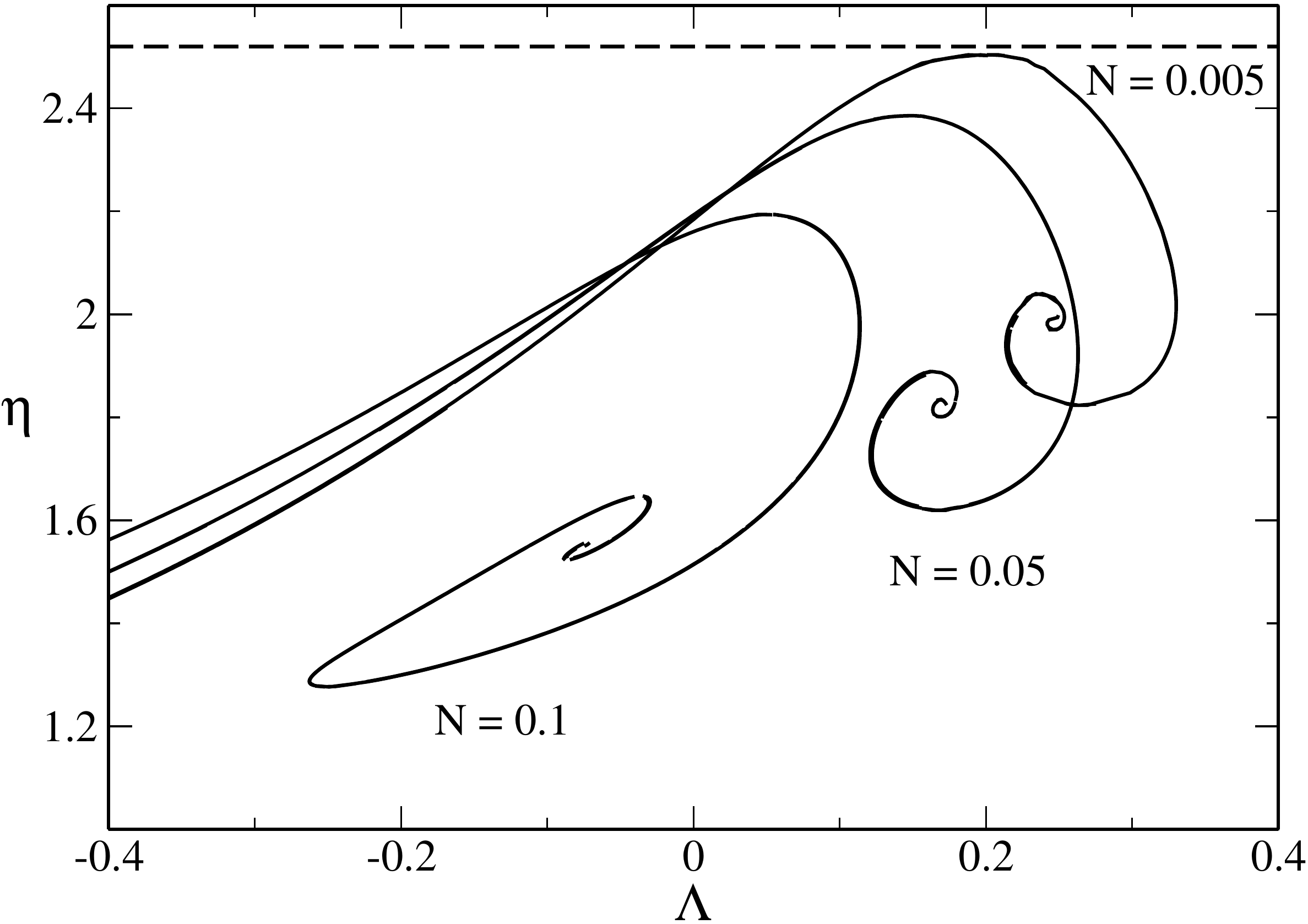}
\caption{Evolution of the
cold spiral as $N$ increases,
starting from the
nonrelativistic spiral ($N\rightarrow 0$). Relativistic effects increase the
critical energy and the critical temperature, making the system less stable.
The dashed line corresponds to $\eta_c=2.52$ (see the discussion in the main
text).}
\label{kcal_1PN_corrections_new_bis}
\end{center}
\end{figure}

In Figs. \ref{total_droite} and \ref{kcal_1PN_corrections_new_bis}, 
 using the dimensionless variables $\eta$ and $\Lambda$, we
show how the caloric curve $\eta(\Lambda)$ evolves with
$N$. Starting from the nonrelativistic caloric curve corresponding to
$N\rightarrow 0$, and increasing $N$, we see that the minimum normalized energy
and the minimum normalized temperature of the cold spiral both increase
($\Lambda_c$
and $\eta_c$ decrease) as $N$
increases.  Therefore, increasing $N$, i.e. increasing the
compactness parameter $\nu$, advances the
destabilization
of the system in the microcanonical and canonical ensembles at low 
energies and low temperatures. Therefore, general relativistic effects
render the system more unstable. Indeed, the instability
occurs sooner than in the nonrelativistic limit.

This result, obtained
from a fully general relativistic
treatment, can be compared with the former result obtained in 
\cite{aarelat1} from a semirelativistic treatment in which the gas is described
by a relativistic
equation of state while gravity is described by Newton's
theory. In
that case, as shown in Fig. 1 of \cite{aarelat1}, the critical energy 
increases with $N$ as in the present work but the critical
temperature remains constant ($\Lambda_c$ decreases while $\eta_c$
remains equal to $2.52$).
Therefore, the  semi relativistic treatment of
\cite{aarelat1} does not correcty account for the increase of the critical
temperature when gravity is described by Einstein's theory.

\begin{figure}
\begin{center}
\includegraphics[clip,scale=0.3]{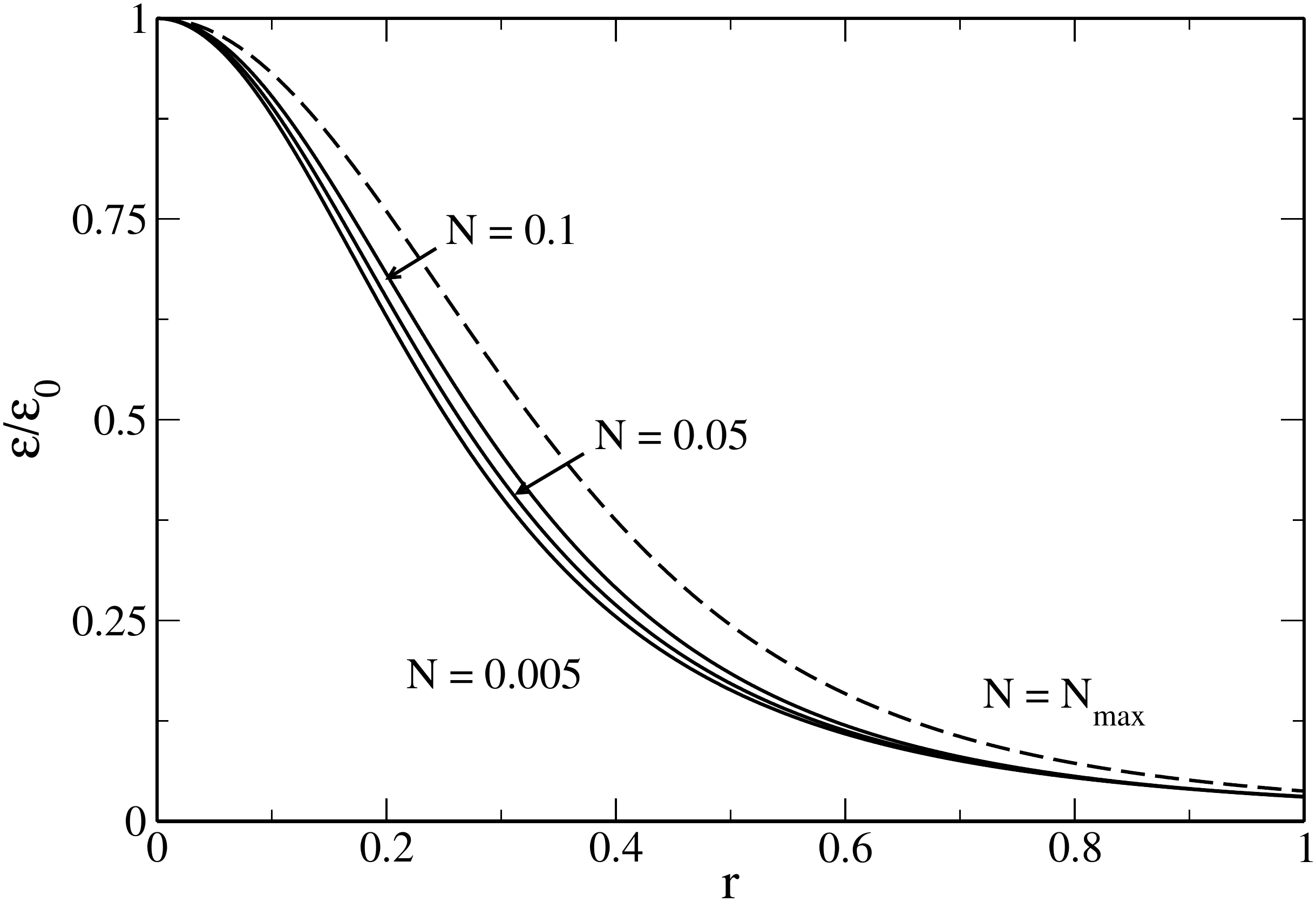}
\caption{
Normalized energy density profiles at  $\eta_c$ for different values of
$N$. The values of the central densities are $\epsilon_0=1.28\times 10^{-2}$
($N=0.005$), $\epsilon_0=0.123$ ($N=0.05$), and $\epsilon_0=0.235$
($N=0.1$). We have added in dashed line the energy density profile at
$N=N_{\rm max}$ ($\epsilon_0=0.386$) for comparison.}
\label{Nrho_rho0_bis.eps}
\end{center}
\end{figure}

In Fig. \ref{Nrho_rho0_bis.eps}, we have plotted the normalized energy
density profile at $\eta_c$ for different values
of $N$. For $N\rightarrow 0$, we recover the nonrelativistic isothermal profile.
As $N$ increases, the energy density profiles remain substantially the
same.

\begin{figure}
\begin{center}
\includegraphics[clip,scale=0.3]{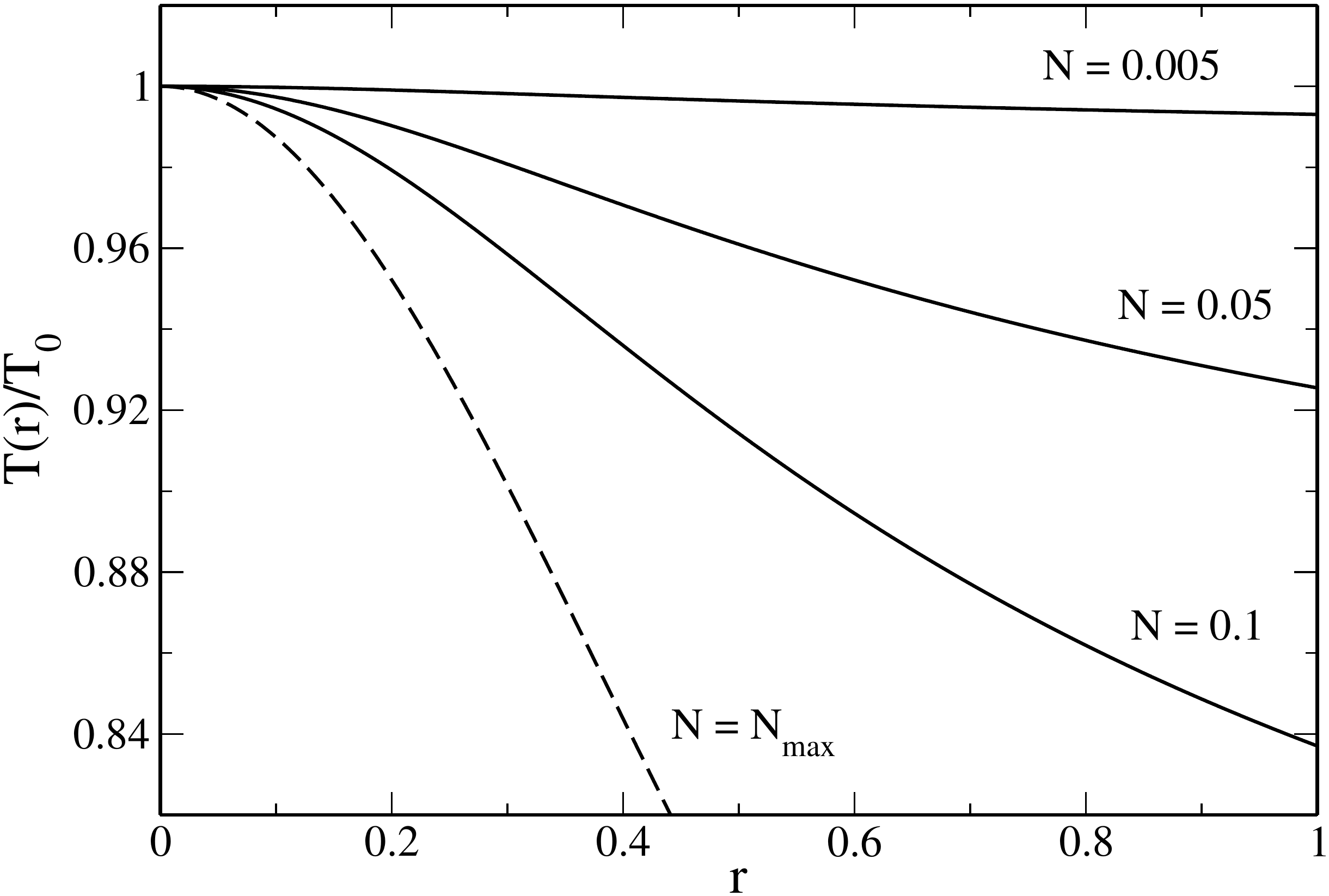}
\caption{Normalized temperature profiles at $\eta_c$ for different values of
$N$. The values of the central temperatures are $T_0
= 2.02 \times 10^{-3}$ ($N=0.005$), $T_0 = 2.39 \times 10^{-2}$ ($N=0.05$),
and $T_0 = 6.09 \times 10^{-2}$ ($N=0.1$). We have added in dashed line the
temperature profile at
$N=N_{\rm max}$ ($T_0=0.316$) for comparison.}
\label{NTr_T0_bis}
\end{center}
\end{figure}

In Fig. \ref{NTr_T0_bis}, we have plotted the normalized temperature profile at
$\eta_c$ for different values of $N$. For $N\rightarrow 0$, the temperature
becomes uniform as it has
to be for a Newtonian self-gravitating system at statistical equilibrium.
As $N$ increases, we see that the temperature becomes
spatially inhomogeneous even though the system is at statistical equilibrium.
This is a manifestation of the Tolman effect \cite{tolman} in general
relativity.

\subsection{The hot spiral}
\label{sec_dhs}

\begin{figure}
\begin{center}
\includegraphics[clip,scale=0.3]{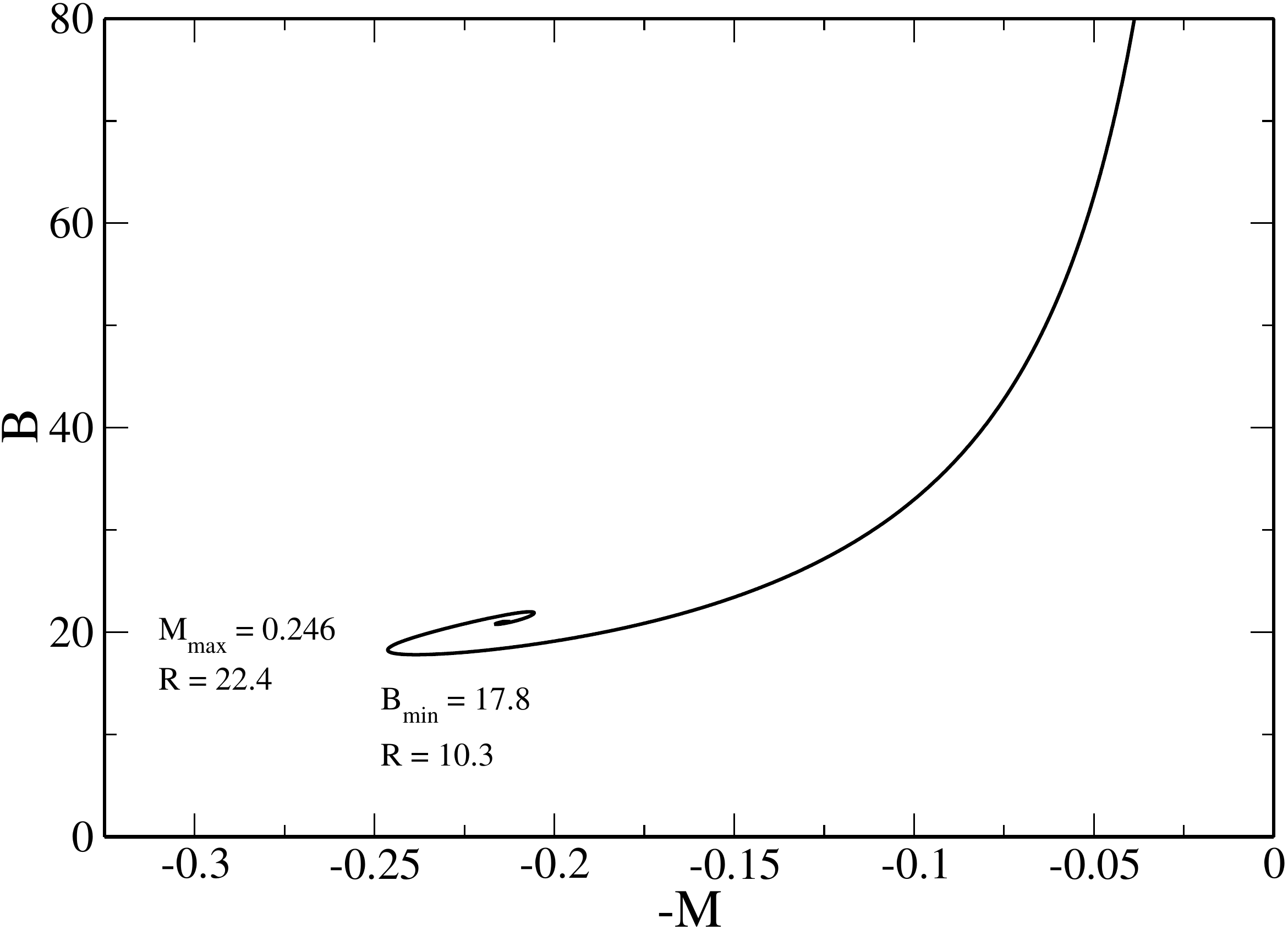}
\caption{Caloric curve ${\cal B}({\cal M})$ of the ultrarelativistic
self-gravitating  classical gas when $N\rightarrow 0$. The manner to obtain this
asymptotic curve is
explained in \cite{rgb}.}
\label{hot}
\end{center}
\end{figure}

For $N\rightarrow 0$, we have seen that the hot spiral of the caloric curve
$\eta(\Lambda)$ is rejected at infinity ($\eta_{\rm min}\rightarrow 0$ and
$\Lambda_{\rm min}\rightarrow -\infty$). In order to investigate the evolution
of the hot spiral  in the  limit
$N\rightarrow 0$ accurately, we need to introduce a
different
normalization of $E$ and
$T_{\infty}$. Indeed, the normalized variables $\Lambda$ and $\eta$ defined by
Eq. (\ref{ble}) are adapted to the nonrelativistic limit $k_B T \ll mc^2$ (cold
spiral) when
$\nu\rightarrow 0$. As shown in \cite{rgb}, the normalized variables adapted to
the ultrarelativistic limit $k_B T \gg mc^2$ (hot spiral) when $\nu\rightarrow
0$ are 
\begin{equation}
{\cal M}\equiv \frac{GM}{Rc^2}=-\Lambda \nu^2+\nu,\qquad {\cal B}\equiv
\frac{Rc^4}{GNk_B T_{\infty}}=\frac{\eta}{\nu^2}.
\end{equation}
Using the dimensionless variables introduced
in Appendix B of \cite{acf}, we get
\begin{equation}
{\cal M}\equiv M=-\Lambda N^2+N,\qquad {\cal B}\equiv
\frac{\beta_{\infty}}{N}=\frac{\eta}{N^2}.
\end{equation} 
The first quantity represents the mass-energy $M$ and the second quantity
represents the
inverse Tolman temperature $\beta_{\infty}$  divided by
the
particle number $N$.\footnote{By contrast, $\Lambda=-(M-Nm)c^2R/GN^2m^2$ (or
$\Lambda=-(M-N)/N^2$)
represents the binding energy divided by $N^2$ and $\eta=\beta_{\infty}
GNm^2/R$ (or $\eta=\beta_{\infty}N$) represents the inverse Tolman temperature
$\beta_{\infty}=1/(k_B
T_{\infty})$ multiplied by $N$.} 
When $N\rightarrow 0$,  the
caloric curve ${\cal B}({\cal M})$ tends to a limit curve corresponding to
the caloric curve of an ultrarelativistic classical self-gravitating gas (see
Fig. \ref{hot}) \cite{rgb}. The cold spiral is
rejected at infinity (${\cal M}_c\rightarrow 0$ and ${\cal B}_c\rightarrow
+\infty$) as discussed in Sec. \ref{sec_ev}.  The caloric curve  ${\cal B}({\cal
M})$ of Fig. \ref{hot} is similar, but not equivalent, to the caloric curve
$1/{\cal T}({\cal M})$ of the self-gravitating black-body radiation represented
in Fig. \ref{radiation} (see \cite{rgb} for a detailed discussion of the
analogies
and differences between the two caloric curves). The maximum mass
and the corresponding energy density
contrast of the hot spiral when $N\rightarrow 0$ are \cite{rgb}
\begin{equation}
{\cal
M}_{\rm max}=\frac{GM_{\rm
max}}{Rc^2}=0.24632, \qquad {\cal
R}_c=22.4.
\end{equation}
These are the same values as for the  self-gravitating black-body radiation
\cite{sorkin,aarelat2} (see Sec. \ref{sec_ur}). On the other hand, the maximum 
temperature and the corresponding energy density
contrast of the hot spiral when $N\rightarrow 0$ are \cite{rgb}
\begin{equation}
{\cal B}_{\rm min}\equiv \frac{Rc^4}{GNk_B (T_{\infty})_{\rm max}}=17.809, 
\qquad
{\cal
R}'_c=10.3.
\end{equation}
These values differ from those of the self-gravitating black-body radiation (see
Sec.
\ref{sec_ur})
for the reasons explained in \cite{rgb}.

\begin{figure}
\begin{center}
\includegraphics[clip,scale=0.3]{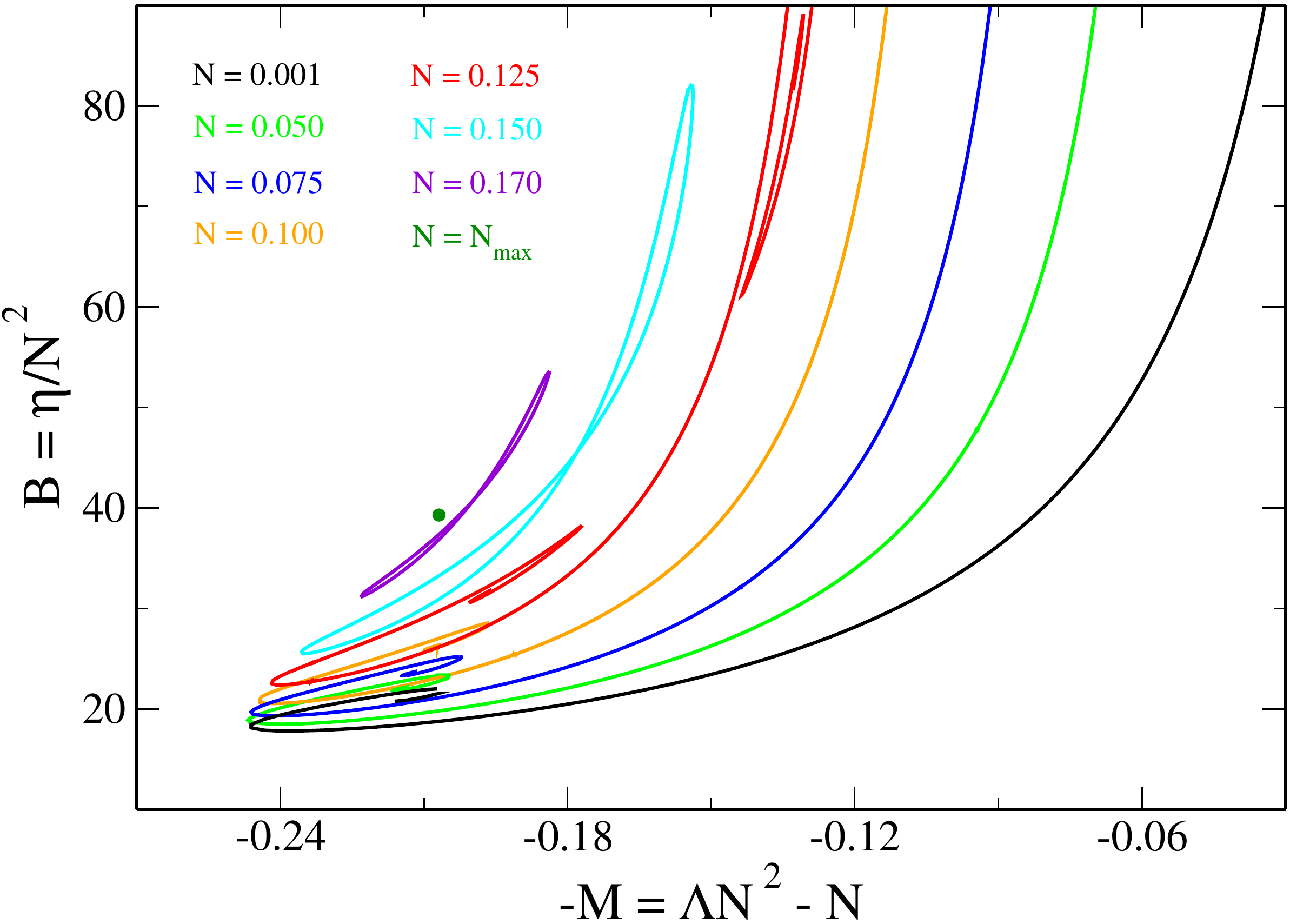}
\caption{Caloric curves ${\cal
B}({\cal M})$ for different values of
$N$ (specifically $N=0.001, 0.05, 0.075, 0.1, 0.125, 0.15, 0.17, N_{\rm
max}$). A zoom of the hot spiral is
provided in Fig.
\ref{deviazione_scalingZOOMPH}.}
\label{deviazione_scalingPH}
\end{center}
\end{figure}

\begin{figure}
\begin{center}
\includegraphics[clip,scale=0.3]{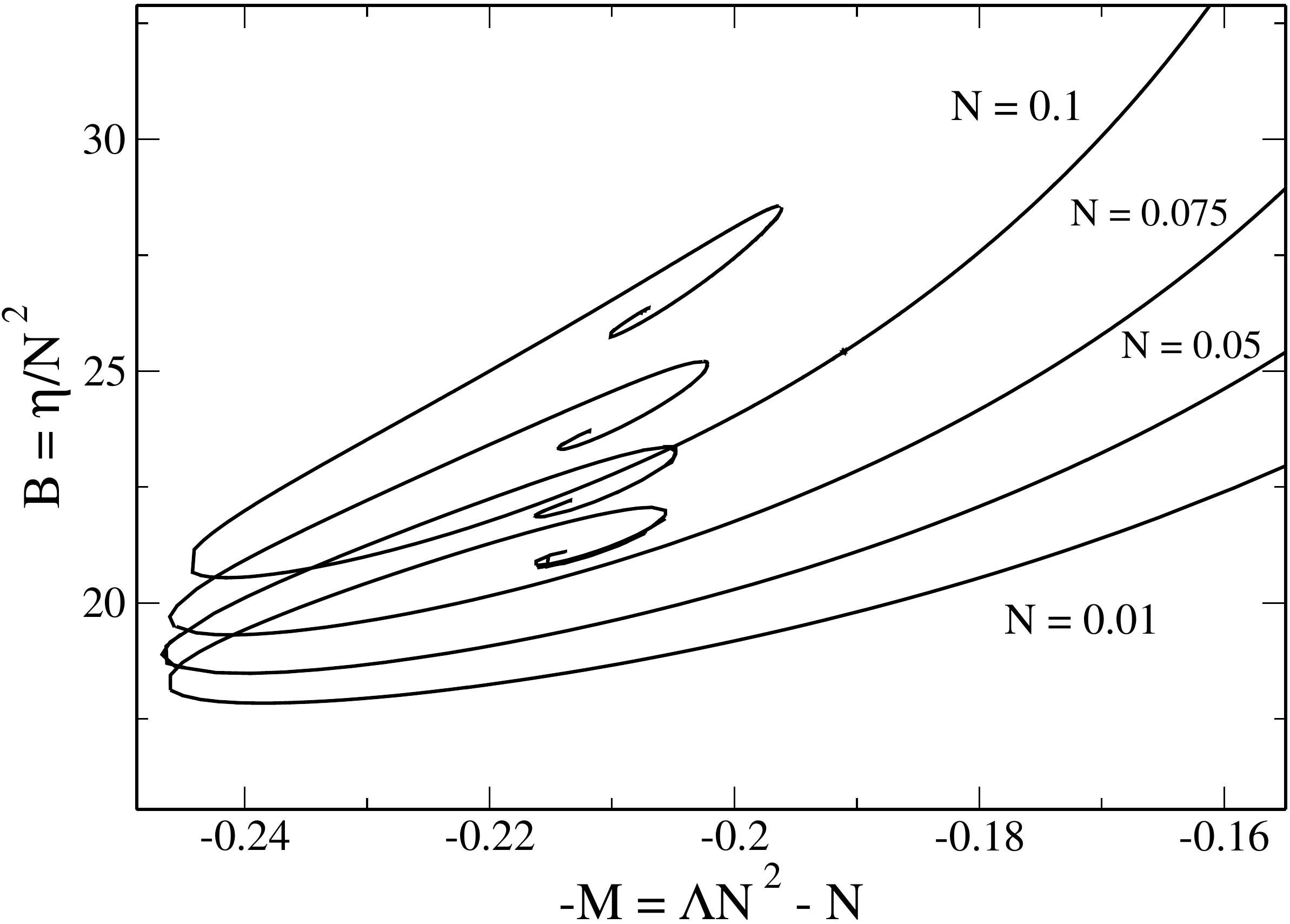}
\caption{Evolution of the hot spiral as $N$ increases,
starting from the ultrarelativistic spiral ($N\rightarrow 0$). Except
for very small values of $N$, increasing $N$ decreases the
critical energy and the critical temperature making the system less stable.}
\label{deviazione_scalingZOOMPH}
\end{center}
\end{figure}

In Figs. \ref{deviazione_scalingPH} and \ref{deviazione_scalingZOOMPH}, using
the dimensionless variables ${\cal B}$ and ${\cal M}$,  we show
how the caloric curve ${\cal B}({\cal M})$ evolves
with $N$. Starting from the ultrarelativistic caloric curve corresponding to
$N\rightarrow 0$, and increasing $N$, we see that the maximum mass ${\cal
M}_{\rm max}$ first slightly increases before  decreasing while the maximum
normalized temperature always decreases (${\cal B}_{\rm min}$ increases).
Therefore, except for very small values of $N$, increasing
$N$, i.e. increasing the compactness parameter $\nu$, advances the 
destabilization of the system
in
the microcanonical and canonical
ensembles at high energies and high temperatures.  Therefore, general
relativistic effects render the system more unstable. Indeed, the
instability occurs sooner than in the ultrarelativistic
limit.\footnote{It is important to realize  that when the compactness
parameter
$\nu$ increases the system is less relativistic on the hot spiral from the
viewpoint of kinematics since the ultrarelativistic limit corresponds to
$\nu\rightarrow 0$. Therefore, the system is more general relativistic but
less special relativistic.} We also note that ${\cal M}\ll 1$
so that isothermal spheres are always far from the black hole
(Schwarzschild) limit ${\cal
M}_{S}=1/2$.

\section{Evolution of the critical points with $N$}
\label{sec_ev}

In this section, we study how the critical points of the caloric
curve $\eta(\Lambda)$ identified in the preceding sections evolve with the
particle number $N$. 

\subsection{Energy and temperature}

\begin{figure}
\begin{center}
\includegraphics[clip,scale=0.3]{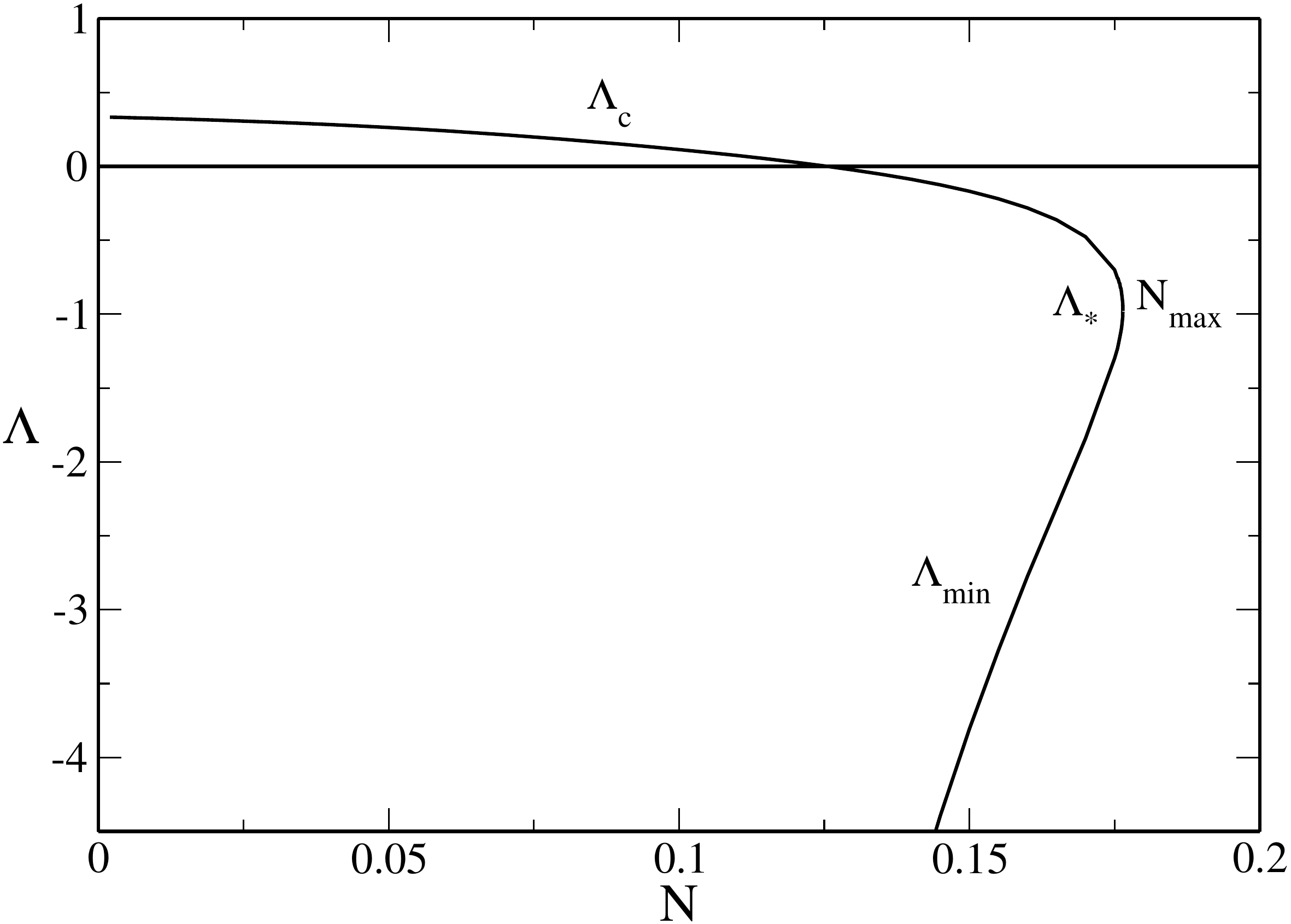}
\caption{Evolution of the critical energies $\Lambda_c$ and
$\Lambda_{\rm min}$ with $N$.}
\label{phase_lambda1}
\end{center}
\end{figure}

In Fig. \ref{phase_lambda1} we have plotted the evolution of the critical
energies $\Lambda_c$ and $\Lambda_{\rm min}$, corresponding to the cold and hot
spirals, as a function of $N$. For $N\rightarrow 0$, we find that $\Lambda_c$
tends
to the Newtonian value $0.335$ \cite{lbw} (see Sec.
\ref{sec_nr}) while
$\Lambda_{\rm min}$ tends
to
$-\infty$ as (see Sec. \ref{sec_dhs})
\begin{equation}
\Lambda_{\rm min}\sim -\frac{0.24632-N}{N^2}.
\end{equation}
As $N$
increases, we see on  Fig. \ref{phase_lambda1} that $\Lambda_c$ decreases
monotonically while $\Lambda_{\rm
min}$
increases monotonically. For
$N\rightarrow N_{\rm max}=0.1764$, they tend to the
common value 
\begin{equation}
\Lambda_*=-0.9829.
\end{equation}
Close
to the maximum particle number, we  have the scaling law\footnote{Here and in
the
following, the prefactors are given at an indicative level because the numerics
is not
very accurate close to $N_{\rm max}$.}
\begin{equation}
\label{}
\Lambda_X - \Lambda_{*} \sim \pm 6.7 \, (N_{\rm max} -
N)^{1/2},
\end{equation}
where $\Lambda_X$ stands for $\Lambda_c$ or $\Lambda_{\rm min}$. We note on
Fig.  \ref{phase_lambda1} 
that
$\Lambda_{\rm min}$ is always negative while $\Lambda_c$ is positive for
$N<N_b=0.1255$ and negative for $N>N_b=0.1255$. A negative value of $\Lambda$,
i.e. a positive
energy $E$, is usually impossible because the system would explode and
disperse
away.\footnote{In Newtonian gravity, there is no equilibrium state for stellar
systems with positive energy. Indeed, according to the equilibrium virial
theorem $2K+W=0$ we have $E=K+W=(1/2)W=-K<0$. In general
relativity, equilibrium states of star clusters may have a positive
energy. This
observation was first made
by Zel'dovich \cite{zeldovichEpos}. However, when the system
is unbounded, the
equilibrium states with
$E>0$ are usually unstable.} However, in our case, stable equilibrium states
with
positive
energies are possible because we work inside a
box so
that the system is confined by the walls of the box.

\begin{figure}
\begin{center}
\includegraphics[clip,scale=0.3]{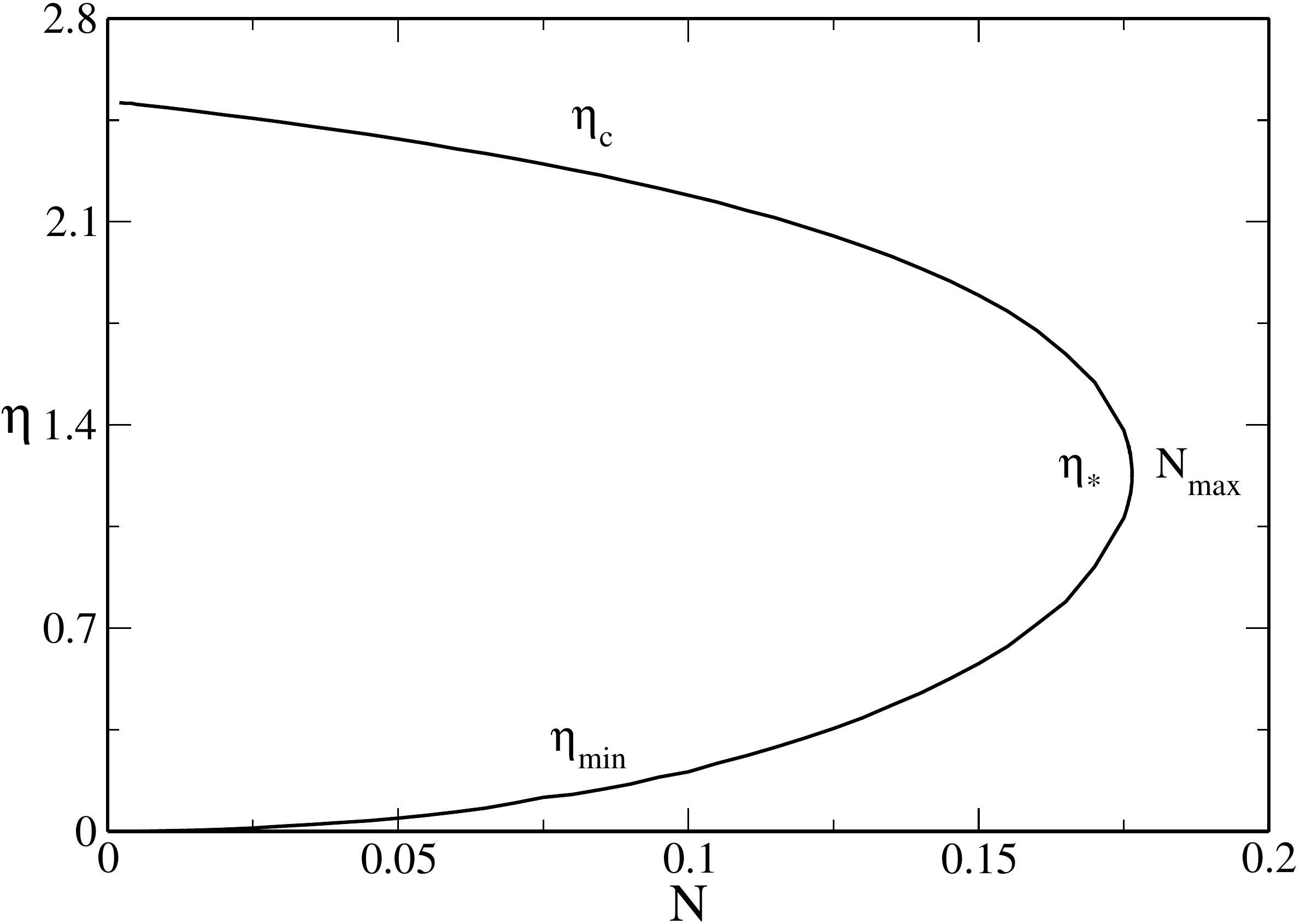}
\caption{Evolution of the critical inverse temperatures $\eta_c$ and
$\eta_{\rm min}$ with $N$.}
\label{phase_eta1}
\end{center}
\end{figure}

In Fig. \ref{phase_eta1} we have plotted the evolution of the critical
temperatures $\eta_c$ and $\eta_{\rm min}$, corresponding to the cold and hot
spirals, as a function of $N$. For $N\rightarrow 0$, we find that $\eta_c$ tends
to the Newtonian value $2.52$ \cite{emden} (see
Sec. \ref{sec_nr}) while $\eta_{\rm min}$ tends to zero as (see Sec.
\ref{sec_dhs})
\begin{equation}
\eta_{\rm min}\sim 17.809\, N^2.
\end{equation}
As $N$
increases, we see on Fig. \ref{phase_eta1} that $\eta_c$ decreases
monotonically while $\eta_{\rm min}$
increases monotonically. For
$N\rightarrow N_{\rm max}=0.1764$, they tend to the
common value 
\begin{equation}
\eta_*=1.2203.
\end{equation}
Close to the
maximum particle number, we  have the scaling law
\begin{equation}
\label{ev1}
\eta_X - \eta_{*} \sim \pm 3.9\, (N_{\rm max} -
N)^{1/2},
\end{equation}
where $\eta_X$ stands for $\eta_c$ or $\eta_{\rm min}$.

As explained in Sec. \ref{sec_dhs}, in order to study the evolution of the hot
spiral with $N$,
it is better to use the dimensionless variables  ${\cal M}$ and
${\cal B}$.

\begin{figure}
\begin{center}
\includegraphics[clip,scale=0.3]{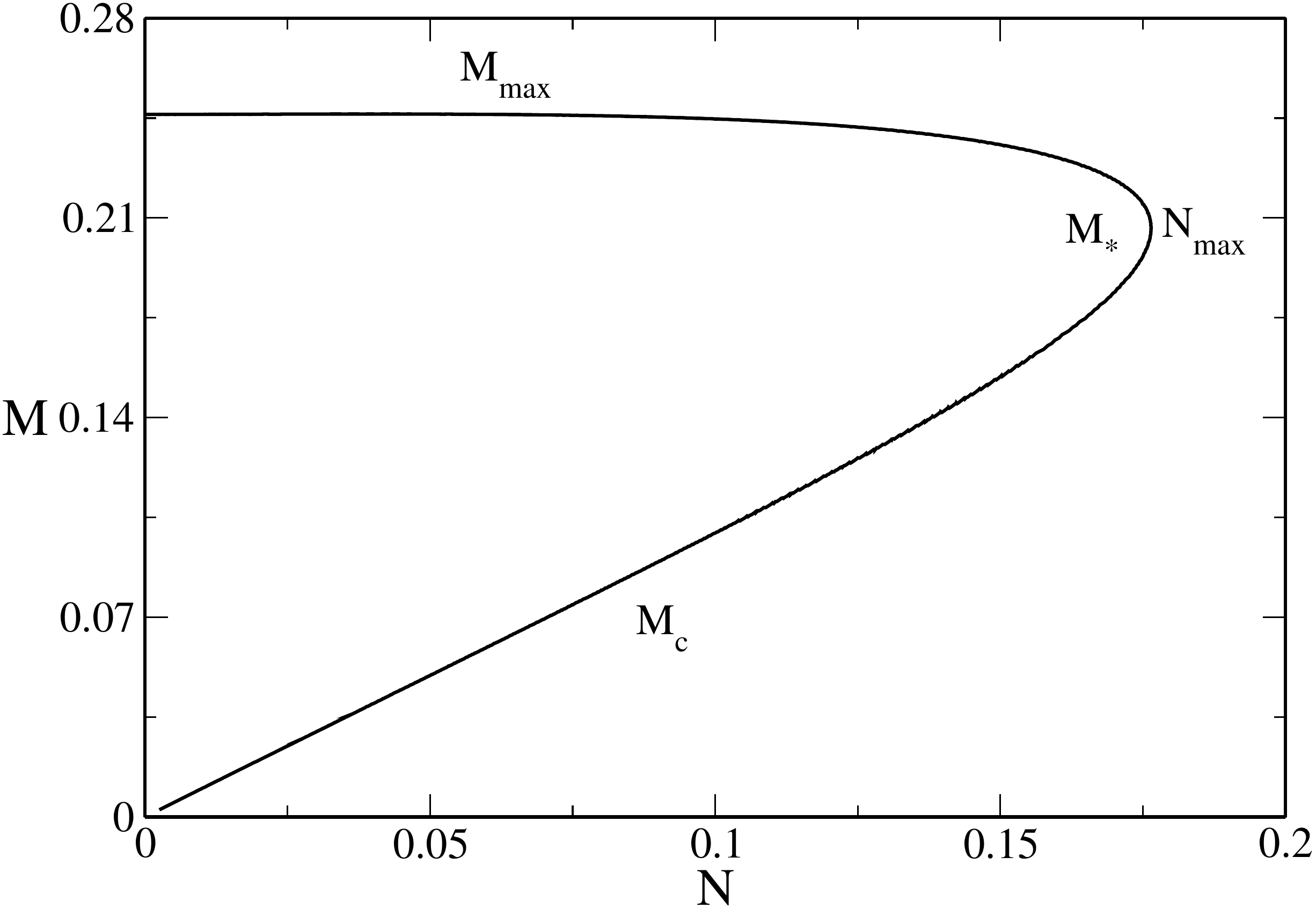}
\caption{Evolution of the critical masses ${\cal M}_c$ and
${\cal M}_{\rm max}$ with $N$.}
\label{Mmax_Nmax}
\end{center}
\end{figure}

In Fig. \ref{Mmax_Nmax}  we have plotted the evolution of the critical
masses ${\cal M}_c$ and ${\cal M}_{\rm max}$, corresponding to the cold and hot
spirals, as a function of $N$. For $N\rightarrow 0$,  ${\cal M}_c$ tends
to zero as (see Sec. \ref{sec_nzerocold}) 
\begin{equation}
{\cal M}_c\simeq N-0.335\, N^2
\end{equation}
while ${\cal
M}_{\rm max}$ tends
to the ultrarelativistic value $0.24632$ \cite{rgb} coinciding with the value
obtained in Refs.
\cite{sorkin,aarelat2} for the self-gravitating black-body radiation (see
Secs. \ref{sec_ur} and \ref{sec_dhs}).  As $N$
increases, we see on Fig. \ref{Mmax_Nmax}  that ${\cal M}_c$ increases
monotonically while ${\cal M}_{\rm
max}$ first increases (zoom not shown), reaches a maximum value $0.24642$ at
$N=0.0418(137)$, and
decreases. For
$N\rightarrow N_{\rm
max}=0.1764$, they
tend
to the common value 
\begin{equation}
{\cal M}_*=0.20703.
\end{equation}
Close to the
maximum particle number, we  have the scaling law
\begin{equation}
\label{mar1}
{\cal M}_X - {\cal M}_{*} \sim \pm 0.24\, (N_{\rm max} -
N)^{1/2},
\end{equation}
where ${\cal M}_X$ stands for ${\cal M}_{\rm max}$ or ${\cal M}_{c}$.

\begin{figure}
\begin{center}
\includegraphics[clip,scale=0.3]{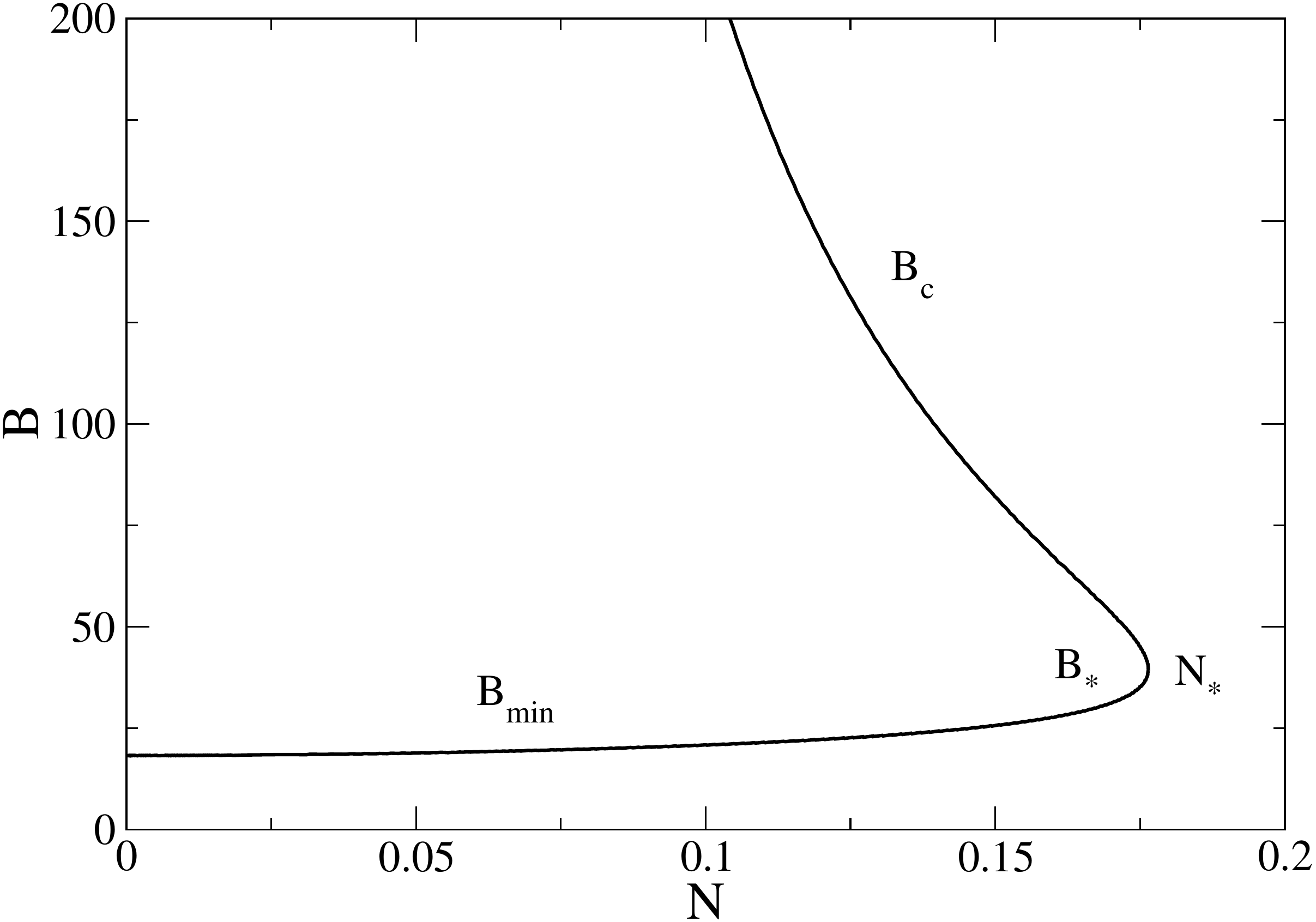}
\caption{Evolution of the critical inverse temperatures ${\cal B}_c$ and
${\cal B}_{\rm min}$ with $N$.}
\label{figureGA}
\end{center}
\end{figure}

In Fig. \ref{figureGA} we have plotted the evolution of the critical
temperatures ${\cal B}_c$ and ${\cal B}_{\rm min}$, corresponding to the cold
and
hot
spirals, as a function of $N$. For $N\rightarrow 0$, ${\cal B}_c$
tends
to $+\infty$ as (see Secs. \ref{sec_nr} and \ref{sec_dhs})
\begin{equation}
{\cal B}_c\sim \frac{2.52}{N^2}
\end{equation}
while ${\cal B}_{\rm
min}$ tends to the ultrarelativistic
value $17.809$ \cite{rgb} (see Sec. \ref{sec_dhs}). As explained in \cite{rgb},
this value
is
different from the value of the
maximum temperature obtained for the self-gravitating black-body radiation (see
Sec.
\ref{sec_ur}). As $N$
increases, we see on  Fig. \ref{figureGA} that ${\cal B}_c$ decreases
monotonically while ${\cal B}_{\rm min}$  increases monotonically. For
$N\rightarrow
N_{\rm max}=0.1764$, they tend to the common value 
\begin{equation}
{\cal B}_*=39.1918.
\end{equation}
{Close to the
maximum particle number, we  have the scaling law
\begin{equation}
\label{mar2}
{\cal B}_X - {\cal B}_{*} \sim \pm 130\, (N_{\rm max} -
N)^{1/2},
\end{equation}
where ${\cal B}_X$ stands for ${\cal B}_{\rm min}$ or ${\cal B}_{c}$.

\subsection{Density contrast}

In Fig. \ref{NLambda_contrast2} we have plotted the evolution of the energy
density
contrast ${\cal R}=\epsilon(0)/\epsilon(R)$ at the critical energies $\Lambda_c$
and $\Lambda_{\rm
min}$ as a function of $N$. For $N\rightarrow 0$, we find that ${\cal
R}(\Lambda_c)$ tends
to the Newtonian value $709$ \cite{antonov} (see
Sec. \ref{sec_nr}) while ${\cal R}(\Lambda_{\rm min})$  tends to the
ultrarelativistic value $22.4$ \cite{rgb}  coinciding with the value
obtained in Ref. \cite{aarelat2} for the self-gravitating black-body
radiation (see Secs. \ref{sec_ur} and \ref{sec_dhs}). As $N$
increases, ${\cal R}(\Lambda_c)$ decreases
monotonically. On the
other hand, ${\cal R}(\Lambda_{\rm
min})$ slightly decreases (zoom not shown),  reaches a
minimum value $21.735$ at $N=0.0785(784)$ and increases.  For
$N\rightarrow N_{\rm max}=0.1764$, we find that  ${\cal R}(\Lambda_*)=27.5$.
Close to the
maximum particle number, we  have the scaling laws
\begin{equation}
\label{dc1}
{\cal R}(\Lambda_c) - {\cal R}(\Lambda_{*}) \sim
106\,  (N_{\rm max} -
N)^{1/2},
\end{equation}
\begin{equation}
\label{dc2}
{\cal
R}(\Lambda_{\rm
min})-{\cal R}(\Lambda_*) \sim -29\, (N_{\rm max} -
N)^{1/2}.
\end{equation}
We note that the prefactors are different on the two branches.

\begin{figure}
\begin{center}
\includegraphics[clip,scale=0.3]{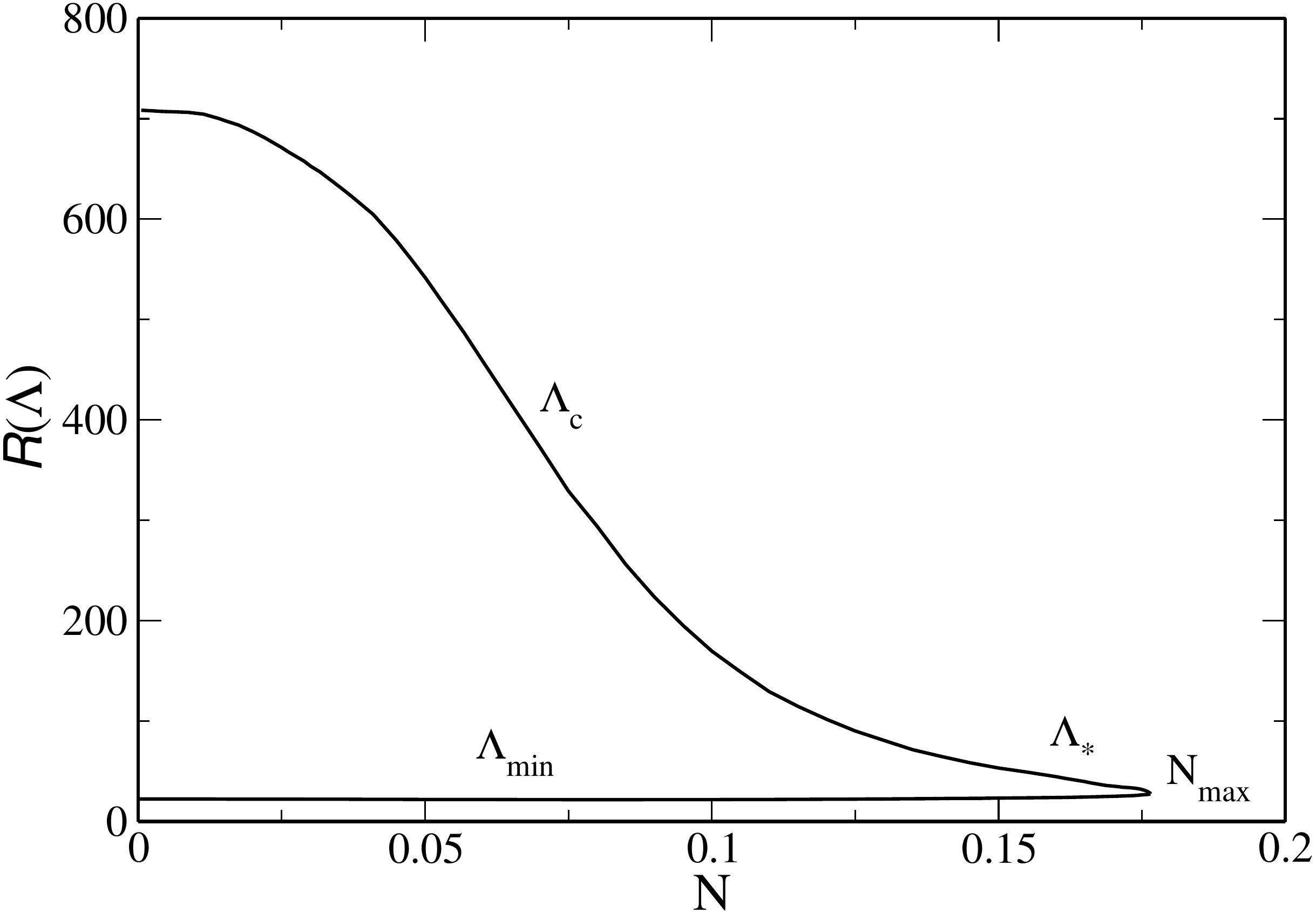}
\caption{Energy density contrast at the microcanonical collapse energies
$\Lambda_c$
and $\Lambda_{\rm min}$ as a function of $N$.}
\label{NLambda_contrast2}
\end{center}
\end{figure}

In Fig. \ref{Neta_contrast2} we have plotted the evolution of the energy density
contrast ${\cal R}=\epsilon(0)/\epsilon(R)$ at the critical inverse temperatures
$\eta_c$ and $\eta_{\rm
min}$ as a function of $N$. For $N\rightarrow 0$, we find that ${\cal
R}(\eta_c)$ tends
to the nonrelativistic value $32.1$ \cite{emden}
(see
Sec. \ref{sec_nr}).
For $N\rightarrow 0$, ${\cal R}(\eta_{\rm min})$  tends to the
ultrarelativistic value $10.3$
\cite{rgb} which is different from the value $1.91$
obtained in Ref. \cite{aarelat2} for the self-gravitating black-body
radiation (see Secs. \ref{sec_ur} and \ref{sec_dhs}). As
$N$
increases, ${\cal R}(\eta_c)$ increases, reaches a
maximum value $36.532$ at $N=0.121(471)$ and decreases. On the other hand,
${\cal
R}(\eta_{\rm
min})$ increases monotonically with $N$.  For
$N\rightarrow N_{\rm max}=0.1764$, we find that  ${\cal R}(\eta_*)=27.5$.
Close to the
maximum particle number, we  have the scaling laws
\begin{equation}
\label{dc3}
{\cal R}(\eta_c) - {\cal R}(\eta_{*}) \sim 39\, (N_{\rm
max} -
N)^{1/2},
\end{equation}
\begin{equation}
\label{dc4}
{\cal R}(\eta_{\rm
min})-{\cal R}(\eta_*) \sim -113\, (N_{\rm max} -
N)^{1/2}.
\end{equation}

We note that there is a sort of antisymmetry in Figs.
\ref{NLambda_contrast2} and \ref{Neta_contrast2} between the critical energies
and critical temperatures on the cold and hot spirals respectively (see in this
connection the Remark at the end of Appendix \ref{sec_anti}). In particular,
${\cal R}(\Lambda_c)$ and ${\cal R}(\eta_{\rm min})$ are monotonic functions of
$N$ while ${\cal R}(\Lambda_{\rm min})$ and ${\cal R}(\eta_{c})$ present an
extremum.

\begin{figure}
\begin{center}
\includegraphics[clip,scale=0.3]{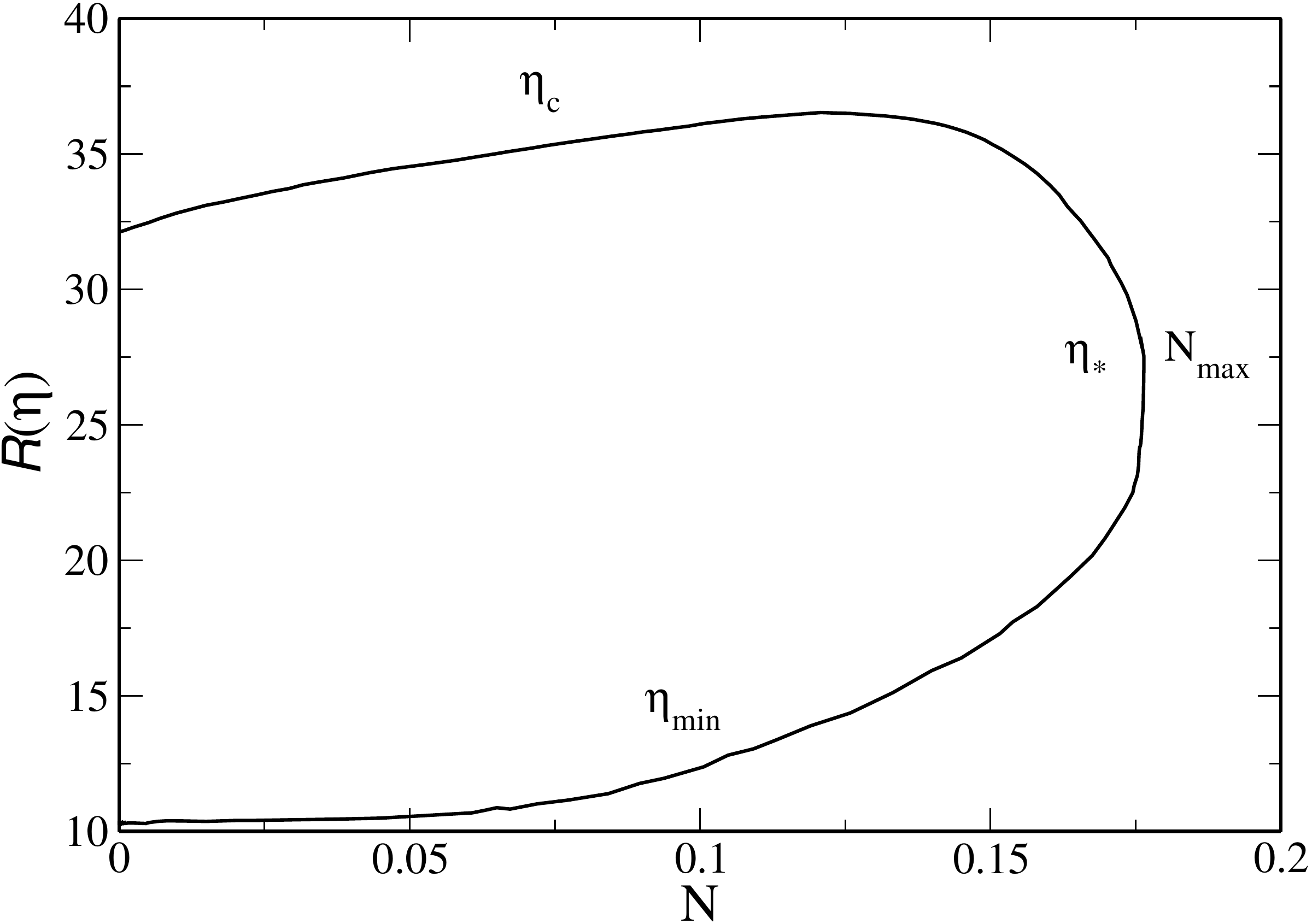}
\caption{Energy density contrast at the canonical collapse temperatures $\eta_c$
and
$\eta_{\rm min}$ as a function of $N$. }
\label{Neta_contrast2}
\end{center}
\end{figure}

Finally, we have plotted in Fig. \ref{Rmin} the minimum value of the
energy density contrast ${\cal R}_{\rm min}$ (see Sec. \ref{sec_edc}) as a
function of $N$. For $N\rightarrow 0$, we find that ${\cal R}_{\rm
min}\rightarrow 1$. For $N\rightarrow N_{\rm max}=0.1764$, we find that
${\cal R}_{\rm
min}\rightarrow 27.5$. We
also find that $\alpha({\cal R}_{\rm
min})\simeq 6.527+\ln N$, $\Lambda({\cal R}_{\rm
min})\sim  -0.43/N$ and  $\eta({\cal R}_{\rm
min})\sim  4.33N$ when $N\rightarrow 0$. The locus of the minimum energy
density contrast is pushed to infinity when $N\rightarrow 0$.

\begin{figure}
\begin{center}
\includegraphics[clip,scale=0.3]{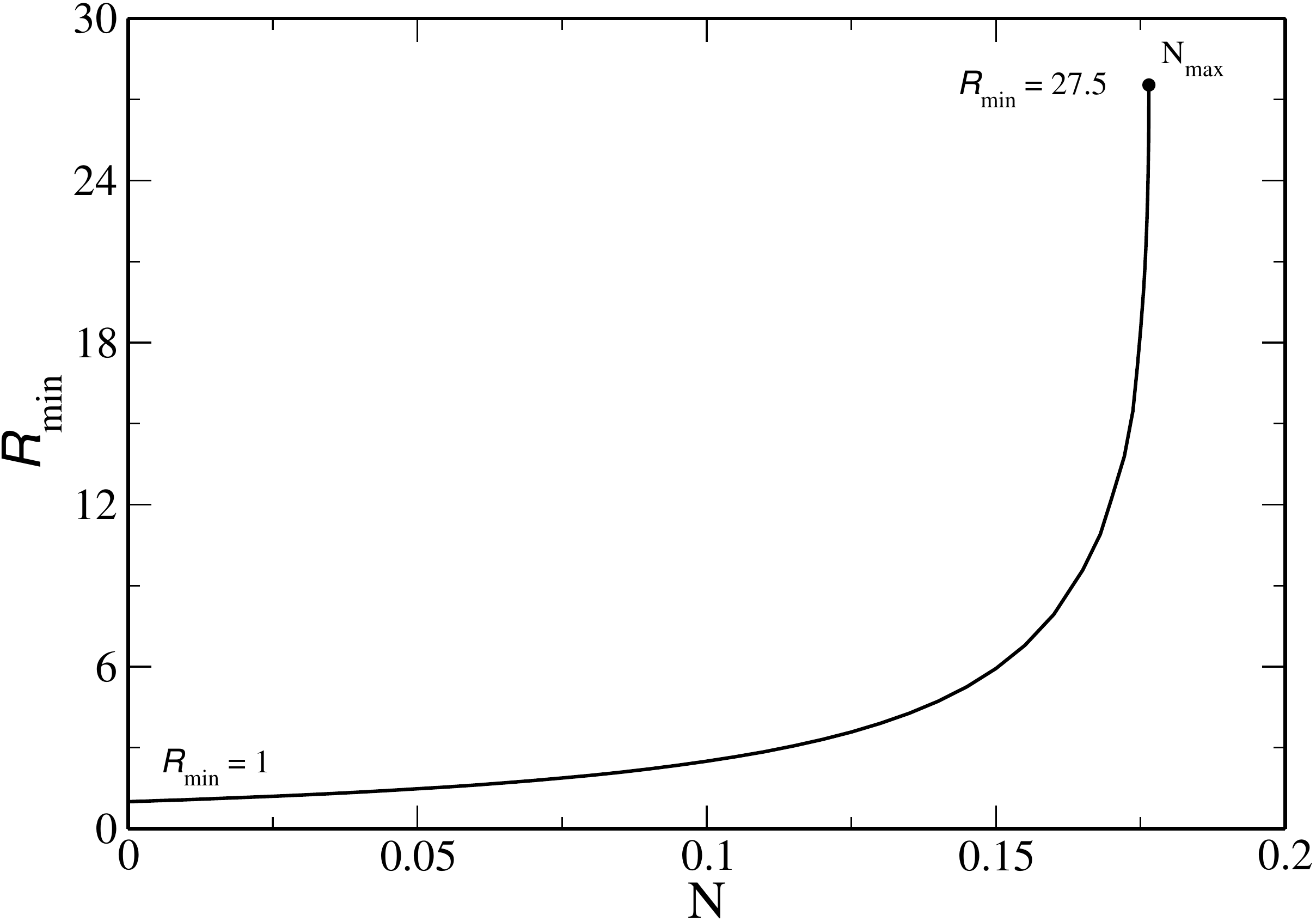}
\caption{Minimum value of the energy density contrast ${\cal R}$ as a
function of $N$.}
\label{Rmin}
\end{center}
\end{figure}

\subsection{Temperature contrast}

In Fig. \ref{b0bR_Lambda} we have plotted the evolution of the temperature
contrast $\Theta=T(0)/T(R)$ at the critical
energies $\Lambda_c$ and
$\Lambda_{\rm
min}$ as a function of $N$. The corresponding central redshift
$z_0=e^{-\nu(0)/2}-1$ is given by
\begin{equation}
\label{shift}
z_0=\frac{\Theta}{\sqrt{1-2{\cal M}}}-1,
\end{equation}
where we have used the Tolman relation $T(r)=T_{\infty} e^{-\nu(r)/2}$ and Eq.
(\ref{b11}). For $N\rightarrow 0$, we find that
$\Theta(\Lambda_c)\rightarrow 1$ since the temperature is uniform in a
nonrelativistic
self-gravitating system at statistical equilibrium while $\Theta(\Lambda_{\rm
min})\rightarrow {\cal R}(\Lambda_{\rm
min})^{1/4}=2.18$ for an ultrarelativistic gas (see Eq. (153)
of \cite{rgb}), corresponding to a central redshift $z_0=2.06$. These results
also apply to the self-gravitating black-body radiation. As
$N$
increases,  $\Theta(\Lambda_c)$ increases monotonically
while $\Theta(\Lambda_{\rm
min})$ decreases monotonically. For
$N\rightarrow N_{\rm max}=0.1764$, we find that 
$\Theta(\Lambda_*)=1.68$, corresponding to a central redshift
$z_0=1.19$.
Close to the
maximum particle number, we  have the scaling laws
\begin{equation}
\label{tc1}
\Theta(\Lambda_c) - \Theta(\Lambda_{*}) \sim -1.37 \,  (N_{\rm max} -
N)^{1/2},
\end{equation}
\begin{equation}
\label{tc2}
\Theta (\Lambda_{\rm
min})-\Theta (\Lambda_*) \sim 1.64 \, (N_{\rm max} -
N)^{1/2}.
\end{equation}

\begin{figure}
\begin{center}
\includegraphics[clip,scale=0.3]{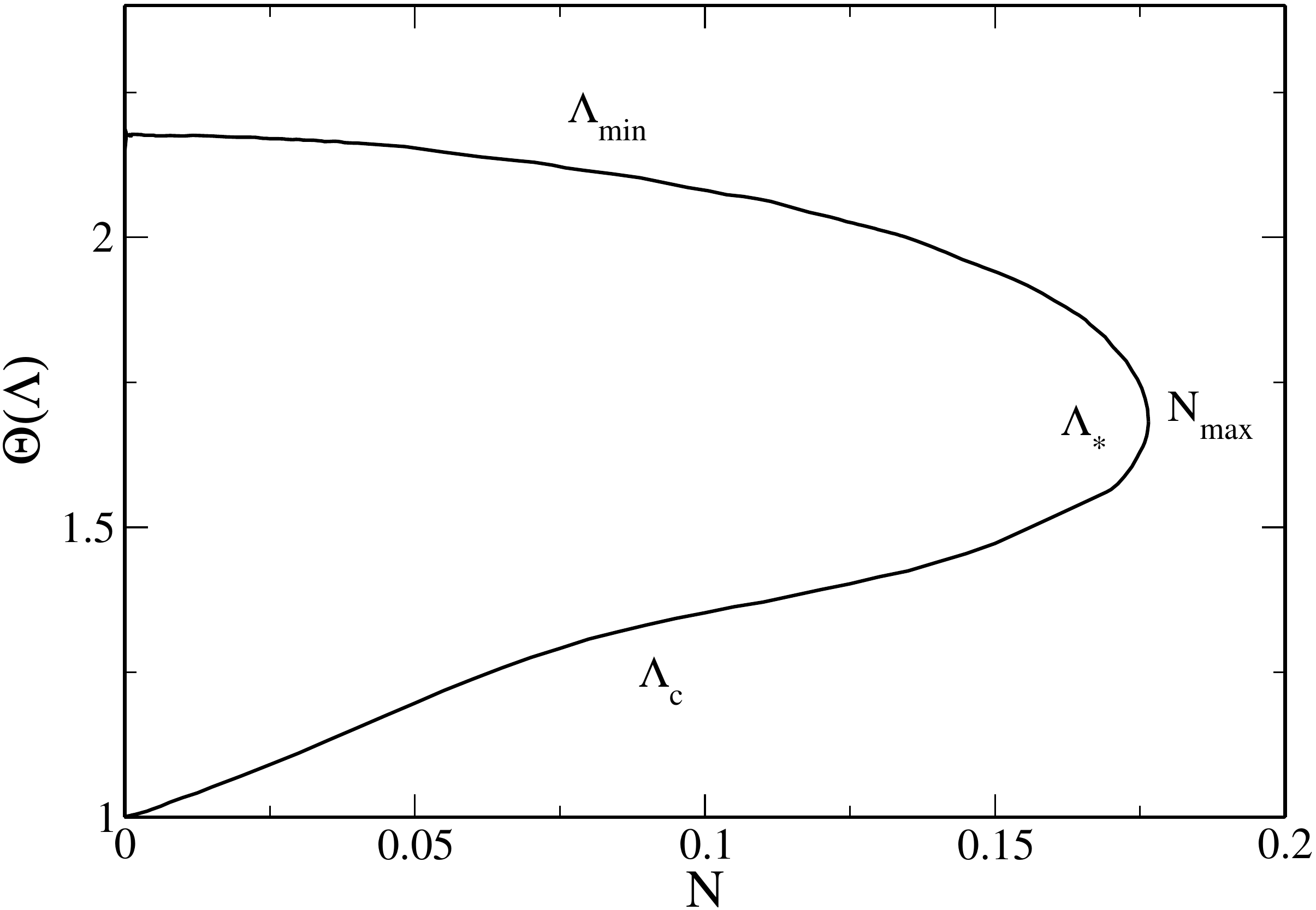}
\caption{Temperature contrast at the microcanonical collapse energies
$\Lambda_c$
and $\Lambda_{\rm min}$ as a function of $N$. }
\label{b0bR_Lambda}
\end{center}
\end{figure}

In Fig. \ref{b0bR_eta} we have plotted the evolution of the temperature
contrast $\Theta=T(0)/T(R)$ at the critical
inverse temperatures $\eta_c$ and
$\eta_{\rm
min}$ as a function of $N$. For $N\rightarrow 0$, we find that
$\Theta(\eta_c)\rightarrow 1$ since the temperature is uniform in a
nonrelativistic
self-gravitating system at statistical equilibrium while $\Theta(\eta_{\rm
min})\rightarrow  {\cal R}(\eta_{\rm
min})^{1/4}=1.79$ for  an ultrarelativistic gas (see Eq. (153)
of \cite{rgb}). As
$N$
increases, $\Theta(\eta_c)$ increases monotonically. On the other
hand, $\Theta(\eta_{\rm
min})$ increases slightly, reaches a
maximum value $1.85$ at $N=0.108$  and decreases. For
$N\rightarrow N_{\rm max}=0.1764$, we find that  $\Theta(\eta_*)=1.68$.
Close to the
maximum particle number, we  have the scaling laws
\begin{equation}
\label{tc3}
\Theta(\eta_c) - \Theta(\eta_{*}) \sim -1.83 \, (N_{\rm
max} -
N)^{1/2},
\end{equation}
\begin{equation}
\label{tc4}
\Theta(\eta_{\rm
min})-\Theta(\eta_*) \sim 0.873 \, (N_{\rm max} -
N)^{1/2}.
\end{equation}

\begin{figure}
\begin{center}
\includegraphics[clip,scale=0.3]{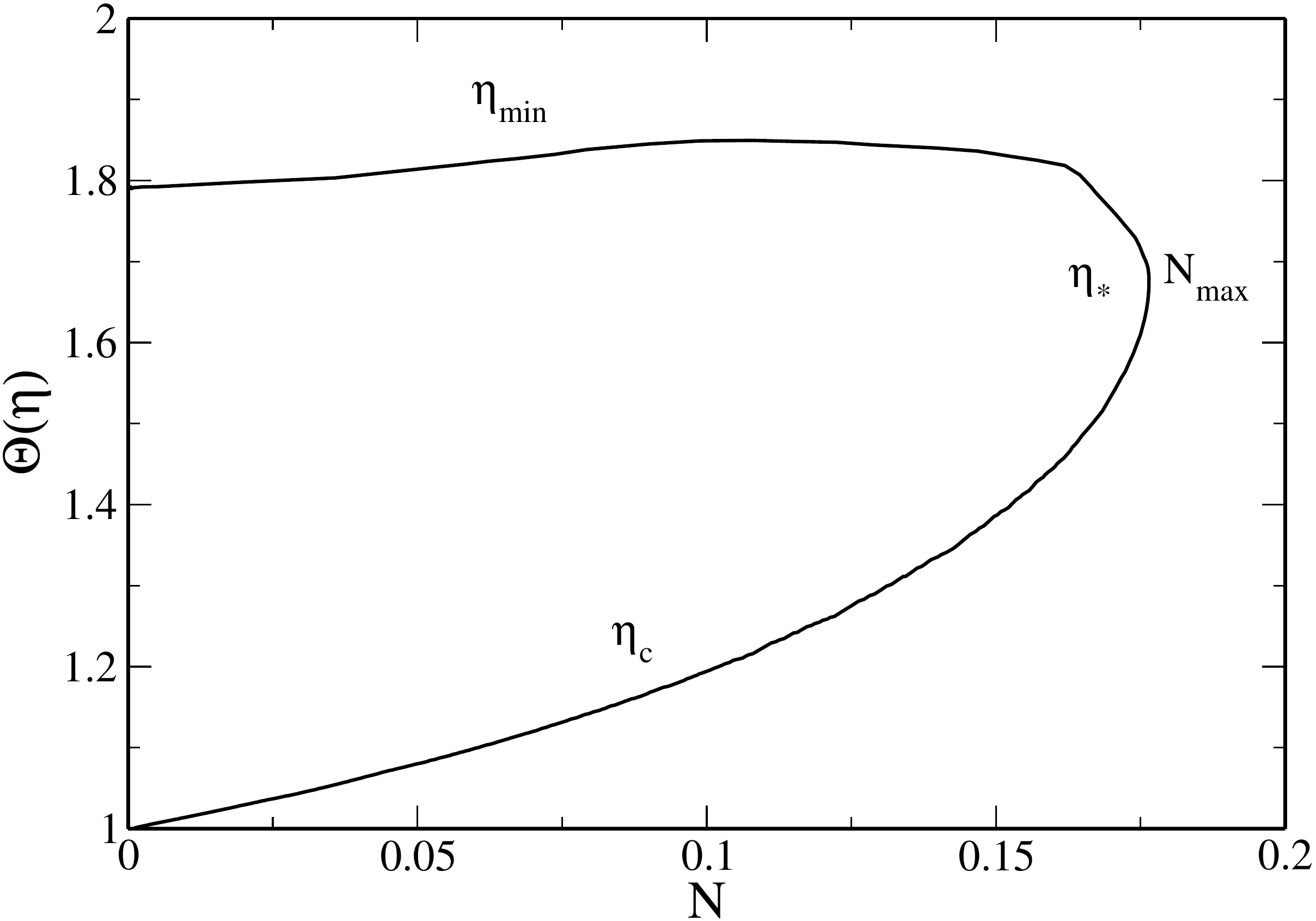}
\caption{Temperature contrast  at the canonical collapse temperatures $\eta_c$
and $\eta_{\rm min}$ as a function of $N$.}
\label{b0bR_eta}
\end{center}
\end{figure}

\subsection{Another normalization}
\label{sec_an}

To plot the caloric curves, we have to introduce convenient normalizations of
the temperature and binding energy. In the preceding sections, we have
considered the normalizations $\eta$ and $\Lambda$ adapted to the
nonrelativistic limit (for $\nu\rightarrow 0$ and $\Lambda,\eta\sim 1$ we are
in the
case $k_B T \ll mc^2$) and the normalizations ${\cal B}$ and ${\cal M}$ adapted
to the ultrarelativistic limit (for $\nu\rightarrow 0$ and ${\cal M}, {\cal B}
\sim 1$ we are
in the case $k_B T \gg mc^2$). We could have also introduced the normalizations
\begin{equation}
\label{nn1}
\frac{1}{b_{\infty}}\equiv \frac{k_B
T_{\infty}}{mc^2}=\frac{\nu}{\eta}=\frac{1}{{\cal B}\nu}
\end{equation}
and
\begin{equation}
\label{nn2}
e\equiv \frac{E}{Nmc^2}=\frac{M}{Nm}-1=-\Lambda\nu=\frac{{\cal M}}{\nu}-1.
\end{equation}
Here, $e$ is called the fractional binding energy. We note that these normalized
variables do not depend on the
box radius $R$. Using the dimensionless
variables introduced in Appendix B of \cite{acf}, we get
\begin{equation}
\frac{1}{b_{\infty}}=T_{\infty}\quad {\rm and} \quad 
e=\frac{E}{N}.
\end{equation}
For $N\rightarrow 0$, the turning points of the cold and hot
spirals behave with $N$ as
\begin{equation}
\label{nn3}
\frac{1}{b_{\infty}^c}=\frac{k_B T_{\infty}^c}{mc^2}\sim \frac{N}{2.52},
\end{equation}
\begin{equation}
\label{nn4}
e_c=\frac{E_c}{Nmc^2}=\frac{M_c}{Nm}-1\sim -0.335\, N,
\end{equation}
\begin{equation}
\label{nn5}
\frac{1}{b_{\infty}^{\rm min}}=\frac{k_B T_{\infty}^{\rm max}}{mc^2}\sim
\frac{1}{17.809\, N},
\end{equation}
\begin{equation}
\label{nn6}
e_{\rm max}=\frac{E_{\rm max}}{Nmc^2}=\frac{M_{\rm max}}{Nm}-1\sim
\frac{0.24632}{N}-1.
\end{equation}
We note
that the caloric curve $b_{\infty}(e)$  does
not tend to a limit when $N\rightarrow 0$. In this sense, the normalizations 
$\eta(\Lambda)$ and ${\cal B}({\cal M})$ seem to be more adapted to our study
than $b_{\infty}(e)$. On the other hand, for
$N\rightarrow N_{\rm max}=0.1764$, we have
\begin{equation}
\label{nn7}
\frac{1}{b_{\infty}^*}=\frac{k_B T_{\infty}^*}{mc^2}=0.1446,
\end{equation}
\begin{equation}
\label{nn8}
e_*=\frac{E_*}{Nmc^2}=\frac{M_*}{Nm}-1=0.1734.
\end{equation}
The order of magnitude of $k_B T_{\infty}^*/mc^2=0.1446$
and the corresponding central redshift
$z_0^*=1.19$ are  in agreement with
the maximum temperature $k_B(T_\infty)_{\rm max}/mc^2=0.273$ and the
corresponding central redshift
$z_0=1.08$ found by
Zel'dovich and Podurets \cite{zp} for heavily truncated isothermal
distributions. However, the correspondance between the two approaches is not
straightforward. We refer to Appendix B
of \cite{rgb} for a detailed comparison between box-confined isothermal
spheres and heavily
truncated isothermal distributions.

\section{Astrophysical applications}
\label{sec_dt}

The caloric curve of the general relativistic classical gas with its
double spiral shape displays different types of instabilities. We discuss here
some astrophysical applications of these instabilities.

\subsection{Instability at $T_c$ for gaseous stars and self-gravitating
Brownian particles}

Let us first consider the instability that takes place at $T_c$ in the CE.  The
CE is appropriate to study a system in contact with a heat bath
fixing the temperature $T$. This could be an isothermal self-gravitating gas
(star) described by the Euler-Poisson equations \cite{penston,pomeau1} or a
system of self-gravitating Brownian particles described by the
Smoluchowski-Poisson equations \cite{crs,sc,post,tcoll,virial1,virial2,zero}.

In that case, the temperature is constant and uniform. The dynamical
evolution of the system is due to a departure from hydrostatic equilibrium. When
$T>T_c$ the system settles on a stable equilibrium state in which the
pressure gradient equilibrates the gravitational attraction. At $T=T_c$
the equilibrium becomes unstable and the system undergoes an isothermal
collapse \cite{aaiso}.  The system
collapses because it is too
cold so the thermal pressure cannot balance the gravitational attraction.  The
thermodynamical stability in the CE coincides with the dynamical stability with
respect to the Euler-Poisson equations \cite{aaiso} and with respect to the
Smoluchowski-Poisson equations
\cite{crs,sc}. This leads to a
fast instability at $T_c$.\footnote{The instability may be slow
if the system is described by the Vlasov-Kramers-Poisson equations with a
small friction coefficient.}

The perturbation $\delta\rho(r)$ that triggers the instability at the critical
temperature $T_c$  has a core structure (one node)
\cite{aaiso,pomeau1} while the velocity perturbation $\delta v(r)$ has an
implosive structure (no node) with $\delta v<0$ (implosion)
\cite{aaiso,pomeau1}.
In view of the form of the marginal mode, the isothermal collapse of a star
corresponds to a pure implosion (the star collapses as a whole) leading to a
Dirac peak containing all the mass \cite{penston,post,zero,pomeau1}. This
structure has an
infinite free energy
($F\rightarrow -\infty$) \cite{sc}. In practice, the collapse stops when quantum
mechanics
(Pauli's exclusion principle for fermions) comes into play. This results in the
formation of a compact object such as a white dwarf or a neutron star containing
most of the mass $+$ a tenuous atmosphere \cite{ijmpb,clm2,acf}.  This is
possible as long as the mass of the star is not too high, i.e., below the
Oppenheimer-Volkoff (OV) limit $M_{\rm OV}$. When $M\gtrsim M_{\rm OV}$ general
relativity must be taken into account and there is no equilibrium state anymore.
In that case, the star collapses towards a black hole.

This phase transition is reminiscent of the hypernova phenomenon for
supermassive stars (above $40\, M_{\odot}$) which shows very intense and
directive gamma ray bursts, but no explosion of matter (or a very faint one)
\cite{pomeau1}.

\subsection{Instability at $E_c$ for globular clusters and dark matter halos}

We now consider the instability that takes place at $E_c$ in the MCE. The MCE is
appropriate to study an isolated system evolving at
fixed energy $E$. We consider here a stellar system such as a globular cluster
described by the Vlasov-Landau-Poisson equation or by the
orbit-averaged-Fokker-Planck equation \cite{cohn}.

In that case, the system is in hydrostatic equilibrium but the temperature
(velocity dispersion) is not uniform. The dynamical evolution of the system is
due to a temperature gradient between the
core and the halo and the fact that the specific heat of the core  is
negative \cite{lbw,lbe,cohn,inagakilb} (see Appendix
\ref{sec_nsh}). When
$E>E_c$
the system settles on a stable equilibrium state in which the temperature in the
core is equal to the temperature in the halo (the temperature is uniform). At
$E=E_c$
the equilibrium becomes unstable and the system undergoes the gravothermal
catastrophe \cite{lbw}. In that case, the temperature in the core increases more
rapidly than the temperature in the halo. This leads to a thermal runaway and
a core collapse \cite{lbw,lbe,cohn,inagakilb}. The thermodynamical stability in
the MCE does not coincide with
the dynamical stability with respect to the (collisionless) Vlasov-Poisson
equations. Indeed,
all isotropic distribution functions are (Vlasov) dynamically
stable \cite{doremus71,doremus73,gillon76,sflp,ks,kandrup91}. This
implies that the instability at $E_c$ is slow (secular) since it has a
thermodynamical origin (see Sec. \ref{sec_dvts}).

The perturbation $\delta\rho(r)$ that triggers the instability at the critical
energy $E_c$ has a core-halo structure (two nodes) \cite{paddy}. In view of the
form of the marginal mode, a globular cluster experiencing the gravothermal
catastrophe has a core-halo structure reminiscent of a red giant \cite{lbw}.
Core
collapse leads to a binary star surrounded by a hot
halo \cite{inagakilb}. This structure has an
infinite entropy ($S\rightarrow +\infty$) at fixed energy \cite{sc}. 

In the case of dark matter halos made of fermions (massive
neutrinos) or bosons (axions), the gravothermal catastrophe can be stopped or
even prevented by quantum mechanics. This leads to a fermion or boson
``ball'' -- forming a bulge -- surrounded by an extended halo (quantum core-halo
structure) \cite{modeldm,mcmh}. Alternatively, during the gravothermal
catastrophe, the system may become relativistic and finally undergo a dynamical
instability of general relativistic origin leading to the formation of a
supermassive black hole as described in \cite{zp,fit,strevue}. The application
of this scenario to dark matter halos has been developed in \cite{balberg}
and advocated in \cite{clm1,clm2,modeldm}.

\subsection{Instability at $E_c$ for gaseous stars and self-gravitating
Brownian particles}

As before we consider the instability at $E_c$ in the MCE but
we focus on a different dynamical model in which the temperature
$T(t)$ is uniform throughout the system but evolves with time
so as to conserve the energy $E$. Specifically, we consider a
self-gravitating gas (star) described by the Euler-Poisson equations
\cite{sb,pomeau2} or a system of self-gravitating Brownian particles described
by the Smoluchowski-Poisson equations \cite{crs,sc} with an additional equation
assuring the conservation of energy.

In this model the temperature $T(t)$ is uniform (albeit not constant in time)
and the dynamical
evolution of the system is due to a departure from hydrostatic
equilibrium.\footnote{This model assumes that thermodynamical
equilibrium is established on a timescale short with respect to the dynamical
time. By contrast, for globular clusters, thermodynamical equilibrium is
established on a timescale long with respect to the dynamical time. This model
also assumes an infinite thermal conductivity so that the temperature is uniform
throughout the system.} When
$E>E_c$
the system settles on a stable equilibrium state in which the
pressure gradient equilibrates the gravitational attraction. At $E=E_c$,
the equilibrium becomes unstable and the system undergoes a form of
gravothermal catastrophe \cite{crs,sc,sb,pomeau2}. The system
collapses because it is too cold (even if the temperature increases with
time) so the thermal pressure cannot balance the gravitational attraction. The
thermodynamical stability in the MCE coincides with the dynamical stability with
respect to the Euler-Poisson equations and with
respect to the Smoluchowski-Poisson equations with a
time-dependent temperature $T(t)$ \cite{crs,sc,sb,pomeau2}.  This leads to a
fast (dynamical) instability at $E_c$ in contrast to the slow (secular)
instability of globular
clusters experiencing the standard gravothermal catastrophe.

The perturbation $\delta\rho(r)$ that triggers the instability at the critical
energy $E_c$ has a core-halo structure (two nodes) \cite{paddy,sc,pomeau2}
while the velocity perturbation $\delta v(r)$ has an implosive-explosive
structure (one node) with $\delta v<0$ in the core (implosion) and $\delta v>0$
in the halo (explosion) \cite{pomeau2}. In view of the form of the marginal
mode, the gravothermal catastrophe of a star (in the sense explained above)
corresponds to an implosion-explosion leading to a collapsing core and an
explosive (hot) atmosphere expanding at large distances \cite{pomeau2}. This
structure has an
infinite entropy ($S\rightarrow +\infty$) at fixed energy \cite{sc}. In
practice, the collapse
stops when
quantum mechanics
(Pauli's exclusion principle for fermions) comes into play. This results in
the formation of a compact object such as a white dwarf or a neutron star
containing a finite fraction ($\sim 1/4$) of the total initial mass $+$ an
extended halo \cite{ijmpb,acf,clm2}. This
is possible as long as the mass of the star is not too high. For $M\gtrsim
4M_{\rm OV}$ the mass of the core passes above the OV limit and there is no
equilibrium state anymore. In that case, the core collapses towards a black
hole.

This phase transition is reminiscent of the red giant structure of stars with
low or intermediate mass (roughly $0.3-8\, M_{\odot}$) in a late phase of
stellar evolution before the white dwarf stage. The
implosion of the core and the explosion of the halo is also similar to the
supernova explosion of massive stars with mass in the range of $8-40\,
M_{\odot}$ resulting in the formation of a neutron star \cite{pomeau2}.

\subsection{Instability at $E_{\rm max}$ for the self-gravitating black-body
radiation}

As a preamble of the following section, let us first consider the general
relativistic instability that takes place at $E_{\rm max}$ in the MCE for the
self-gravitating black-body radiation \cite{sorkin,aarelat2}. When $E<E_{\rm
max}$ the system settles on a stable equilibrium state.
At $E=E_{\rm max}$ the equilibrium becomes unstable and the system collapses
because it is too hot and feels ``the weight of heat'' (Tolman's effect)
\cite{tolman}. Indeed, energy is mass so that it
gravitates.
The thermodynamical stability in the MCE coincides with the dynamical stability
with respect to the Euler-Einstein equations  \cite{sorkin,aarelat2}. As a
result, this is a fast
instability.

The perturbation $\delta\epsilon(r)$ that triggers the instability at the
critical energy $E_{\rm max}$ has a core structure (one node)
\cite{aarelat1}. 
In view of the form of the marginal mode, the collapse of  the self-gravitating
black-body radiation at $E=E_{\rm max}$ corresponds to a pure implosion (the
photon star collapses as a whole) leading to a black hole containing all the
mass.

{\it Remark:} We note that the self-gravitating black-body radiation in the MCE
behaves similarly to a nonrelativistic self-gravitating  classical isothermal
gas in the
CE and differently from  a nonrelativistic self-gravitating  classical
isothermal gas in the
MCE. Indeed, for the self-gravitating black-body radiation
in the MCE: (i) the thermodynamical stability coincides with the dynamical
stability with respect to the Euler-Einstein equations governing the evolution
of a gas with a linear
equation of state $P=\epsilon/3$; (ii) the marginal mode $\delta\epsilon(r)$ at
$E_{\rm max}$ has a core structure; (iii) the instability is fast (dynamical).
Similarly, for a nonrelativistic self-gravitating  classical isothermal gas in
the
CE: (i) the thermodynamical stability coincides
with the dynamical stability with respect to the Euler-Poisson
equations  governing the evolution of a gas with a linear
equation of state $P=\rho k_B T/m$;  (ii) the
marginal mode $\delta\rho(r)$ at $T_{c}$ has
a core structure; (iii) the instability is fast (dynamical). By contrast, for a
nonrelativistic self-gravitating classical isothermal gas in the
MCE: (i) the thermodynamical stability does not coincide with the dynamical
stability with respect to the Vlasov-Poisson
equations;  (ii) the
marginal mode $\delta\rho(r)$ at $E_{c}$ has
a core-halo structure; (iii) the instability is slow/secular
(thermodynamical).

\subsection{Instability at $E_{\rm max}$ for a relativistic classical
isothermal gas}

We finally consider the general relativistic instability that takes place at
$E_{\rm
max}$ in the MCE
for a self-gravitating classical isothermal gas.  The MCE is the relevant
ensemble to describe relativistic star clusters that are isolated. However, the
consideration of very high (positive)
energies is possible only if we confine the system within a cavity. This is 
 a 
very artificial situation.\footnote{The CE may be appropriate to describe
relativistic
stars in
contact with a heat bath. In that case, the box may be less artificial as it
can mimic the pressure exerted by an external medium.}
When $E<E_{\rm max}$ the system settles on a stable equilibrium state. At
$E=E_{\rm max}$  the equilibrium becomes
unstable and the system collapses. The thermodynamical stability in the MCE
coincides with the dynamical stability with respect to the Vlasov-Einstein
equations \cite{ipser80}. As a result, this is a
fast (dynamical) instability. This is different from the case of nonrelativistic
systems
experiencing the slow (thermodynamical) gravothermal catastrophe at $E_c$ (see
Sec.
\ref{sec_dvts}).

The perturbation $\delta\epsilon(r)$ that triggers
the instability at the critical energy $E_{\rm max}$ has
not been determined. By analogy with the self-gravitating  black-body
radiation, we suspect that it has a core structure. If confirmed,
the collapse of a relativistic classical isothermal gas
at $E=E_{\rm max}$ would correspond to a pure implosion (the
system collapses as a
whole) leading to a black hole containing all the mass. Quantum mechanics cannot
stop
the collapse in this highly relativisitic situation \cite{acf,rc}.

{\it Remark:} owing to the analogy between a strongly relativistic classical
isothermal gas and the self-gravitating black-body radiation, we understand
why thermodynamical stability coincides with dynamical stability in
general relativity  in the MCE similarly to Newtonian systems
in the CE and differently from 
Newtonian systems in the MCE. We also understand why the marginal mode of
strongly relativistic
systems at $E_{\rm max}$ in the MCE has (presumably) a core structure
similarly to the marginal mode of nonrelativistic systems at $T_c$ in the CE 
and differently from the   marginal mode of nonrelativistic systems at
$E_{c}$ in the MCE (that has a core-halo structure).
The self-gravitating black-body radiation is a particular
system where Ipser's conjecture \cite{ipser80} (see Sec. \ref{sec_dvts}) can be
easily
demonstrated. However, it is important to
realize that, for the self-gravitating black-body
radiation,  dynamical stability
refers to the Euler-Einstein equations while, for a classical
isothermal system, it refers to the Vlasov-Einstein
equations.   This is a  difference of fundamental importance.

\section{Conclusion}

In this paper, we have studied the statistical equilibrium states of
box-confined classical self-gravitating systems in general
relativity and we have determined their caloric curves. This is a
generalization of the problem introduced by Antonov \cite{antonov} and
Lynden-Bell and Wood \cite{lbw} for nonrelativistic classical
stellar systems. This also corresponds to the nondegenerate limit of the
self-gravitating Fermi gas studied by Hertel and Thirring \cite{ht}, Bilic and
Viollier \cite{bvn} and Chavanis \cite{pt,ijmpb} in the nonrelativistic limit
and by Bilic and Viollier \cite{bvr}, Alberti and Chavanis \cite{acf} and Roupas
and Chavanis \cite{rc} in general relativity. 

The caloric curves of the general relativistic classical self-gravitating gas
depend on the compactness parameter $\nu=GNm/Rc^2$. They typically have the form
of a double spiral. The system undergoes a gravitational collapse at both
low and high energies and temperatures. At low temperatures
the gas collapses because
it is too cold and the thermal pressure
cannot balance the gravitational attraction. At high energies the
gas collapses because it is too hot
and feels ``the weight of heat''
\cite{tolman} (energy is mass so that it gravitates). The two spirals approach
each other as the compactness parameter $\nu=GNm/Rc^2$ increases, indicating
that general relativistic effects advance the onset of gravitational collapse.
There is a maximum compactness $\nu_{\rm max}=0.1764$ above which no equilibrium
state exists whatever the values of energy and temperature. 

As mentioned in the
Introduction, similar results have been obtained by Roupas \cite{roupas}. Below,
we make a brief comparison between the two studies in order to  show that they
are complementary to each other:

(i) We have made a detailed history of the statistical mechanics of
self-gravitating systems in Newtonian gravity and general relativity (see
also \cite{acf,rgf,rgb,prd1}). In
particular, we have stressed the important works of Zel'dovich and
Podurets \cite{zp}, Ipser
\cite{ipser69b,ipser80} and Katz and Horwitz   \cite{kh2,hk} related to the
general relativistic classical isothermal gas.  We have
discussed the analogies and the differences between the maximum temperature
$k_B(T_\infty)_{\rm
max}/mc^2=0.273$  (and the corresponding central redshift
$z_0=1.08$) of heavily
truncated isothermal distributions discovered by Zel'dovich and Podurets
\cite{zp} and the maximum temperature  $k_B
T_{\infty}^*/mc^2=0.1446$ (and the corresponding central
redshift $z_0^*=1.19$) of box-confined isothermal systems at $\nu_{\rm
max}=0.1764$. This comparison
is further developed in Appendix B of \cite{rgb}.

(ii) We have shown that the ``hot spiral'' of the caloric curve of
general relativistic classical self-gravitating
systems has some similarities with the caloric curve of the self-gravitating
black-body radiation represented in Fig 15 of \cite{aarelat2} that
also displays a striking spiral structure. In
particular, the values of the maximum mass-energy $GM_{\rm max}/Rc^2=0.24632$
(and the
corresponding density contrast ${\cal R}_{\rm MCE}=22.4$) are the same. This is
no true, however, for the maximum temperature because the two systems
have a different physical nature. This comparison is further developed in Sec.
VI of \cite{rgb}.

(iii) We have used a different numerical method to obtain the caloric curves of
the general relativistic classical gas which
is based on the algorithm proposed by Bilic and Viollier \cite{bvr} for
self-gravitating fermions. This method (described in Appendix \ref{sec_cons})
allows us to understand clearly the origin of the critical parameters $N'_S$,
$N_S$,
$N_{\rm max}$, $\eta_c$, $\Lambda_c$, $\eta_{\rm min}$, $\Lambda_{\rm min}$...
appearing in
our study and to determine them from simple graphical constructions. 

(iv) We have shown that the curves $\Lambda({\cal R})$ and  $\eta({\cal
R})$ close themselves at high density contrasts and that they contain a
``relevant'' part and an ``irrelevant'' part. The ``irrelevant'' part
(corresponding to unstable equilibrium states with a very high central density)
may be of
interest to mathematicians in order to understand better the rich and complex
structure of the isothermal TOV equations presented in Sec. \ref{sec_b}. 

(v) We have plotted the curves $S(E)$ and $F(T)$ (see
Figs. \ref{SLambda_N01b_unified} and \ref{Feta_N01c_unified}) and explained why
they display spikes at the critical points.

(vi) We have discussed the thermodynamical and dynamical stability of
isothermal star clusters in general relativity (see Sec. \ref{sec_dvts}). We
have
argued that the instability
that occurs at low energies (cold spiral) is slow/secular, like in the case of
globular clusters experiencing the gravothermal catastrophe, because it has a
thermodynamical origin while the 
instability that
occurs at high
energies (hot spiral) is fast because it has a dynamical
origin.\footnote{If we
extrapolate the conjecture of Ipser \cite{ipser80} to box-confined systems,
and argue that thermodynamical  and dynamical stability coincide in general
relativity, we are led to conclude that general relativistic systems
essentially experience a (fast) dynamical instability. As a result, the
collapse at $E_{\rm max}$ (hot spiral -- strongly relativistic
configurations) occurs quickly, on a dynamical timescale. This is very
different
from the gravothermal catastrophe of nonrelativistic systems
which is a (slow) thermodynamical instability. As a result, the
collapse at $E_c$ (cold spiral -- weakly relativistic
configurations) occurs secularly, on a relaxation timescale of the order of
the age of the Universe. In this sense, the collapse at high
energies is very different from the collapse at low energies.} This is
corroborated by the
fact that  the marginal mode of
instability at $E_{\rm max}$ (hot spiral -- strongly relativistic
configurations) presumably has a ``core'' structure, like for
the self-gravitating black-body radiation \cite{aarelat1,aarelat2}, contrary to
the marginal mode of
instability at $E_{c}$ (cold spiral -- weakly relativistic
configurations) that has a ``core-halo'' structure \cite{paddy} (see
the discussion
Sec. \ref{sec_dt}).

(vii) We have made a detailed study of the caloric curve close to $N_{\rm
max}=0.1764$ showing that it has a complex structure giving rise to
amputed
spirals, and a loop, before disappearing (see
Figs. \ref{kcal_N0130_linked} and \ref{cal0p15_new}). This complex
topological structure may be of interest to mathematicians. Note that
the stable part of the caloric curve remains however ``simple''.

(viii) We have compared the caloric curves of general relativistic classical
systems for $N\rightarrow 0$ (see Fig. \ref{kcal_1PN_corrections_new_bis})  with
the caloric curves obtained in the case
of Newtonian self-gravitating classical systems described by a relativistic
equation of state (see Fig. 1 of \cite{aarelat1}). We have shown
that
the critical energy $E_c$ increases with $N$ in the two cases. By contrast, the
critical
temperature $T_c$ increases with $N$ for general relativistic systems while it
remains constant for
relativistic Newtonian systems \cite{aarelat1}. On the other hand, the hot
spiral is absent for Newtonian systems.

(ix) Following \cite{rgb}, we have introduced a new normalization  of the energy
and temperature, ${\cal
M}=GM/Rc^2$ and ${\cal B}=Rc^4/GNk_B T_{\infty}$, adapted to the
ultrarelativistic limit. Using this normalization, the
caloric curve ${\cal B}({\cal M})$ tends to an asymptotic curve when
$\nu\rightarrow 0$ (see Figs. \ref{hot} and \ref{deviazione_scalingPH}). This
asymptotic curve (hot spiral) has been theoretically
characterized in \cite{rgb}. We have compared this caloric curve, corresponding
to the ultrarelativistic limit of the classical self-gravitating gas, with the
caloric curve of the self-gravitating black-body radiation (see
Fig. \ref{radiation}). This comparison is further developed in Sec. VI of  
\cite{rgb}.\footnote{To investigate the ultrarelativistic
limit, Roupas \cite{roupas} uses the normalized temperature $k_B
T_{\infty}/mc^2$ (see his Fig. 7(a)). However,  as explained in Sec.
\ref{sec_an}, this normalization does not allow one to obtain an asymptotic
caloric curve when $\nu\rightarrow 0$. This is apparent on his Fig. 7(a).}

(x) We have performed a detailed study of the evolution of the critical points
of the general relativistic classical gas with $N$ obtaining their asymptotic
behaviors when $N\rightarrow 0$ and $N\rightarrow N_{\rm max}$.

(xi) We have discussed (see also
\cite{pomeau2,acf,ca}) some analogies between
the implosion-explosion instability that occurs at low energies (cold spiral)
in certain models of stars \cite{pomeau2} and the
supernova phenomenon. A different scenario of
supernova explosion
connected to the gravitational collapse occurring at very high
positive energies (hot spiral) is developed by Roupas
\cite{roupasnew1,roupasnew2}.

\appendix

\section{The very early history of the theory of globular clusters}
\label{sec_hgc}

In this Appendix, we briefly retrace the very early history of the theory of
globular clusters. This completes the review of the
statistical mechanics of self-gravitating systems made in
the introduction.

Inspired by an idea of Lord Kelvin, von Zeipel \cite{zeipel} modeled the density
distribution of stars
in a globular cluster by analogy with the density profile of a 
self-gravitating isothermal gas and
found a good agreement with observations in the central parts
of the cluster. Plummer \cite{plummer}, on the other hand, modeled the
density distribution of stars in a
globular cluster by
analogy with the density profile of a self-gravitating polytropic gas
in convective equilibrium whose
general theory had been elaborated by Lord Kelvin \cite{kelvin,kelvin2},
Lane \cite{lane}, Ritter \cite{ritter} and
Emden \cite{emden}. He
considered analytical polytropic models: the polytrope of index $n=1$ found by
Ritter  \cite{ritter} and the polytrope of index $n=5$ found by Schuster
\cite{schuster}. He noted that the polytrope of index 
$n=5$, whose spatial
density decreases at large distances as $r^{-5}$,
provides a good agreement with observations in the outer parts of the cluster.
This is
now called the ``Plummer distribution''.  At the end of his paper, Plummer
proposed to model the density of stars in a globular cluster by an isothermal
distribution in the center and by a polytropic distribution of index $n=5$  in
the envelope.
In other words, he assumed that globular clusters have a central core in
isothermal equilibrium and an outer envelope
in convective equilibrium. Eddington \cite{eddington1,eddington2}
argued that globular clusters are collisionless stellar systems (instead of
collisional gases as assumed by  von Zeipel \cite{zeipel} and Plummer 
\cite{plummer}) and derived a
self-consistent quasisteady state with a Schwarzschild velocity
distribution containing the Maxwell distribution as a special case.
Jeans \cite{jeansth} determined the general form of steady state
solutions of the collisionless Boltzmann equation (deriving the so-called
``Jeans theorem'' \cite{bt}) and pointed out that 
the Maxwell-Boltzmann distribution is a particular case of his general
theorem. Jeans \cite{jeanskin2} mentioned that the effect of
``collisions'' (encounters) between stars would tend to establish an isothermal
(Maxwell-Boltzmann) distribution. However, since the density of an
isothermal sphere decreases as $r^{-2}$ at large distances (implying an
infinite mass) the collisions would lead to the complete disintegration of the
cluster if they were allowed to act for a sufficient time. As a result, in a
subsequent paper,  Jeans
\cite{jeansr4} modeled globular clusters as collisionless stellar systems with
a distribution function of the form $f=f(\epsilon)$ satisfying the condition
$f(\epsilon)=0$ when
$\epsilon>0$ (for a
permanent cluster the energy of every star must be negative). He found that
 the density
decreases at large distances as $r^{-4}$. He
argued that this law provides a better description of globular clusters than
the Plummer law.

Eddington \cite{eddington3} reconsidered the results of von Zeipel and
Plummer from the point of view of collisionless stellar systems instead of
collisional gases. The connection is made through the so-called ``Eddington
formula'' \cite{bt} which relates the distribution function $f(\epsilon)$ of a
spherical
stellar system to the density $\rho(\Phi)$  of the corresponding barotropic
gas. In particular, he introduced the stellar polytropic
distributions $f=A(\epsilon_m-\epsilon)^{n-3/2}$
associated with the polytropic
gaseous spheres. Eddington tried to find physical arguments to justify the
Plummer
distribution
(polytrope $n=5$)  but did not arrive at a justification for this particular
law. Following Jeans, he insisted on the fact that the distribution function
$f(\epsilon)$ must vanish for positive energies ($\epsilon\ge 0$) in order to
account for the escape of high energy stars and avoid an infinite mass. He
introduced 
the truncated
Boltzmann distribution
$f=Ae^{-\beta\epsilon}$ if $\epsilon\le 0$ and $f=0$ if $\epsilon\ge
0$ that will be studied in detail later by Woolley
\cite{woolley}.\footnote{In addition, the paper of Eddington \cite{eddington3}
contains many insight
and intuitions about: (i) the escape of stars studied lated by Ambartsumian 
\cite{ambartsumian} and Spitzer \cite{spitzer}: ``the
system will settle down to a state which may be considered steady, since it is
only slowly altered by the encounters of the stars. Stars with high velocities
may be lost in the process; it is highly probable that some will escape
 since the
condition for a steady state $T=\frac{1}{2}\Omega$, shows that on the average
the mean kinetic energy of the stars is as much as a quarter of the kinetic
energy required for escape.''; (ii) the violent relaxation of collisionless
stellar systems studied
later by Lynden-Bell \cite{lb}: ``The system might oscillate at first; but,
since the periods of the
individual stars are not isochronous, the oscillations would die out rapidly.
The question arises whether the ultimate steady state may not be practically
independent of the initial conditions; and if so, can we show that it is the
state which the actual clusters assume?''; (iii) the King \cite{king} model:
``The law of
distribution should therefore be approximately the Maxwellian law for small
velocities, but modified for the larger velocities so as to fall to zero at or
before the limit $\sqrt{2\Phi}$''; (iv) the maximum entropy state and the
$H$-theorem
studied
lated by Ogorodnikov \cite{ogo1,ogo2}, Antonov \cite{antonov} and Lynden-Bell
and Wood \cite{lbw}: ``We
assume then that the most probable distribution is that which leads to the
lowest value of $H$, subject to the condition that the mass is given and the
energy is not greater than a fixed value''.}

Heckmann and Siedentopf \cite{hs} treated globular clusters as collisional
stellar systems
described by the Boltzmann equation. By
cancelling simultaneously
the advection term and the collisional term in the Boltzmann equation, they
derived from it the
Maxwell-Boltzmann distribution function $f({\bf r},{\bf v})$ and the 
Boltzmann density distribution $\rho({\bf r})$ coupled to the Poisson equation.
On the other hand, Mineur \cite{mineur}, Dicke  \cite{dicke} and Spitzer
\cite{spitzer} assumed that globular clusters have reached a state of
statistical equilibrium and that they are described by the Boltzmann-Poisson
equations applying to isothermal self-gravitating systems. The next developments
in the study of globular clusters and in the statistical mechanics of
classical self-gravitating systems are reviewed in the Introduction.

\section{Negative specific heats and ensembles inequivalence}
\label{sec_nsh}

In this Appendix, we show how the concepts of negative specific heats and
ensembles inequivalence for systems with long-range interactions emerged in
physics and astrophysics. 

The notion of negative specific heats was known to astronomers since the
19th century. Indeed, according to the virial
theorem, when a star or a star cluster
loses energy its temperature increases \cite{eddingtonbook,eddingtoncneg}.

In statistical physics, Landau and Lifshitz \cite{ll} (P. 62)
mentioned that inhomogeneous
systems whose energy is nonadditive, such as self-gravitating systems, may have
negative specific heats $C=dE/dT<0$. As a result, the body gets hotter when its
energy decreases. They stressed that this property is not in contradiction with
the laws of thermodynamics.

Lynden-Bell and Wood \cite{lbw} proved from the virial theorem that
self-gravitating systems have negative specific heat (they also showed that
the presence of a box in their model does not alter this result provided the
system is sufficiently centrally condensed). By losing heat they grow hotter.
The  evolution is thus away from equilibrium: If two systems with negative
specific heat are put in contact, the hotter loses heat and gets hotter while
the colder gains heat and gets colder. For the same
reason, no equilibrium is
possible for a system of negative specific heat in contact with a heat bath
(canonical ensemble). If the system is cooler than the bath it will accept
heat and grow cooler. Inversely, if the system is hotter than the bath it will
lose heat and become hotter. 
Finally, isolated systems (microcanonical ensemble) may also experience a
similar instability. If they are sufficiently concentrated, they take a
core-halo structure with a temperature gradient between the core and the halo
(the core being hotter than the halo). The inner parts being self-gravitating
have negative specific heat $C_c<0$; by losing heat they shrink and grow hotter.
The outer parts (which may be held by the walls of a box) have positive specific
heat $C_h>0$. On receiving heat they grow hotter (and expand if they are not
artificially confined). If $C_h<|C_c|$ the temperature increases faster in the
halo than in the core and the system as a whole can reach an equilibrium state.
If $C_h>|C_c|$ the temperature difference increases and no equilibrium is
possible. This is the gravothermal catatrophe. According to the previous
considerations, it occurs when the total specific heat $C=-|C_c|+C_h$ passes
from negative to positive
values (this is in accordance with the Poincar\'e turning point
criterion). A system of negative heat capacity can only
be in equilibrium with a system of
positive heat capacity provided that the sum of the heat capacities is
negative.
Lynden-Bell and Wood \cite{lbw}
related the core-halo structure of a system undergoing the gravothermal
catastrophe to the onset of red-giant structure in stellar evolution. They also
anticipated the existence of microcanonical phase transitions in a
self-gravitating gas of fermions or hard spheres (see their Appendix IV).

Thirring \cite{thirringgrg} mentioned that stars have negative specific
heat
$C=dE/dT<0$. They become hotter when energy is lost. However, a body with
negative specific heat cannot exist in thermal balance with a heat bath. It
undergoes a first order phase transition to a new state where the body has a
positive specific heat. This happens, for example, for sufficiently massive
white dwarfs (above the Chandrasekhar \cite{chandra31} limit) when the
zero-point
pressure of the electrons are no longer able to
counterbalance the gravitational attraction. This leads to one of those cosmic
catastrophes which we see as supernovae. The result is a pulsar (rotating
neutron star) surrounded by a cloud like the crab nebula. Thirring 
\cite{thirringgrg}  mentioned
that a sufficiently massive
neutron star (above the Oppenheimer-Volkoff \cite{ov} limit) may itself contract
to a ``mathematical point'' but considered this possibility as ``science
fiction''. This ``mathematical point'' is now interpreted as a black hole.

In a subsequent paper, Thirring \cite{thirring} refined his arguments. Referring
to Landau and Lifshitz \cite{ll} and
ter Haar \cite{haar}, he mentioned that self-gravitating systems may have
negative
specific heat for some energies. In contact with a heat bath a system
with $C<0$ creates an instability leading to a phase
transition. He argued that supernovae are, in essence, a phase transition of
this origin. He showed from
the virial theorem that $C=-(3/2)Nk_B<0$ in the microcanonical ensemble but
recalled
that, in the canonical ensemble, the specific heat is necessarily
positive.\footnote{In the microcanonical ensemble, using the virial theorem
$2K+W=0$ with
$K=(3/2)Nk_B T$ we get $E=K+W=-K=-(3/2)Nk_B T$ implying
$C=dE/dT=-(3/2)Nk_B<0$. In the canonical ensemble, using the partition function,
one can show
that $C=d\langle E\rangle/dT=\beta^2(\langle
E^2\rangle-\langle E\rangle^2)>0$ so the specific heat is a measure of the
variance of the fluctuations of energy. In general, equivalence of the
canonical and microcanonical descriptions is obtained whenever fluctuations
of energy are small. The canonical and microcanonical descriptions will be
inequivalent if the fluctuations are large $(\langle
E^2\rangle-\langle E\rangle^2)/\langle
E^2\rangle\gg 1$ that is to say when $C\rightarrow
+\infty$ corresponding to the turning point of temperature on the caloric
curve.}
Therefore,
the statistical ensembles are not equivalent.\footnote{To illustrate these
properties,
Thirring \cite{thirring} considered the statistical mechanics of
one particle in a Coulombian potential. This can be seen as a preamble to the
binary star model developed later by Padmanabhan \cite{paddy}. On the
other hand, Thirring \cite{thirring} and Hertel and Thirring
\cite{httoy} developped an
analytical toy model which presents two phases, one gaseous, the other
consisting of gas and one cluster of condensed matter. In the
microcanonical ensemble, there is a region of negative specific heat and
a first order phase transition marked by a jump of temperature at some energy.
In the canonical ensemble the region of negative specific heat is replaced by a
first order phase transition marked by a jump of energy at some
temperature. This is similar to the phase transitions in a self-gravitating gas
of fermions \cite{ht} (see a detailed history in
\cite{ijmpb,acf}) or hard spheres \cite{ah}. This is also similar to phase
transitions in the van der
Waals gas \cite{vk}.}
He argued that the region $C<0$ is jumped by a phase
transition of
the first kind. He argued that stars act like systems with $C<0$. At the end of
their life, when no more nuclear fuel is available, the star takes a core-halo
structure similar to a red giant or a supernova. These events reflect the
instability of systems with negative specific heat. Similarly, stellar
systems may separate into a collapsing core and a halo. However, the timescale
governing these phase transitions is very different in the two cases. For
supernovae where the energy is carried quickly by neutrinos they are fast (a
few days), but
for stellar systems they are very slow (of the order of the age of the
Universe).

In later years, Carlitz \cite{carlitz} showed that negative specific heat and
ensembles inequivalence also occur in hadronic matter at high density. On the
other
hand, Bekenstein \cite{bekenstein} and Hawking \cite{hawking1,hawking2} showed
that the thermodynamics of black holes involves negative specific heats. The
negative specific heat paradox was
further discussed by Lynden-Bell and Lynden-Bell \cite{lblb} and Padmanabhan
\cite{paddy} with the aid of analytical toy models. Some reviews about negative
specific heats and ensembles inequivalence in self-gravitating systems are
provided by Padmanabhan \cite{paddy}, Lynden-Bell \cite{lbphysicaA}, Katz
\cite{found} and Chavanis \cite{ijmpb}. More general results valid for
arbitrary systems with long-range interactions are given in
\cite{houches,cdr,campabook}.

\section{Construction of the caloric  curves}
\label{sec_cons}

In order to obtain the caloric curves $\eta(\Lambda)$ of a classical
self-gravitating gas in general relativity, we have followed the method of Bilic
and Viollier \cite{bvr}.

\subsection{The principle of the method}

Let us recall the procedure and illustrate it with a simple example. 
To construct the caloric curve $\eta(\Lambda)$ corresponding
to $N=0.1$ (see Fig. \ref{kcal_N01_linked_colors}), we proceeed as follows.
We first fix a value of $\alpha$. For this
value of $\alpha$, we can solve the TOV equations (\ref{b7}) and (\ref{b6}) by
prescribing the value
$\Phi_0$ of the potential at the origin [we work in terms of $\Phi(r)$ instead
of $b(r)$ using Eq. (\ref{b14})]. We stop the integration at
$r=R=1$ and compute the number of particles $N$, the
mass $M$ and the Tolman temperature $T_{\infty}$  from
Eqs. (\ref{b9})-(\ref{b11}). Using Eq. (\ref{ble}) we obtain $\Lambda$
and $\eta$. We then vary $\Phi_0$ from
$-1$ to $+\infty$ and plot $N$ as a
function of $\Phi_0$ for the initially fixed value of $\alpha$. An example of
curve
$N_{\alpha}(\Phi_0)$ is shown in Fig. \ref{line_level_Phi0}. This curve displays
damped
oscillations for large values of $\Phi_0$.\footnote{For very large values of
$\Phi_0$ the oscillations are
revived and the curve at high fields mirrors the curve at low fields. This is,
however, essentially a mathematical curiosity because the solutions associated
to these revived oscillations are unstable. They lead to the right parts of the
curves of Figs. \ref{ZLambda_R_N01a_unified} and \ref{Zeta_R_N01a_unified} that
are unstable (see footnote 35). Therefore, we shall not pay too
much attention on this part of the curve that we call ``irrelevant''. In the
following, we
shall focus essentially on the ``relevant'' part of the curve corresponding to
``low''
values of $\Phi_0$, up to the end of the damped
oscillations.}

\begin{figure}
\begin{center}
\includegraphics[clip,scale=0.3]{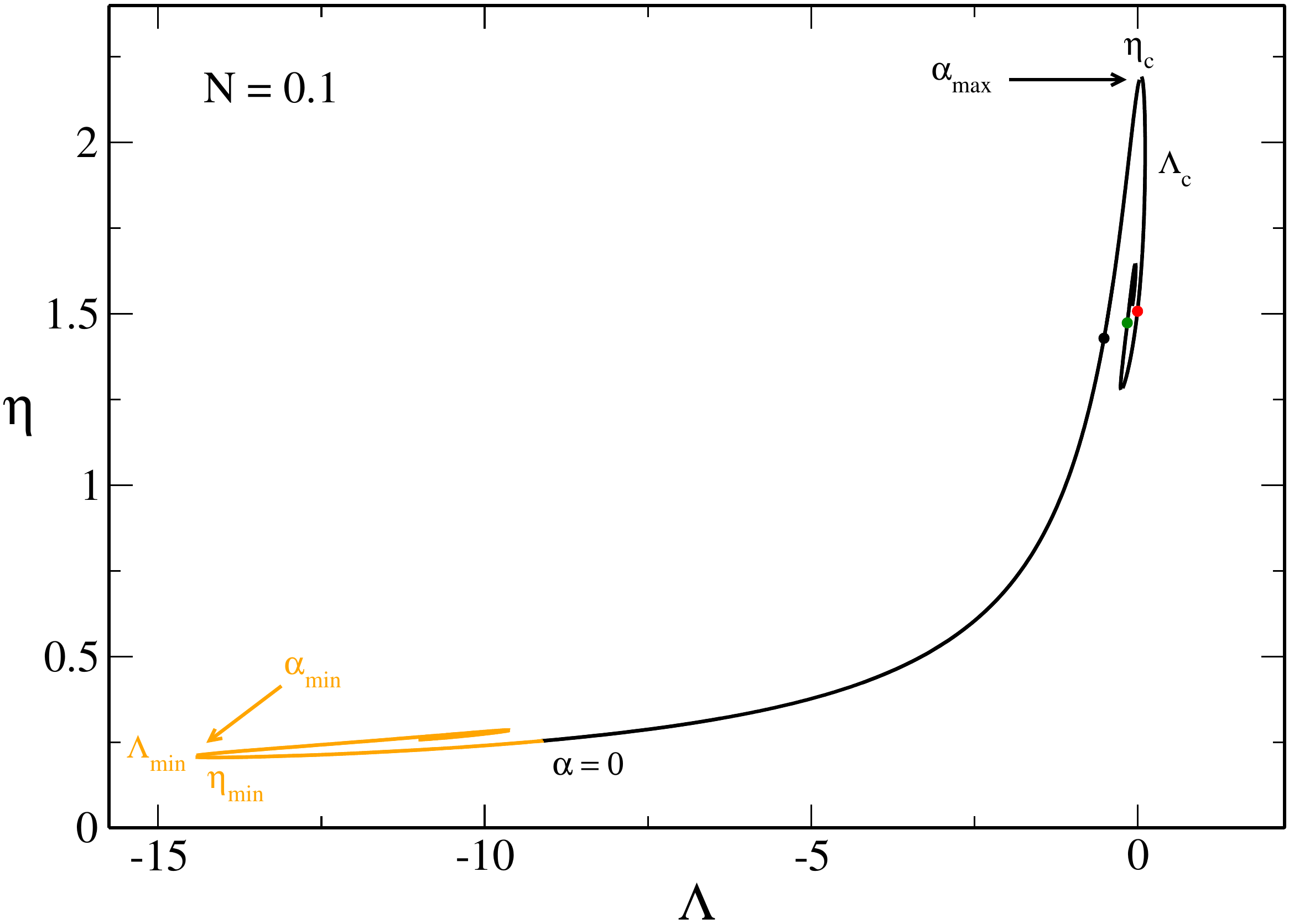}
\caption{Caloric curve $\eta(\Lambda)$ for $N=0.1$. The black part corresponds
to $\alpha>0$ (positive chemical potential) and the orange part corresponds to
$\alpha<0$ (negative chemical potential).}
\label{kcal_N01_linked_colors}
\end{center}
\end{figure}

\begin{figure}
\begin{center}
\includegraphics[clip,scale=0.3]{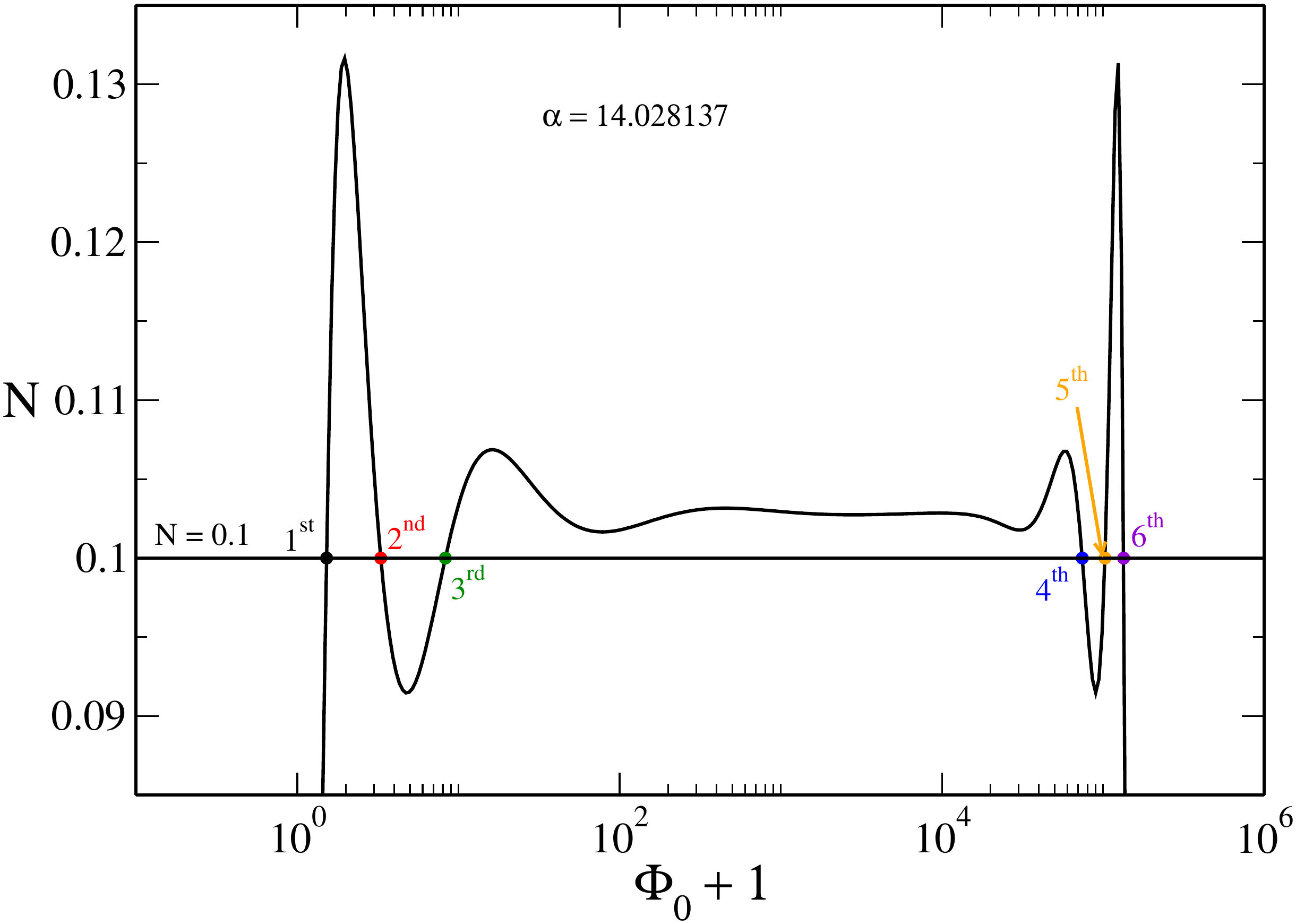}
\caption{The curve $N_{\alpha}(\Phi_0)$ for a fixed value of $\alpha$
(here $\alpha=14.028137$) illustrating the principle of the method detailed
in the text. We have selected $N=0.1$. The first three intersections correspond
to equilibrium states with $(\Phi_{0},\Lambda,\eta)_1=(0.5207,
-0.5164, 1.4286)$, $(\Phi_{0},\Lambda,\eta)_2=(2.3008, -3.663 \times
10^{-3}, 1.5076)$ and $(\Phi_{0},\Lambda,\eta)_3=(7.3043,  -0.1595,
1.4736)$. They have been localized on the caloric curve of Fig.
\ref{kcal_N01_linked_colors} by using the same color convention (black, red,
green). The subsequent intersections are ``irrelevant'' as explained in footnote
51.}
\label{line_level_Phi0}
\end{center}
\end{figure}

We now consider the possible intersection(s)
between the curve
$N_{\alpha}(\Phi_0)$ and the line level $N$.
Depending on the value of
$N$, there can be zero, one, or several
(up to an infinity) intersections.
In the example shown in Fig.  \ref{line_level_Phi0}, corresponding to
$\alpha=14.028137$ and $N=0.1$, there are three
intersections in the relevant part of the curve (see footnote 51). These
intersections
correspond to equilibrium states that have the same values of $N$ and $\alpha$
but that are characterized by different
values of $\Phi_0$ that we denote by $(\Phi_0)_{i\in\lbrace 1,...,n\rbrace}$.
In the present example, $n=3$.
The corresponding profiles of $\Phi(r)$ and $\epsilon(r)$ are
represented in Figs. \ref{Phi_profiles_example_intersections} and
\ref{epsilon_profiles_example_intersections} for illustration. These solutions
have different masses $M$ and different Tolman temperatures
$T_{\infty}$, hence different
values of $\Lambda$ and $\eta$. For the selected value of $N$ and for
the initially fixed value of $\alpha$ we
can report these solutions on the caloric curve $\eta(\Lambda)$. In our example,
this
defines three points (black, red and green) in Fig.
\ref{kcal_N01_linked_colors}. By varying $\alpha$ these
points form $n$ branches in the caloric curve
$\eta(\Lambda)$. The branches $B_1$ and $B_2$ corresponding to the first and
second intersections have been represented in color (black and red) in Fig.
\ref{branches_new2}.

In conclusion, the complete caloric curve
$\eta(\Lambda)$ for the selected value of $N$ is obtained by determining the
intersections between  the curve $N_{\alpha}(\Phi_0)$ and the line level $N$
for ``all'' values of $\alpha$ ranging from $-\infty$ to $+\infty$. We
can then redo this work for different values of $N$ in order to see how the
caloric
curve $\eta(\Lambda)$ changes with the number of particles. This leads to the
caloric curves presented in Sec. \ref{sec_gc}.

\begin{figure}
\begin{center}
\includegraphics[clip,scale=0.3]
{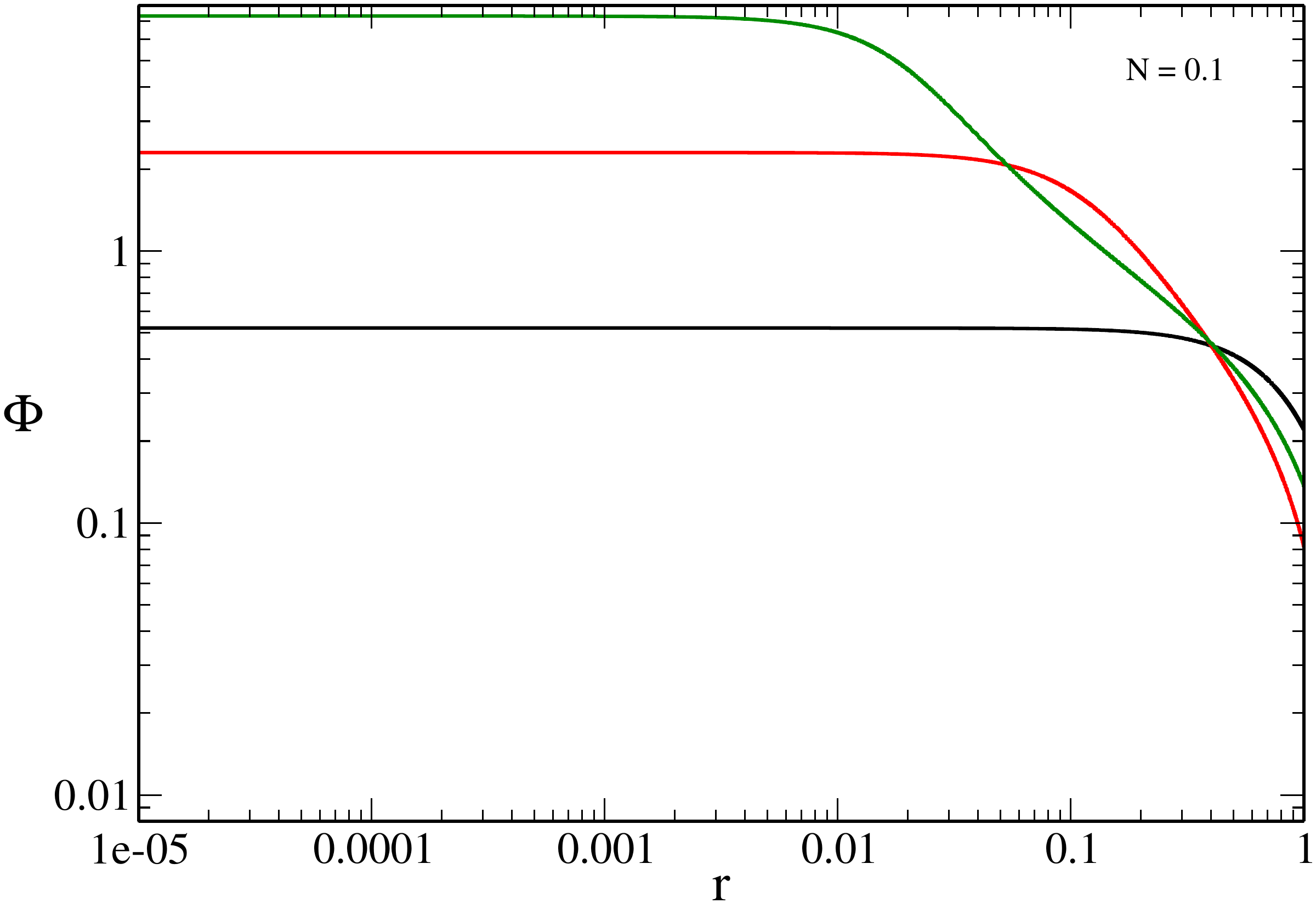}
\caption{Gravitational potential $\Phi(r)$
corresponding to the three intersections displayed in Fig.
\ref{line_level_Phi0}.}
\label{Phi_profiles_example_intersections}
\end{center}
\end{figure}

\begin{figure}
\begin{center}
\includegraphics[clip,scale=0.3]
{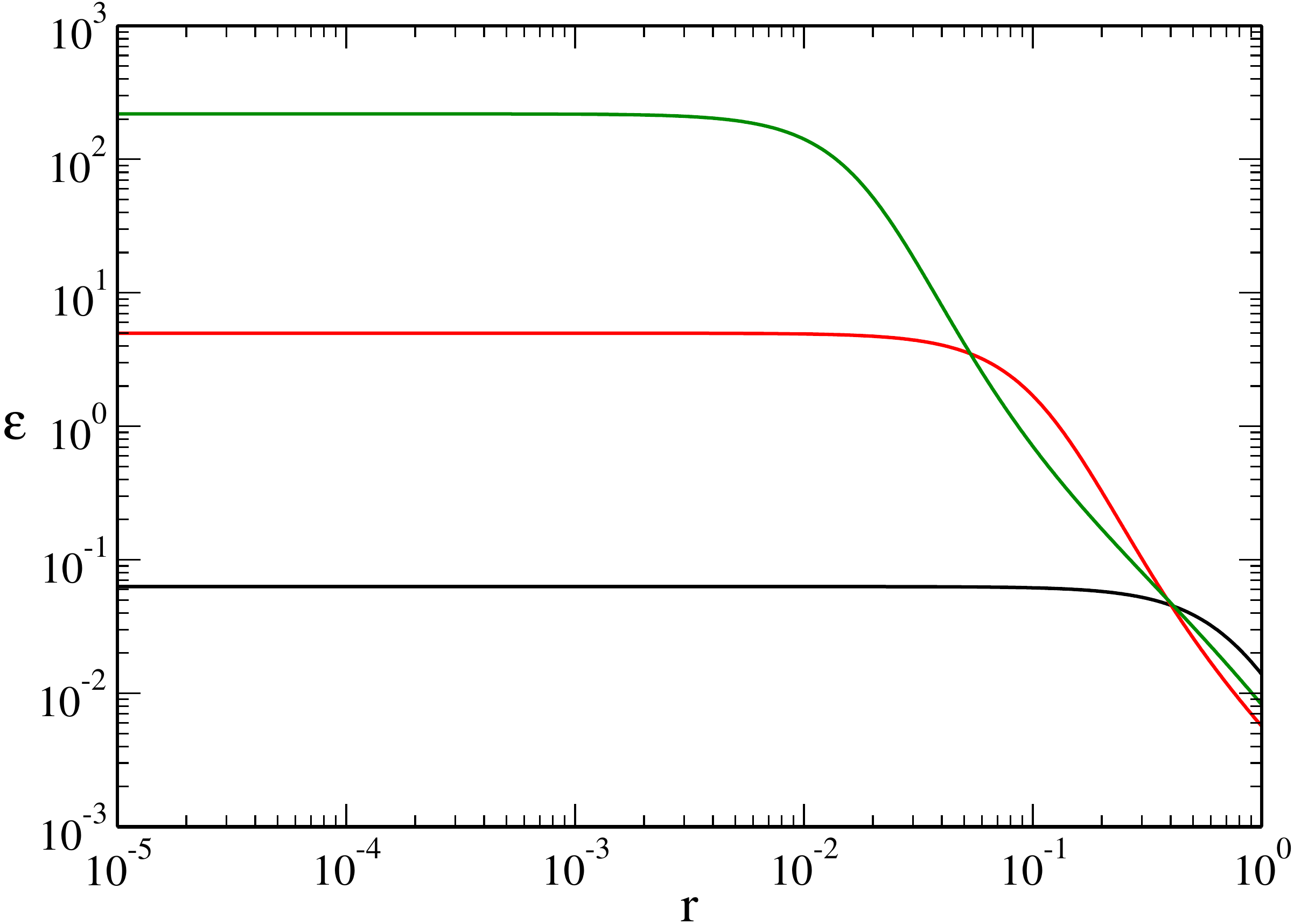}
\caption{Energy density profile $\epsilon(r)$ corresponding to the three
intersections displayed in Fig. \ref{line_level_Phi0}.}
\label{epsilon_profiles_example_intersections}
\end{center}
\end{figure}

\begin{figure}
\begin{center}
\includegraphics[clip,scale=0.3]{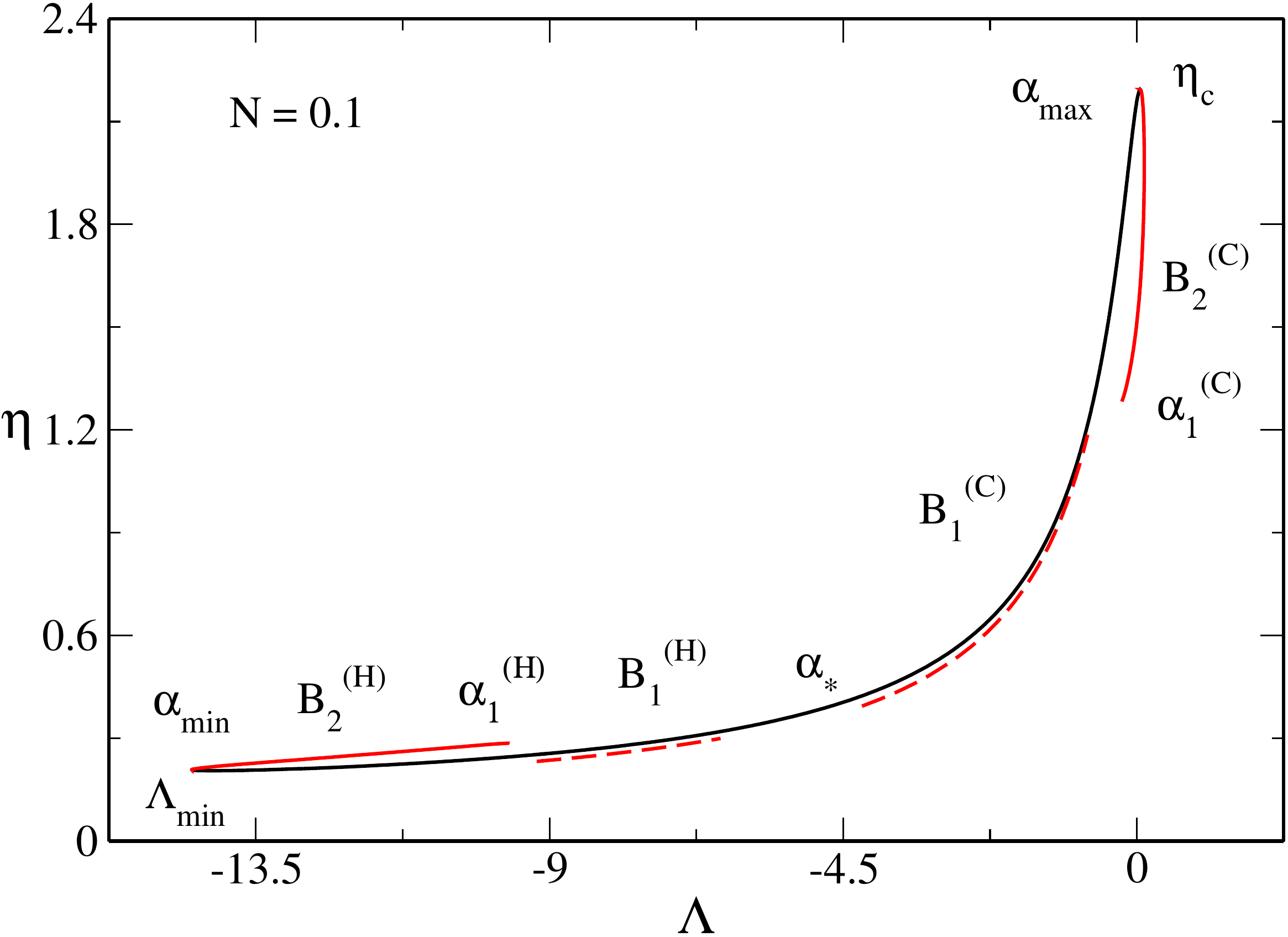}
\caption{Caloric curve $\eta(\Lambda)$ for
$N=0.1$ showing just
the branches $B_1$ and $B_2$ corresponding to the first two intersections in
Fig.
\ref{line_level_Phi0} (extended to all values of $\alpha\in [\alpha_{\rm
min},\alpha_{\rm max}]$). The branches $B_1^{(H)}$ and $B_1^{(C)}$
(black) correspond to the ensemble of the first intersections. They give rise to
the main
branch.  The branch $B_1^{(H)}$
corresponds to $\alpha_{\rm min}\le \alpha\le \alpha_*$ and the  branch
$B_1^{(C)}$ corresponds to
$\alpha_*\le \alpha\le \alpha_{\rm max}$. They connect each other at
$\alpha=\alpha_*$.
The branches $B_2^{(H)}$ and $B_2^{(C)}$ (red) correspond to the ensemble of
the second
intersections. They give rise to the begining of the hot (H) and cold (C)
spirals. The branch
$B_2^{(H)}$
corresponds to $\alpha_{\rm min}\le\alpha\le \alpha_1^{(H)}$. It connects the
branch $B_1^{(H)}$ 
at $\alpha_{\rm min}$ and stops at $\alpha_1^{(H)}$. The branch
$B_2^{(C)}$
corresponds to $ \alpha_1^{(C)}\le \alpha\le \alpha_{\rm max}$. It
starts at $ \alpha_1^{(C)}$ and connects the branch $B_1^{(C)}$ 
at $\alpha_{\rm max}$. Next order intersections (not represented) form the 
other branches of the  spirals. We observe that a part of the second
branches $B_2$ is
superimposed on the main branch $B_1$ (we have slightly
shifted the red curves lying on the black curve for a better visualization).
They correspond to different equilibrium states with the same energy and the
same temperature but a different density contrast (see the right parts in red of
Figs. \ref{ZLambda_R_N01a_unified} and \ref{Zeta_R_N01a_unified}). These
solutions which have a very high central density (see Fig.
\ref{profiles_isoerg2})
are irrelevant because they are unstable (see footnote 51). We
note that the branch $B'_2$ (dashed line) corresponds to the branch of last
intersections that becomes a branch of second intersections when the branch
$B_2$
disappears.}
\label{branches_new2}
\end{center}
\end{figure}

\begin{figure}
\begin{center}
\includegraphics[clip,scale=0.3]{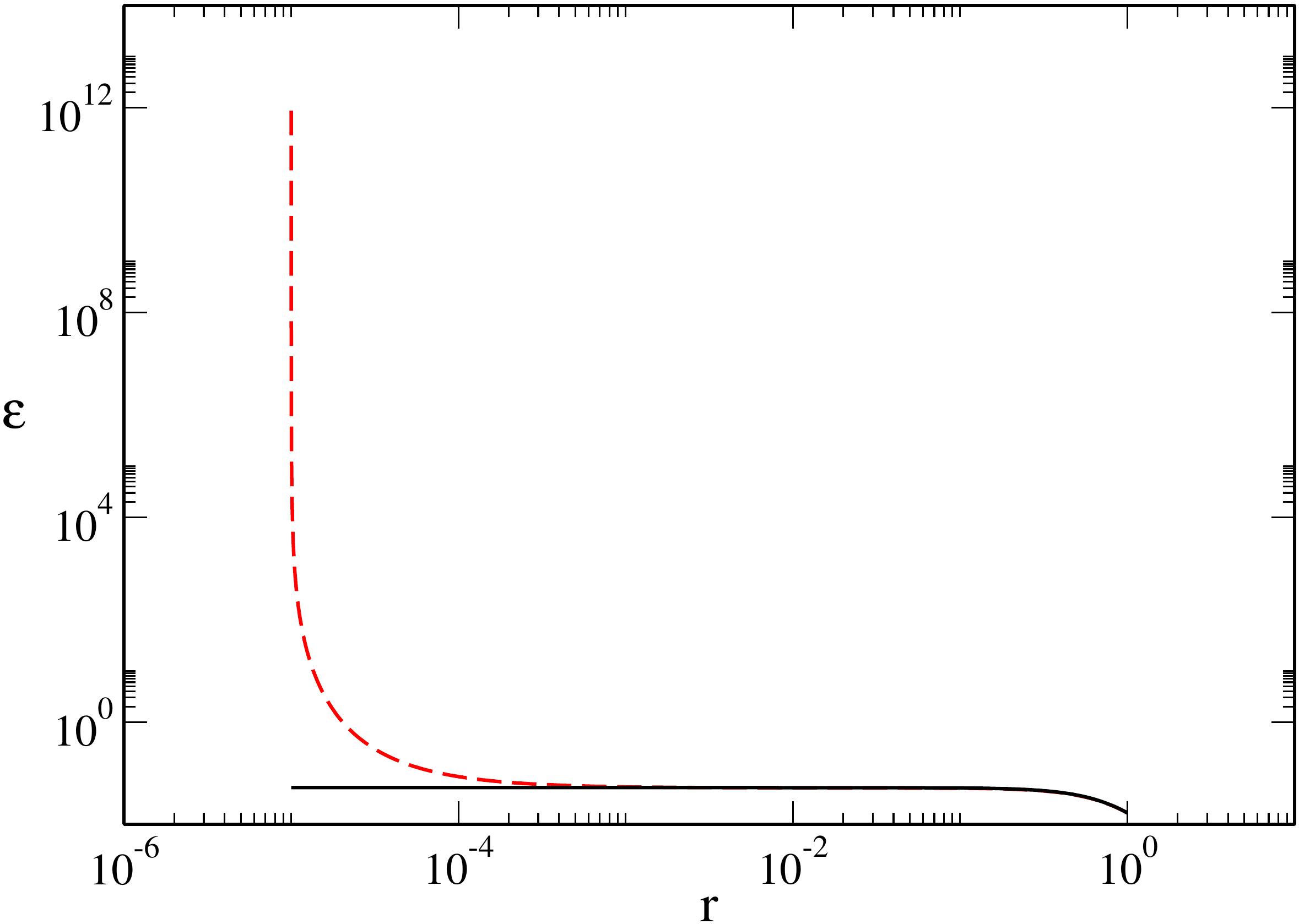}
\caption{Energy density profile of a stable
solution (black solid line) and of an 
unstable solution (red dashed line) with the same value of energy and
temperature (specifically $\alpha=9.77$, $\eta=1.01$ and
$\Lambda=-1.09$). These solutions are superimposed on the main 
branch of the caloric curve (see Fig. \ref{branches_new2}). We see that their
profiles coincide except at the very center where the unstable solution has a
very high density.}
\label{profiles_isoerg2}
\end{center}
\end{figure}

\subsection{Evolution of $N_{\alpha}(\Phi_0)$ with $\alpha$}

We now describe how the curve $N_{\alpha}(\Phi_0)$ changes with
$\alpha$ (as explained in
footnote 51, we only consider the relevant part of the 
curve $N_{\alpha}(\Phi_0)$). To facilitate the discussion, we introduce some
notations. We call
${\cal N}(\alpha)$ the maximum
value of the curve $N_{\alpha}(\Phi_0)$ and we denote by $\Psi(\alpha)$ the
value
of the
central potential $\Phi_0$ corresponding to this maximum. 

We note that the curve $N_0(\Phi_0)$ corresponding to $\alpha=0$ is singular
because $\alpha$ appears explicitly in the relation 
$T_0=\frac{1}{|\alpha|}\sqrt{\Phi_0+1}$ between $T_0$ and $\Phi_0$ [see Eq.
(\ref{b14})]. Therefore, when $\alpha\rightarrow 0$ the curves
$N_{\alpha}(\Phi_0)$ are ``squeezed'' near $\Phi_0=-1$. This is, however, just
an apparent singularity that would have not occurred if we had chosen to plot
$N_{\alpha}$ as a function of $b_0$ instead of $\Phi_0$. 

\begin{figure}
\begin{center}
\includegraphics[clip,scale=0.3]{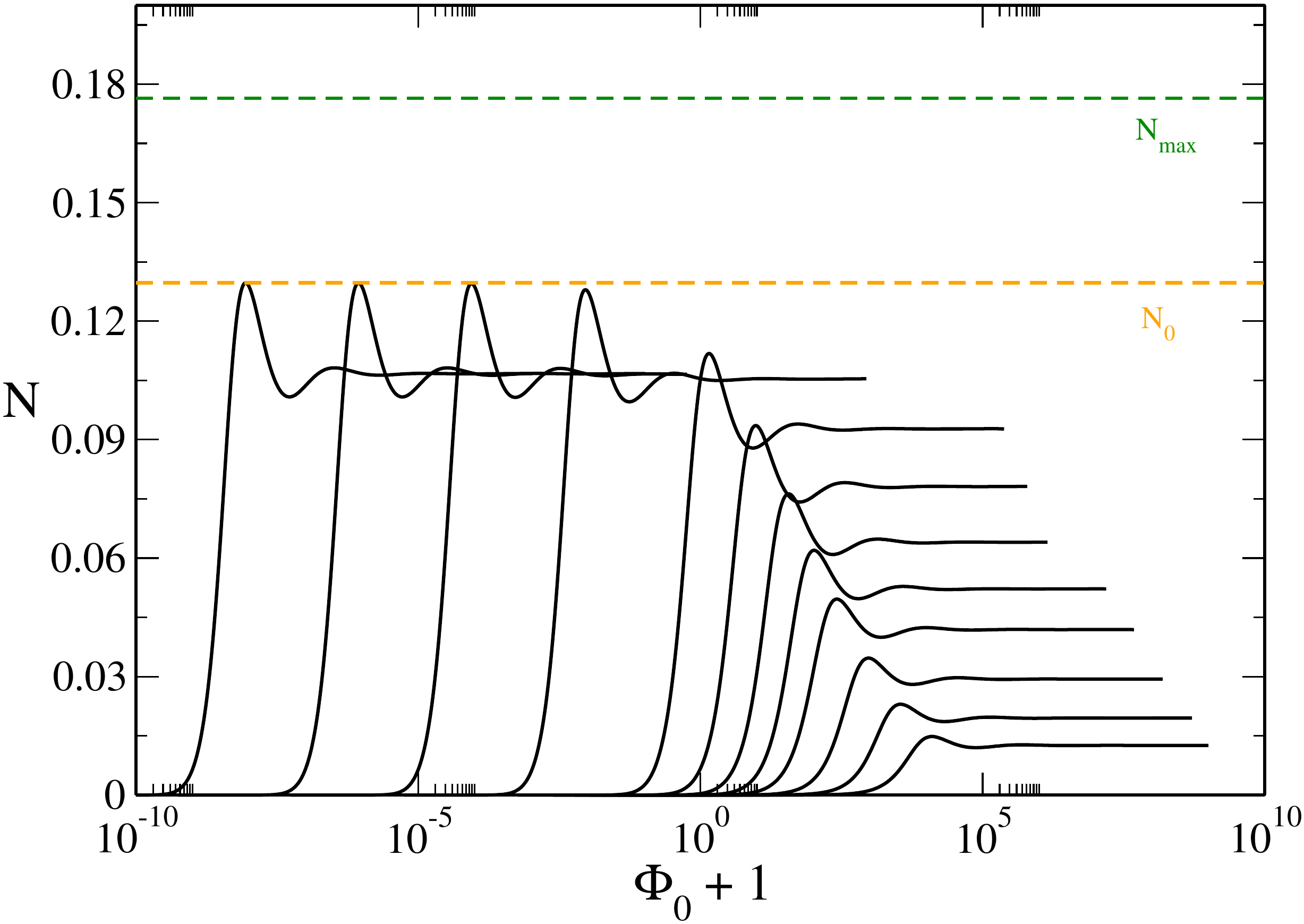}
\caption{Evolution of the curve $N_{\alpha}(\Phi_0)$ for different
values of $\alpha<0$ (for illustration, the curves go from $\alpha = -10$ to
$\alpha = -10^{-6}$). We have indicated different characteristic values of $N$:
$N_0=0.1297$ and $N_{\rm max}=0.1764$.}
\label{NPhi0_neg_Boltzmann2}
\end{center}
\end{figure}

Let us first consider the case $\alpha<0$ (see Fig. \ref{NPhi0_neg_Boltzmann2}).
When $\alpha\rightarrow -\infty$, we find that ${\cal N}(\alpha)\rightarrow 0$
and 
$\Psi(\alpha)\rightarrow +\infty$.  This corresponds to the
ultrarelativistic regime that gives rise to the hot spiral in the limit
$N\rightarrow 0$ (see Sec. \ref{sec_dhs}).  As
$\alpha$ increases, ${\cal N}(\alpha)$ increases and 
$\Psi(\alpha)$ decreases: the peak of the curve $N_{\alpha}(\Phi_0)$ grows
and moves to the left. When
$\alpha\rightarrow 0$, we find that ${\cal
N}(\alpha)\rightarrow N_0=0.1297$ and  $\Psi(\alpha)\rightarrow -1$: the peak
of the curve $N_{\alpha}(\Phi_0)$  is squeezed near $\Phi_0=-1$.

\begin{figure}
\begin{center}
\includegraphics[clip,scale=0.3]{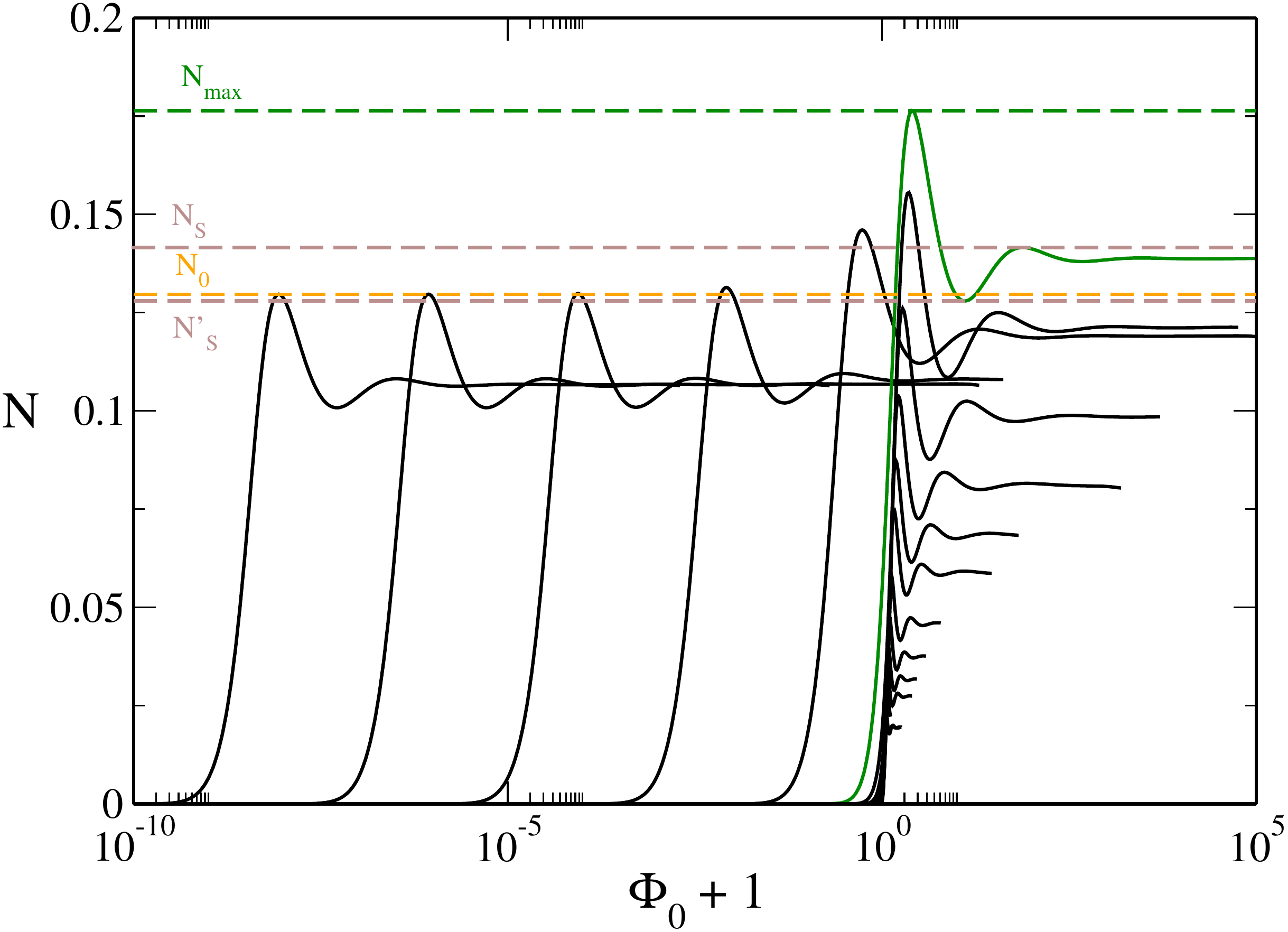}
\caption{Evolution of the curve $N_{\alpha}(\Phi_0)$ for different
values of $\alpha>0$ (for illustration, the curves go from $\alpha
= 10^{-6}$ to $\alpha=100$). We have indicated different characteristic values
of $N$:
$N_0=0.1297$, $N_{\rm max}=0.1764$, $N'_S=0.128$ and 
$N_S=0.1415$. }
\label{huit}
\end{center}
\end{figure}

We now turn to the case $\alpha>0$ (see Fig. \ref{huit}).
When $0<\alpha\le \alpha_*=5.012$, we find that ${\cal N}(\alpha)$ and
$\Psi(\alpha)$ both
increase: the peak of the curve $N_{\alpha}(\Phi_0)$ grows
and moves to the right. When $\alpha=\alpha_*$, ${\cal
N}(\alpha)$ reaches its maximum value $N_{\rm
max}=0.1764$ at $\Psi_*=1.51157$. The
first minimum and the second maximum of the curve
$N_{\alpha_*}(\Phi_0)$ will play a particular role in the  interpretation of
the caloric curves (see below). The values of $N$ at these
points are $N'_S=0.128$ and 
$N_S=0.1415$. When $\alpha\ge \alpha_*$,
${\cal
N}(\alpha)$ and $\Psi(\alpha)$ both decrease: the peak of the curve
$N_{\alpha}(\Phi_0)$ decays and moves to the left. When
$\alpha\rightarrow +\infty$, we find that ${\cal
N}(\alpha)\rightarrow 0$ and  $\Psi(\alpha)\rightarrow 0$. This corresponds to
the
nonrelativistic regime that gives rise to the cold spiral in the limit
$N\rightarrow 0$ (see Sec. \ref{sec_nzerocold}).

\subsection{Extremal values of $\alpha$ for a given $N$}

Let us select a value of $N$ and progressively increase $\alpha$, starting from
$\alpha\rightarrow -\infty$ (see Fig. \ref{Nb0_b}).

\begin{figure}
\begin{center}
\includegraphics[clip,scale=0.3]{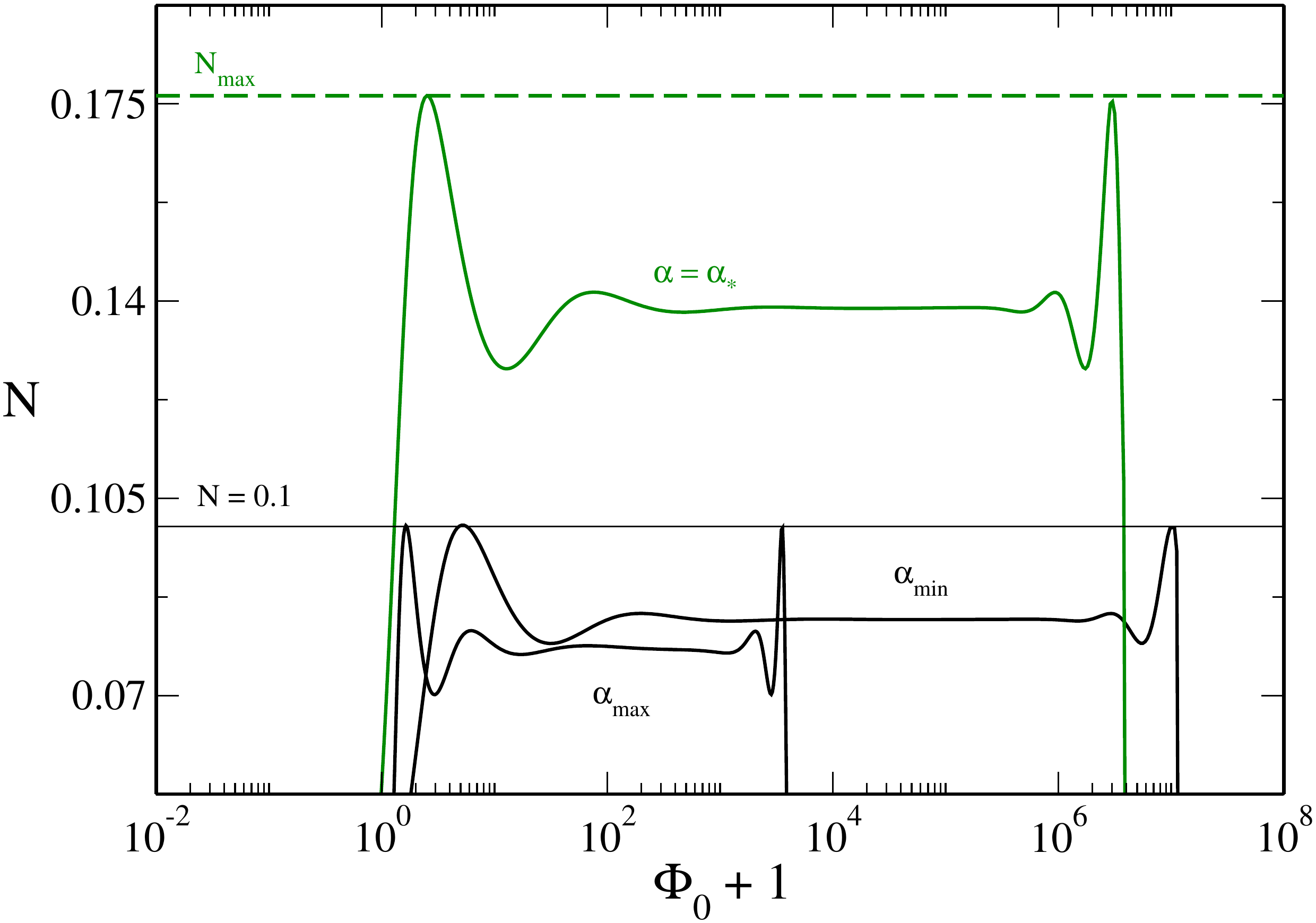}
\caption{Evolution of the curve $N_{\alpha}(\Phi_0)$ with $\alpha$ illustrating
the first intersection with the line level $N$ occuring at
$\alpha_{\rm min}(N)$ and the last intersection occuring at  $\alpha_{\rm
max}(N)$.
There is no intersection below $\alpha_{\rm min}(N)$ or above  $\alpha_{\rm
max}(N)$. For illustration we have taken $N=0.1$ for which $\alpha_{\rm
min}=-1.641$ and 
$\alpha_{\rm max}=20.989$. We have also represented the curve
$N_{\alpha_*=5.012}(\Phi_0)$ whose peak ${\cal
N}(\alpha_*)$ reaches the maximum value $N_{\rm max}=0.1764$. 
}
\label{Nb0_b}
\end{center}
\end{figure}

For small values of $\alpha$, there is no intersection between the line level
$N$ and the curve $N_{\alpha}(\Phi_0)$. However, as
the peak
${\cal N}(\alpha)$ grows as $\alpha$ increases, some intersections become
possible. The first intersection
with the line level $N$
occurs for $\alpha=\alpha_{\rm min}(N)$. For $\alpha>\alpha_{\rm min}(N)$ the
peak  ${\cal N}(\alpha)$ continues to grow, reaches a maximum $N_{\rm
max}=0.1764$
at $\alpha_*=5.012$, then
decreases. The last intersection with the line level $N$ occurs for
$\alpha=\alpha_{\rm max}(N)$. For $\alpha>\alpha_{\rm max}(N)$, there
is no intersection.

\begin{figure}
\begin{center}
\includegraphics[clip,scale=0.3]{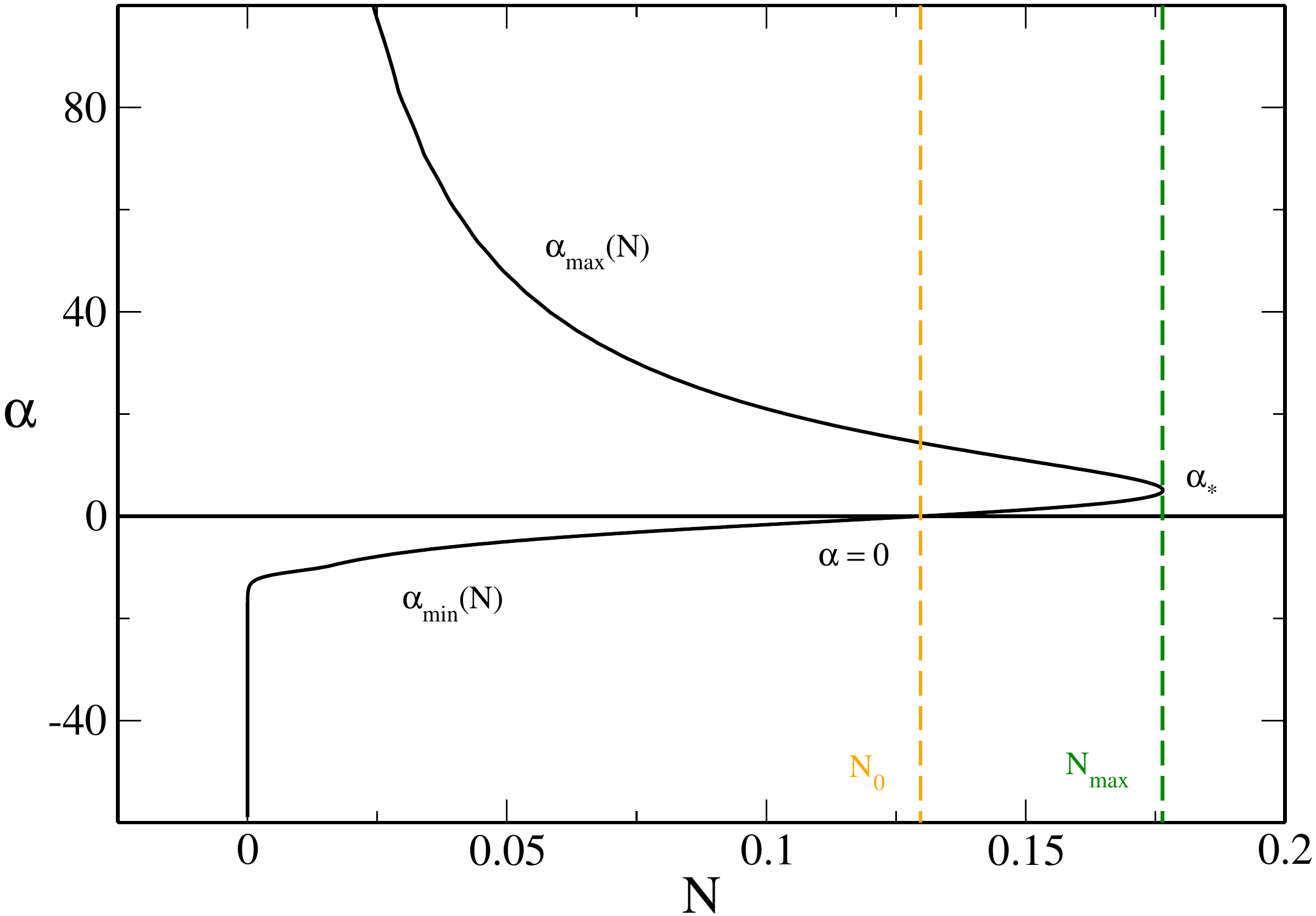}
\caption{Evolution of $\alpha_{\rm min}$ and $\alpha_{\rm
max}$ with $N$. We find that $\alpha_{\rm min}\simeq
-6.28+0.90\ln N$ and $\alpha_{\rm max}\sim 2.40/N$ for $N\rightarrow 0$.}
\label{alpha_N2}
\end{center}
\end{figure}

In Fig. \ref{alpha_N2}, we have ploted $\alpha_{\rm min}$ and $\alpha_{\rm
max}$ as a function of $N$. Let us mention some characteristic values. 
For  $N=N_{\rm max}=0.1764$ we have $\alpha_{\rm min}=\alpha_{\rm
max}=\alpha_*=5.012$. On the other hand, we find that $\alpha_{\rm min}=0$ for
$N=N_0=0.1297$. Therefore, when $N>N_0$, the caloric curve is made exclusively
of equilibrium states with $\alpha>0$ (positive chemical potential). Apart from
this property, the
value $N_0=0.1297$ does not seem to play a special role in the problem.

\subsection{Relation with the caloric curves for different values of $N$}

We are now ready to discuss the relation between the topological properties of
the curves $N_{\alpha}(\Phi_0)$ and the
caloric curves $\eta(\Lambda)$ analyzed in Sec. \ref{sec_gc}.

\subsubsection{$N<N'_S$}
\label{sec_anti}

Let us consider the case $N<N'_S=0.128$ (for illustration we
take $N=0.1$).

We first consider the possible intersections between the curve
$N_{\alpha}(\Phi_0)$ and the line level $N$ when $\alpha\le \alpha_*$. As we
shall see, this range of $\alpha$ is associated with the left part of the main
branch $+$ the ``hot spiral''
corresponding to the strongly relativistic gas.

For $\alpha<\alpha_{\rm min}(N)$, there is no intersection.
For $\alpha$ just above $\alpha_{\rm min}(N)$, two intersections appear (see
Fig. \ref{NPhi0_N01_new_ter_CUT}). If we keep increasing  $\alpha$, we 
successively find
 more and more intersections, then less and less intersections, as
the oscillations of the curve $N_{\alpha}(\Phi_0)$ traverse the line level $N$.
For
even larger values of $\alpha$, i.e. $\alpha>\alpha_1^{\rm (H)}$, the
oscillations of the curve
$N_{\alpha}(\Phi_0)$ have passed above the line level $N$ so there is only one
(relevant) intersection. We can see these different intersections, as a function
of
$\alpha$, in  Figs. \ref{ZPhi0_alpha_N01_unified}
and \ref{Zb0_alpha_N01b_unified} (left side). Each
set
of intersections, as we vary $\alpha$, defines a branch 
${B}^{\rm (H)}_i=\lbrace \Lambda_{i}(\alpha),
\eta_{i}(\alpha)\rbrace$ of the caloric curve
$\eta(\Lambda)$: ${B}^{\rm (H)}_1$ is the branch corresponding to the first
intersections (black), ${B}^{\rm (H)}_2$ is the branch corresponding to
the second intersections (red), etc. The ensemble of
these branches forms the left part of the main branch and the ``hot spiral''. We
see in  Figs.
\ref{ZPhi0_alpha_N01_unified}
and \ref{Zb0_alpha_N01b_unified} (left side) that the
typical values of $\Phi_0$ and  $T_0$ are
large so the system is strongly relativistic on this part of the caloric curve.
By
plotting these branches in different colors on the density contrast versus
energy curve of Fig. \ref{ZLambda_R_N01a_unified} or on the caloric curve of
Fig. \ref{branches_new2}, and considering the hot spiral, we observe that two
successive branches
merge at a turning point of energy (we do not have a mathematical proof for
that). In particular, the branches ${B}^{\rm (H)}_1$ and ${B}^{\rm
(H)}_2$ corresponding to the first two
intersections (plotted in black and red) merge, for $\alpha=\alpha_{\rm
min}(N)$, at the
critical point $\Lambda_{\rm min}$ corresponding to the maximum energy.
Similarly, the branches ${B}^{\rm (H)}_2$ and ${B}^{\rm (H)}_3$
corresponding to
the second and third intersections (red and green) merge at the second turning
point of energy, and so on... As a result, for $\alpha\le \alpha_*$, the
ensemble of the first intersections ${B}^{\rm (H)}_1$ forms the left part
of the main branch of the caloric curve and the ensemble of the subsequent
intersections ${B}^{\rm (H)}_2$, ${B}^{\rm (H)}_3$... form the ``hot
spiral''.

\begin{figure}
\begin{center}
\includegraphics[clip,scale=0.3]{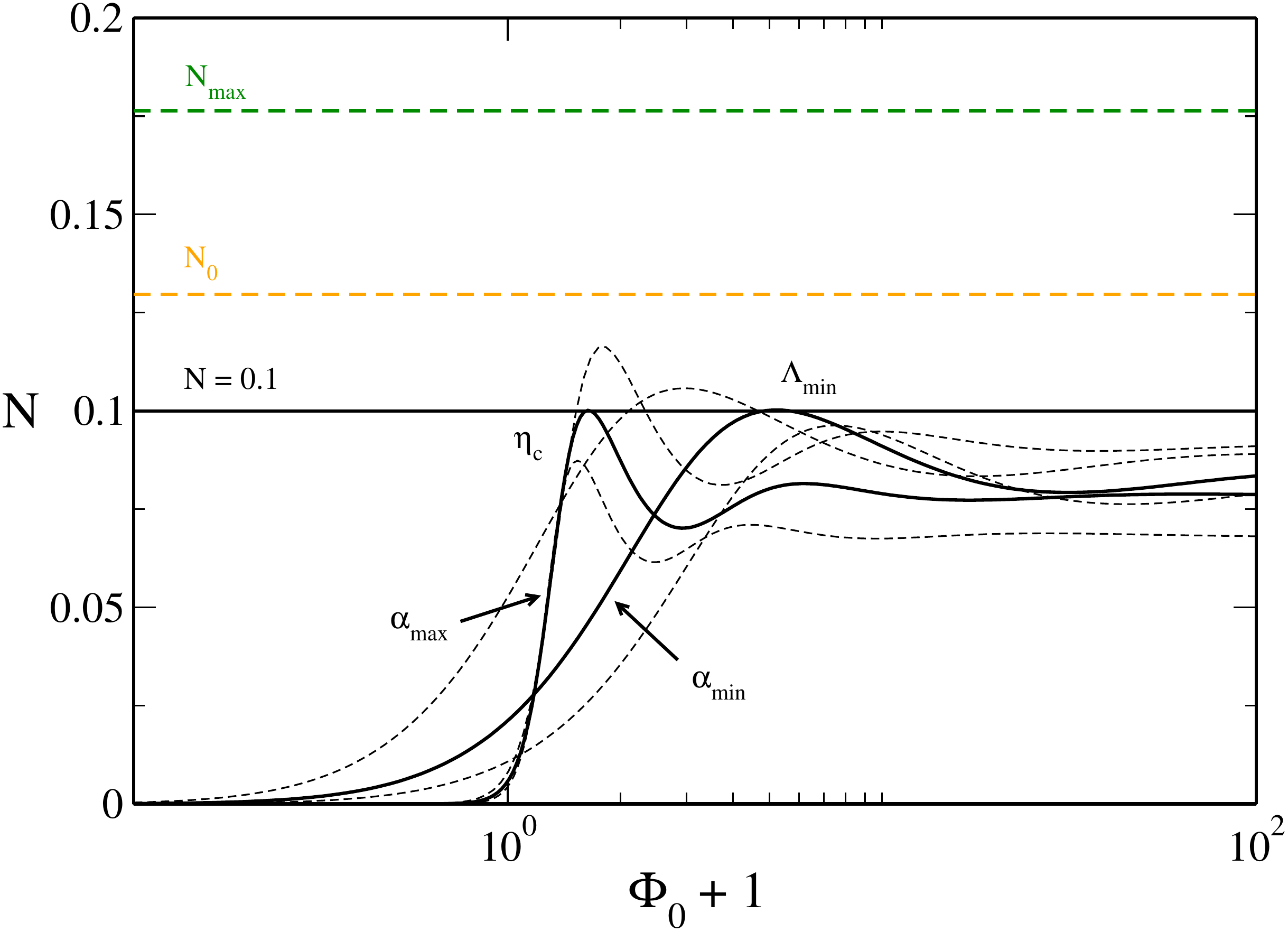}
\caption{(i) Appearance of two intersections
when $\alpha$ passes above
$\alpha_{\rm min}$ (the full line corresponds to $\alpha=\alpha_{\rm min}$ and
the dashed lines correspond to values of $\alpha$ slightly below or above 
$\alpha_{\rm min}$). They are associated with the first turning point of
energy $\Lambda_{\rm min}$ of the hot spiral ($\Phi_0$ large). (ii)
Disappearance of two
intersections when $\alpha$ passes above $\alpha_{\rm max}$ (the full
line corresponds to $\alpha=\alpha_{\rm max}$ and
the dashed lines correspond to values of $\alpha$ slightly above or below 
$\alpha_{\rm max}$). They
are
associated with the first turning point of temperature $\eta_{c}$ of
the cold spiral ($\Phi_0$ small).
}
\label{NPhi0_N01_new_ter_CUT}
\end{center}
\end{figure}

\begin{figure}
\begin{center}
\includegraphics[clip,scale=0.3]{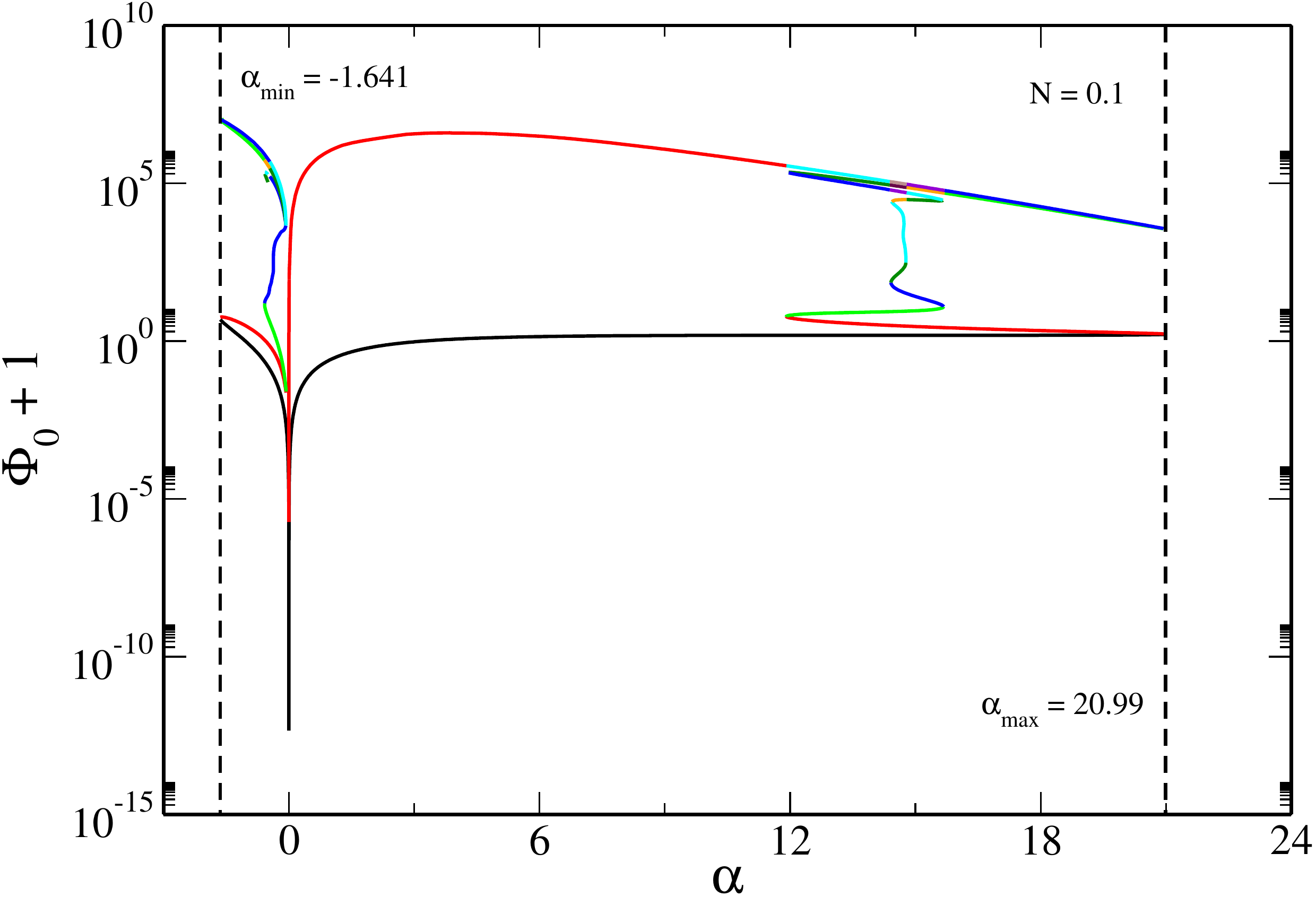}
\caption{Central value of the
gravitational potential
$(\Phi_0)_{i=1,2,...}$, corresponding
to the different intersections between the curve $N_{\alpha}(\Phi_0)$ and
the line level $N$ (here $N=0.1$), as a function of $\alpha$. These curves
clearly displays
the bound $\alpha_{\rm min}$ and
$\alpha_{\rm max}$ between which equilibrium states exist. Each
intersection is plotted with a different color, the first one corresponding to
the black curve, the second to the red curve, the third to the green
curve etc. We are essentially
interested in the first two intersections because the other ones correspond
to unstable equilibrium states.}
\label{ZPhi0_alpha_N01_unified}
\end{center}
\end{figure}

\begin{figure}
\begin{center}
\includegraphics[clip,scale=0.3]{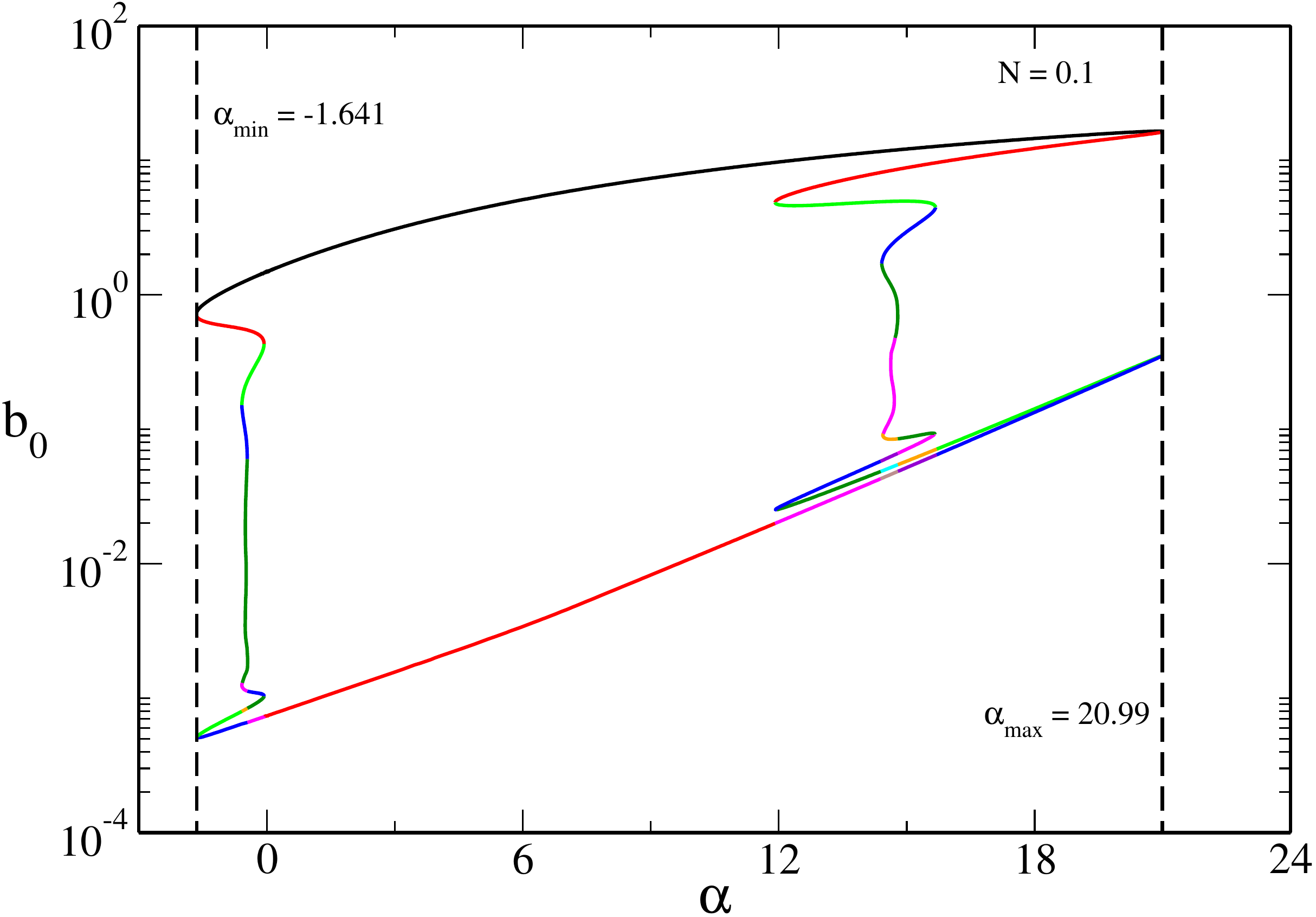}
\caption{Same as Fig. \ref{ZPhi0_alpha_N01_unified} except that we have plotted
the inverse central temperature $b_0$ instead of the central potential $\Phi_0$
for a better visualization.}
\label{Zb0_alpha_N01b_unified}
\end{center}
\end{figure}

We now consider the possible intersections between the curve
$N_{\alpha}(\Phi_0)$ and the line level $N$ when $\alpha\ge \alpha_*$. As we
shall see, this range of $\alpha$ is associated with the right
part of the main
branch $+$  the ``cold
spiral'' corresponding to the weakly relativistic gas. To
make the discussion
symmetric with respect to the previous one, we start from $\alpha\rightarrow
+\infty$ and progressively decrease its value.

For $\alpha>\alpha_{\rm max}(N)$, there is no intersection.
For $\alpha$ just below $\alpha_{\rm max}(N)$, two intersections appear (see
Fig. \ref{NPhi0_N01_new_ter_CUT}). If we
keep decreasing  $\alpha$, we successively find more and more intersections,
then less and
less intersections, as the oscillations of the curve $N_{\alpha}(\Phi_0)$
traverse
the line level $N$. For
even lower values of $\alpha$, i.e. $\alpha<\alpha_1^{\rm (C)}$, the
oscillations of the curve
$N_{\alpha}(\Phi_0)$ have passed above the line level $N$ so there is only one
(relevant) intersection. We can see these different intersections, as a function
of
$\alpha$, in  Figs. \ref{ZPhi0_alpha_N01_unified}
and \ref{Zb0_alpha_N01b_unified} (right side). Each set
of intersections, as we vary $\alpha$, defines a branch  ${B}^{\rm
(C)}_i=\lbrace \Lambda_{i}(\alpha),\eta_{i}(\alpha)\rbrace$  of the caloric
curve
$\eta(\Lambda)$: ${B}^{\rm (C)}_1$ is the branch corresponding to the first
intersections (black), ${B}^{\rm (C)}_2$ is the branch corresponding to
the second intersections (red), etc. The ensemble of
these branches forms the right part of the main branch and the ``cold spiral''.
We see in  Figs.
\ref{ZPhi0_alpha_N01_unified}
and \ref{Zb0_alpha_N01b_unified} (right side) that the
typical values of $\Phi_0$ and $T_0$ are small so
the system is weakly
relativistic on this part of the caloric curve. By plotting these branches in
different colors on the density
contrast versus temperature curve of Fig. \ref{Zeta_R_N01a_unified} or on the
caloric curve of Fig. \ref{branches_new2}, and considering the cold spiral,
we observe that two successive branches
merge at a turning point of temperature (we do not have a mathematical proof for
that). In particular, the branches ${B}^{\rm
(C)}_1$ and ${B}^{\rm (C)}_2$
 corresponding to the first two
intersections (plotted in black and red) merge, for $\alpha=\alpha_{\rm
max}(N)$, at the
critical point $\eta_{c}$ corresponding to the minimum temperature.
Similarly, the branches ${B}^{\rm (C)}_2$ and ${B}^{\rm (C)}_3$
corresponding to
the second and third intersections (red and green) merge at the second turning
point of temperature, and so on... As a result, for $\alpha\ge \alpha_*$, the
ensemble of the first intersections ${B}^{\rm (C)}_1$ forms the right part
of the main
branch of the caloric curve and the ensemble of the secondary
intersections ${B}^{\rm (C)}_2$, ${B}^{\rm (C)}_3$... form the
``cold spiral''.

We can make the following comments:

(i) The roles of $\Lambda$  and $\eta$ are  reversed for the hot
and cold spirals. For the hot spiral, the first two branches ${B}^{\rm (H)}_1$
and ${B}^{\rm (H)}_2$ merge at the first turning point of energy $\Lambda_{\rm
min}$. By contrast, for the cold
spiral, the first two branches ${B}^{\rm (C)}_1$
and ${B}^{\rm (C)}_2$ merge at the first turning point of temperature
$\eta_c$. 

(ii) The branch  ${B}^{\rm (H)}_1$ corresponding to the
ensemble of the first intersections for $\alpha_{\rm min}\le \alpha\le
\alpha_*$ forms the left part of the main branch of the caloric curve
$\eta(\Lambda)$ while the branch  ${B}^{\rm (C)}_1$ corresponding
to the
ensemble of the first intersections for $\alpha_{*}\le \alpha\le
\alpha_{\rm max}$ forms the right part of the main branch of the
caloric curve $\eta(\Lambda)$. They are represented in black in Fig.
\ref{branches_new2}. These two branches meet at $\alpha=\alpha_*$. We
have numerically observed,
however, that $\alpha_*$ does not correspond to the
point at which the density contrast is minimum (see
Figs. \ref{ZLambda_R_N01a_unified} and \ref{Zeta_R_N01a_unified}).

(iii) Only the lower part of Fig.  \ref{ZPhi0_alpha_N01_unified}  and only the
upper part of Fig.  \ref{Zb0_alpha_N01b_unified} are
``relevant'' in the sense explained in footnote 51. We note that a part
of the second intersection (plotted in red) lies in the irrelevant region. This
corresponds to the dashed red curves that
superimpose the main  (black) branch of the caloric curve on Fig.
\ref{branches_new2} (we also see this irrelevant red branch on the right side
of
Figs. \ref{ZLambda_R_N01a_unified} and \ref{Zeta_R_N01a_unified}). 
The corresponding energy density profiles coincide except at the very center
where the irrelevant (unstable) solution has a
very high density (see Fig. \ref{profiles_isoerg2}).

{\it Remark:} We have seen [see comment (i) above] that the maximum of the
curve 
$N_{\alpha}(\Phi_0)$ with $\alpha<\alpha_*$ determines the point
$\Lambda_{\rm min}(N)$ of the caloric curve $\eta(\Lambda)$ corresponding to
$N={\cal
N}(\alpha)$. Antisymmetrically, the maximum of the curve  $N_{\alpha}(\Phi_0)$
with $\alpha>\alpha_*$ determines the point $\eta_{c}(N)$ on the caloric
curve  $\eta(\Lambda)$ corresponding to  $N={\cal N}(\alpha)$. Therefore, it
is very easy
to obtain the curves $\Lambda_{\rm
min}(N)$ and $\eta_c(N)$ plotted in Figs. \ref{phase_lambda1} and
\ref{phase_eta1}. For each value of $\alpha$, we 
determine the values of $N$, $\eta$ and $\Lambda$ corresponding to the maximum
of the curve $N_{\alpha}(\Phi_0)$. For $\alpha<\alpha_*$,  they
determine  $N(\alpha)$ and  $\Lambda_{\rm min}(\alpha)$. For
$\alpha>\alpha_*$, they determine $N(\alpha)$ and 
$\eta_c(\alpha)$. By running
$\alpha$ from $-\infty$ to $\alpha_*$  we obtain the curve $\Lambda_{\rm
min}(N)$. On the other hand, by running
$\alpha$ from $\alpha_*$ to $+\infty$  we obtain the curve  $\eta_c(N)$.
Unfortunately, it does not seem possible to obtain the values
of $\eta_{\rm min}$ and $\Lambda_c$ by a simple graphical construction based on
the curve $N_{\alpha}(\Phi_0)$. Therefore, the curves $\eta_{\rm
min}(N)$ and $\Lambda_c(N)$  plotted in Figs. \ref{phase_lambda1} and
\ref{phase_eta1} must be obtained by plotting
the caloric curve $\eta(\Lambda)$ for any value of $N$ and determining the
values of $\eta_{\rm min}$ and $\Lambda_c$ ``by hand'' (in practice
numerically) from these curves.

\subsubsection{$N'_S<N<N_S$}

Let us now consider the case $N'_S<N<N_S$. We recall that $N'_S$ corresponds to
the first minimum of $N_{\alpha_*}(\Phi_0)$. The difference with the previous
case is that the second and third intersections can never merge. As a result,
the spirals will not be complete. They will be amputed (truncated) and touch
each other as shown in Fig. \ref{kcal_N0130_linked}.

\subsubsection{$N_S<N<N_{\rm max}$}

Let us finally consider the case $N_S<N<N_{\rm max}$.  We recall that $N_S$
corresponds to
the second maximum of $N_{\alpha_*}(\Phi_0)$. The difference with the previous
case is that there can be at most two
intersections between the curve $N(\Phi_0)$ and the line level $N$. As a result
there is no spiralling behavior. This is why the caloric curve $\eta(\Lambda)$
looks like a loop resembling the symbol $\infty$ as in Fig. \ref{cal0p15_new}.

\end{document}